\documentclass[aps,prl,amsmath,amssymb,twocolumn, showpacs, superscriptaddress,10pt]{revtex4-1}

\usepackage[utf8]{inputenc}

\usepackage{amsmath}
\usepackage{hyperref}
\usepackage{graphicx}
\usepackage{amsfonts}
\usepackage{amsthm}
\usepackage{cases}
\usepackage{bm}
\usepackage{comment}

\usepackage{color}
\definecolor{Blue}{rgb}{0.00, 0.00, 1.00}
\definecolor{Red}{rgb}{1.00, 0.00, 0.00}
\newcommand{\red}{\color{Red}}

\hypersetup{
    colorlinks=true,       % false: boxed links; true: colored links
    linkcolor=red,          % color of internal links (change box color with linkbordercolor)
    citecolor=blue,        % color of links to bibliography
    filecolor=magenta,      % color of file links
    urlcolor=cyan           % color of external links
}

\newcommand{\nn}{\nonumber}
\newcommand{\be}{\begin{equation}}
\newcommand{\ee}{\end{equation}}
\newcommand{\bea}{\begin{eqnarray}}
\newcommand{\eea}{\end{eqnarray}}

\newcommand{\beq}{\begin{equation}}
\newcommand{\eeq}{\end{equation}}
\newcommand{\beqn}{\begin{eqnarray}}
\newcommand{\eeqn}{\end{eqnarray}}

\newcommand{\as}{_{\alpha}}
\newcommand{\bs}{_{\beta}}

\newcommand{\kap}{_{\kappa}}

\newcommand{\intp}{{\int_{\Lambda_\ell e^{-d\ell}}^{\Lambda_\ell}{d^Dp\over(2\pi)^D}}}
\newcommand{\intqq}{{\int_{\Lambda_\ell e^{-d\ell}}^{\Lambda_\ell}{d^Dq\over(2\pi)^D}}}

\newcommand{\h}{{\vec h}}

\newcommand{\xb}{{\bf x}}

\newcommand{\qb}{{\bf q}}
\newcommand{\kb}{{\bf k}}
\newcommand{\pb}{{\bf p}}
\newcommand{\q}{{\bf q}}

\newcommand{\eps}{\epsilon}

\newcommand{\pt}{{\partial}}

\newcommand{\Gc}{{\cal G}}

\newcommand{\p}{{\bf p}}

\begin{document}

\title{Thermal buckling transition of crystalline membranes in a field}

%\author{David S. \surname{Dean}}
%\affiliation{Univ. Bordeaux and CNRS, Laboratoire Ondes et Mati\`ere  d'Aquitaine
%(LOMA), UMR 5798, F-33400 Talence, France}
%\author{Pierre Le Doussal}
%\affiliation{CNRS-Laboratoire de Physique Th\'eorique de l'Ecole Normale Sup\'erieure, 24 rue Lhomond, 75231 Paris Cedex, France}
\author{Pierre Le Doussal}
\affiliation{Laboratoire de Physique de l'Ecole Normale Sup\'erieure, ENS, Universit\'e PSL, CNRS, Sorbonne Universit\'e, Universit\'e de Paris, 75005 Paris, France}
\email{ledou@lpt.ens.fr}
\author{Leo Radzihovsky} 
\affiliation{Department of Physics,
  University of Colorado, Boulder, CO 80309}
%\affiliation{Kavli Institute for Theoretical Physics, University of California, Santa Barbara, CA 93106}
\email{radzihov@colorado.edu}

\date{\today}

\begin{abstract}
  Two dimensional crystalline membranes in isotropic embedding space
  exhibit a flat phase with anomalous elasticity, relevant e.g., for
  graphene.  Here we study their thermal fluctuations in the absence
  of exact rotational invariance in the embedding space.  An example
  is provided by a membrane in an orientational field, tuned to a
  critical buckling point by application of in-plane stresses. Through
  a detailed analysis, we show that the transition is in a new
  universality class.  The self-consistent screening method predicts a
  second order transition, with modified anomalous elasticity
  exponents at criticality, while the RG suggests a weakly first order
  transition.
\end{abstract}

%\pacs{05.40.-a, 02.10.Yn, 02.50.-r}
\pacs{64.60Fr,05.40,82.65Dp}

%05.40.-a: Fluctuation phenomena, random processes, noise, and Brownian motion 
%02.10.Yn	Matrix theory
%02.50.-r	Probability theory, stochastic processes, and statistics 

\maketitle

{\em Introduction and background.} Experimental realization of freely suspended graphene
\cite{suspendGrapheneNature2007} and other exfoliated
crystals,
%and exfoliated Van der Waals crystals,
following the 2004 pioneering works of Geim and Novoselov
\cite{Geim2004}, launched extensive research in electronic and
mechanical properties of two-dimensional crystalline
membranes\cite{GeimMacDonald,reviewRMPGraphene}. This led to a
renaissance in the statistical mechanics of fluctuating elastic
membranes, first studied in the context of soft and biological matter
three decades ago
\cite{NP,AL,CrumplingBucklingGuitter,GDLP,LRprl,LRrapid,GuitterMC,Jerusalem,Bensimon,RTtubule,LRReview}. Theoretical
interest is also motivated by the opportunity to explore
%opportunity to explore statistical mechanics of extended object,
%displaying 
the nontrivial and rich interplay between field theory and geometry \cite{Jerusalem}.

The most striking prediction is the existence of a low-temperature stable ``flat'' phase of a
tensionless {\em crystalline} membrane \cite{NP}, that spontaneously
breaks rotational symmetry of the embedding space. This is 
in 
%This is a
%highly nontrivial prediction, that, in 
stark contrast to canonical
two-dimensional field theories 
for which the 
%is seemingly in conflict with
Hohenberg-Mermin-Wagner theorems\cite{Hohenberg,MerminWagner,Coleman},
%that 
preclude spontaneous breaking of a continuous symmetry in two
dimensions. 

In such elastic membranes, in a spectacular phenomenon of
order-from-disorder, thermal fluctuations instead stiffen the
long-wavelength ($k^{-1}$) bending rigidity
$\kappa_0 \rightarrow \kappa_0 k^{-\eta}$, $\eta>0$, via a universal
power-law ``corrugation'' effect, with membrane roughness scaling as
$h_{\text rms}\sim L^\zeta$, with
$\zeta=(4-D-\eta)/2$\cite{NP,Jerusalem}, where $D$ is membrane's
internal dimension, with $D=2$ for the physical case. The resulting
anomalous elasticity is characterized by universal exponents,
$\eta,\zeta$ and $\eta_u=4-D-2 \eta$ determined exactly by the
underlying rotational invariance, with a scale dependent Young modulus
$K_0 \to K_0 q^{\eta_u}$. This was predicted, together with the values
of the exponents, by a variety of complementary methods
\cite{NP,AL,GDLP,LRprl,LRReview}. It was verified in numerical
simulations \cite{simulationsGraphene} and continues to be explored
experimentally \cite{experimentElasticModuli}.
%
%This striking phenomenon was first proposed based on a simple
%self-consistent one-loop theory\cite{NP}, and then derived in a
%controlled $\epsilon=4-D$\cite{AL} and $1/d$\cite{GDLP} expansions,
%with $D$ an internal dimension of a generalized elastic manifold,
%embedded in $d$ dimensions. A self-consistent screening approximation
%(SCSA), that builds on these expansions is believed to give most
%accurate predictions, with $\zeta \approx 0.59, \eta \approx 0.821$,
%and a universal {\em negative} Poisson ratio\cite{Poisson} of $nu
%\approx -1/3$ for a physical ($D=2,d=3$) membrane\cite{LRprl}. This
%highly-fluctuating 'flat' state is thus a ``critical phase'', where
%Goldstone mode nonlinearities are crucial to its stability, with
%critical universal properties controlled by a nontrivial infra-red
%attractive fixed point.
%
%\subsection{Stress and embedding space anisotropy}

Most theoretical studies to-date have focused on stress-free
fluctuating membranes in an isotropic embedding environment
\cite{NP,AL,GDLP,LRprl,LRrapid,RTtubule,LRReview,Gazit,MirlinPoisson1,Mouhanna1,MouhannaCrumpling,MouhannaTwoLoopFlat},
as appropriate for e.g., soft matter realizations of a membrane in an
isotropic fluid (though see interesting generalizations for spherical
shells\cite{PouloseNelsonPNAS2012,Kosmrlj}).
%Motivated by soft matter realizations, where a membrane is typically
%unconstrained, 
%fluctuating in an isotropic fluid, and by the
%aforementioned rich phenomenology of a tensionless membrane, 
%most
%theoretical studies to-date have focussed on stress-free membranes in
%an isotropic embedding environment
%\cite{NP,AL,GDLP,LRprl,LRRapid,RTtubule}.
However, many experiments on graphene and other solid-state membranes
(even some suspended ones) may be subjected to embedding space
anisotropy and/or external stresses due to the presence of a substrate
\cite{GuineaPuddlesSubstrate,GuineaPinningSubstrate,GuineaPLDSubstrate},
clamping\cite{Bowick2020Buckling,Bowick2017Clamped,
  MorshedifardKosmrljBuckling2021}, or electric and magnetic
fields\cite{BoothGeimNanoLett2008,BleesMcEuenKirigamiNature2015}. Orientational
fields could also be imposed by suspending the membrane in a nematic
solvent\cite{LehenyDiskNematic}. This was realized in Barium hexaferrite platelets by the Ljubljana
group\cite{nematicSheetsCopic,nematicSheetsSmalyukh,grapheneNematicClark}
showing that they form a ferromagnetic nematic, with membranes'
normals aligning with the nematic director and manipulatable by an
external magnetic field.  It is interesting to consider for instance
the case of an uniaxial easy axis field tending to order the
membrane's normal, and/or the application of a boundary stress
$\sigma$.
%\textcolor{red}{[remove since repeats:] In theoretical descriptions to date, rotational invariance in the embedding space is assumed.} 
%As a result, 
In all previous theoretical descriptions, the rotational invariance in
the embedding space was assumed and the response found to be
controlled by the thermal tensionless membrane fixed
point\cite{AL}. The case of weak field or stresses is treated by
simply introducing a cutoff for the isotropic critical fluctuations,
beyond a large scale $\xi \sim (\kappa/\sigma)^\nu$, that diverges
with a vanishing $\sigma$, where $\nu$ is a universal exponent that we
compute below.  Such perturbations then lead to an anomalous response,
that in the context of tension predicts a non-Hookean stress-strain
relation $\varepsilon \sim\sigma^\alpha$, with
$\alpha = (D-2+\eta)/(2-\eta) =_{D=2} \eta/(2-\eta)$.
\cite{CrumplingBucklingGuitter,GDLP,ML,RTtubule,LRReview,Mirlin,MirlinPoisson1,MirlinPoisson2}.

\begin{figure}[h]
\includegraphics[width=0.9\linewidth]{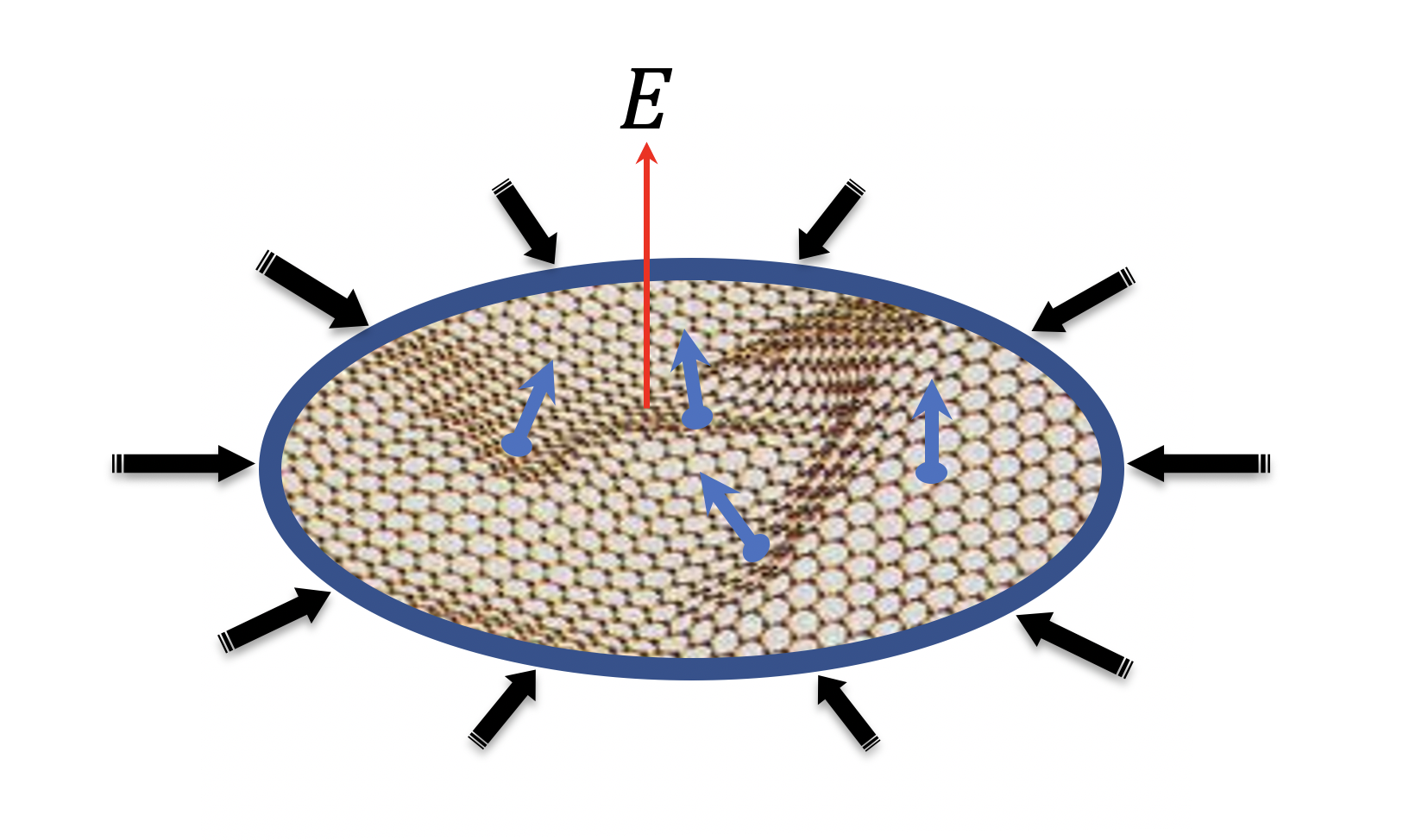}
\caption{A schematic illustration of a  critical membrane tuned to a
  buckling transition, subjected to an external in-plane isotropic
  stress $\sigma_{ij} = \frac{1}{2}\sigma\delta_{ij}$, stabilized and balanced by an external field $\vec E$, which tends to align the normals (blue vectors).}
\label{membraneBucklingTransitionFig} 
\end{figure}

In this Letter we describe such experimental geometries, illustrated
in Fig.\ref{membraneBucklingTransitionFig}, where the imposed stress
and {\em anisotropy} lead to qualitatively richer and {\em universal}
buckling phenomenology.  Generic buckling is a complex out-of-plane
instability of a sheet subjected to compression, that results in a
strongly distorted, non-perturbative state.  Recently, there has been
significant interest and progress in the study of {\em isotropic}
buckling, with focus on effects of thermal fluctuations on the
classical problem of Euler buckling, stabilized only by finite size
effects.\cite{MorshedifardKosmrljBuckling2021} Instead, here we focus
on a gentler, continuous {\em anisotropic} form of this transition,
where the instability is controlled by a stabilizing external
field. Specifically, we consider an externally oriented membrane tuned
to a buckling transition by a compressional boundary stress applied
within the plane explicitly selected by the orientational field
\cite{gentle}.  The compressive stress can be tuned to a critical
value, $\sigma_c$ to cancel out at quadratic order the
(embedding-space) rotational symmetry breaking fields. Our key
observation is that at this new buckling critical point (to which the
isotropic flat membrane critical point\cite{AL} is unstable), although
at harmonic order the membrane {\em appears} to be rotationally
invariant and stress-free, thus exhibiting strong thermal
fluctuations, it admits new important elastic nonlinearities that are
{\em not} rotationally invariant.  These lead to a critical membrane,
tuned to the buckling point, that is, thus qualitatively distinct from
the conventional tensionless membrane\cite{footnote1}.

%\subsection{Outline of the manuscript}
%
%We conclude this Introduction with the summary of our key results and
%an outline of the remainder of the paper. In Sec. 2 we introduce a
%model of a field-oriented elastic membrane in the presence of an
%in-plane stress. We demonstrate in Sec. 3 that at the buckling
%critical point the elastic nonlinearities distabilize the harmonic
%description and assess their role using a momentum-shell
%renormalization group (RG) analysis, relegating detailed treatment to
%an Appendix. In Sec. 4, we complement this RG analysis with the SCSA
%for the buckling critical point to which the stress free thermal fixed
%point is unstable, extracting the corresponding critical exponents. In
%Sec. 5, we compare our analytical predictions with the numerical study
%(about which we learned after our analysis was completed) of this very
%same buckling criticality and find excellent agreement.  We conclude
%with the summary and discussion of future directions in Sec. 6.
%
%\subsection{Results at the buckling transition}

{\em Results.} Subjecting a crystalline membrane to 
%an orienting field $E$ and 
a lateral compressive isotropic boundary stress $\sigma$, tuned
to a critical tensionless buckling point $\sigma_c$ and stabilized by an orienting field,
we find a new buckling universality class, distinct from the isotropic tensionless membrane\cite{NP,AL,GDLP,LRprl,LRReview}. 
We propose a model based on symmetry arguments, supported by more
detailed considerations. We use two complementary approaches to analyze the properties of the resulting critical state.
The first is the self-consistent screening approximation (SCSA) which was found to provide an accurate description for the isotropic case
\cite{LRprl,LRReview}. Thermal fluctuations and
elastic nonlinearities at the buckling transition lead to a {\em universal} anomalous elasticity with exponent
\begin{eqnarray} \label{eta} 
  \eta^{\rm anis}&=& 0.754,
  \end{eqnarray}
  characterizing the divergence of the effective length-scale
  dependent bending rigidity $\kappa(k)\sim k^{-\eta}$. The in-plane
  elastic moduli remain finite at the critical point, i.e.,
  $\eta_u^{\rm anis}= 0$ \cite{footnote2}. This is despite the fact
  that the five eigen-couplings $w_i(q)\sim q^{4-D-2\eta}$ renormalize
  nontrivially, vanishing in the long wavelength limit.
%{\blue [P: $\eta_u$ is not defined]}
  This is at variance with the tensionless {\em isotropic} membrane
  for which SCSA predicts universal exponents $\eta \approx 0.821$,
  $\eta_u \approx 0.358$ \cite{LRprl}. The corresponding roughness
  $h_{\text{rms}}\sim L^\zeta$ of the critically buckled membrane is
  characterized by a {\em universal} roughness exponent
\begin{equation}
\zeta^{\rm anis} = 0.623,
\end{equation}
and it is thus rougher than a tensionless isotropic membrane, with a roughness exponent $\zeta \approx 0.59$ \cite{LRprl}.

We complement this SCSA calculation by an RG analysis in an expansion in $\epsilon=4-D$.
It confirms the instability of the standard anomalous elasticity fixed
point of the isotropic, tensionless membrane, under breaking of the embedding space
rotational symmetry. Let us recall that for the isotropic membrane the elastic
nonlinearities destabilize the harmonic theory (i.e., the Gaussian fixed point) beyond the
length scale $\xi^{\rm iso}_{\rm NL} \sim (\frac{\kappa^2}{T K_0})^{\frac{1}{4-D}}$. If the anisotropy perturbation is very 
weak, e.g., $w \sim \mu_{1,2}, \lambda_{1,2} \ll K_0$ (see below for definitions of these
anisotropy parameters), the membrane still experiences the standard {\em isotropic} anomalous 
elasticity up to scales $\xi^{\rm iso}_{\rm NL}$, crossing over to the 
new anisotropic critical behavior beyond the crossover length
\begin{equation}
\xi^{\rm anis}_{NL} = \xi^{\rm iso}_{\rm NL} \left(\frac{K_0}{w}\right)^{1/\rho}\ , \quad
\rho = \frac{\epsilon d_c}{d_c+24} + O(\epsilon^2), 
\end{equation}
where $\rho$ is the crossover exponent obtained from linearization of the RG flow 
around the isotropic
fixed point. If the anisotropy perturbation is stronger, the thermal fluctuations 
and elastic nonlinearities directly destabilize the harmonic theory
at scales of order $\xi^{\rm iso}_{\rm NL}$. Beyond these scales,
the RG flows to a new stable buckling critical point, which, within the $\epsilon$-expansion,
is however accessible only for space codimension $d_c=d-D > 219$, analogous to the
crumpling transition found by Paczuski, et al.\cite{PKN}.
For the physical case, $d_c = 1$, we interpret the resulting runaway flows 
as a weakly first order transition, as for
the standard crumpling transition. 
We note that the SCSA is exact for the large $d_c$ limit, and confirm that the two methods match in  their common regime of validity.

%
%
%
%We have also analyzed the buckling critical point in a large
%co-dimension $d_c$ expansion (in powers of $1/d_c$), as well as a
%momentum-shell RG controlled by an  expansion. The
%former predicts $\eta^{1/dc} = ..., \eta_u^{1/dc} = ..., \zeta^{1/dc}
%= ...$, not too quantitativelyl trustworthy for the physical case of
%$d_c = 1$. 
%
%We also find XYZ and ABC.

%
%Here we study the fluctuations of a cristalline membrane in the absence of exact rotational
%invariance in the embedding space. We focus on the flat phase, and on 
%the critical point which describes the so-called buckling transition. 
%Our main result is that it is not described by the standard non-trivial thermal fixed
%point of the rotationally invariant membrane. 

{\em Model of anisotropic membrane buckling.} The coordinates of the atoms in the $d$-dimensional embedding space are denoted 
%{\bf [LR: let's us bf x to be more explicit. i defined slash xb command ]} 
$\vec r(\xb) \in \mathbb{R}^d$,
with the atoms labeled by their position ${\bf x} \in \mathbb{R}^D$ in the internal space.
For graphene $D=2$, and  atoms span a triangular lattice, described here in the continuum limit.
The deformations with respect to the flat sheet are described by $D$ phonon fields 
$u_\alpha(\xb)$, and $d_c=d-D$ height fields $\vec h \in \mathbb{R}^{d_c}$ (orthogonal to the $\vec e_\alpha$) 
as $\vec r(\xb) = (x_\alpha +u_\alpha(\xb)) \vec e_\alpha + \vec h(\xb)$,
where the $\vec e_\alpha$ are a set of $D$ orthonormal vectors.
%
%We first recall the rotationally invariant theory of a crystalline membrane
%of internal dimension $D$, embedded in dimension $d$. 
%
%The atoms are labeled 
%by their position $x \in \mathbb{R}^D$ in the internal space: for graphene $D=2$, and they span
%a triangular lattice, described here in the continuum limit. The coordinates of the
%atoms in the embedding space are denoted $\vec r(x) \in \mathbb{R}^d$. 
%In the flat phase the tangent fields $\vec t_\alpha(x) = \partial_\alpha \vec r(x)$ acquire {\it spontaneously} 
%a nonzero
%expectation value $\langle \vec t_\alpha(x) \rangle= \zeta_T \vec e_\alpha$, $\alpha=1,\dots,D$, 
%where the $\vec e_\alpha$ are a set of $D$ orthonormal vectors, and $\zeta_T$ the preferred
%extension.
%The deformations with respect to the ground state $\zeta_0=1$ are described by $D$ phonon fields 
%$u_\alpha(x)$, and $d_c=d-D$ height fields $\vec h \in \mathbb{R^{d_c}}$ (orthogonal to the $\vec e_\alpha$) 
%as $\vec r(x) = (x_\alpha +u_\alpha(x)) \vec e_\alpha
%+ \vec h(x)$. 
While the physical case corresponds to $d=3$ and $d_c=1$,  it is useful to study the
theory for a general $d_c$. The nonlinear strain tensor measures the deformation
of the induced metric relative to the preferred flat metric, $u_{\alpha \beta}  = \frac{1}{2} (\partial_\alpha \vec r \cdot \partial_\beta \vec r - \delta_{\alpha \beta}) \simeq 
\frac{1}{2} (\partial_\alpha u_\beta 
+ \partial_\beta u_\alpha + \partial_\alpha \vec h \cdot
\partial_\beta \vec h)$ to the accuracy needed here, 
with the $O((\partial u)^2)$ phonon nonlinearities irrelevant and therefore neglected (see below). 
The tensor $u_{\alpha \beta}$ encodes 
full rotational invariance in the embedding space, its approximate form being invariant under
infinitesimal rotations by $\theta$, i.e., the $O(\theta^2)$ term vanishes under 
the (apparent) distortion $u_1=x_1(\cos \theta-1)$, $h_1=x_1 \sin \theta$,
which corresponds to a rigid rotation, with the corresponding vanishing of the 
exact strain tensor.

Here we build on the model of a rotationally invariant tensionless membrane. Its 
Hamiltonian is the sum of curvature energy and
in-plane stretching energy
\be \label{F1} 
{\cal F}_1[\vec h,u_\alpha] = \int d^D x \, \left[ \frac{\kappa}{2}  (\partial^2 \vec h)^2 +
\tau u_{\alpha \alpha} + \mu (u_{\alpha \beta})^2 + \frac{\lambda}{2} (u_{\alpha \alpha})^2\right]
\ee
where $\kappa$ is the bending modulus, $\lambda,\mu$ the in-plane Lam\'e elastic constants.
The parameter $\tau$ controls the preferred extension of the membrane in the
$\vec e_\alpha$ plane. 

%
%arises when the extension $\zeta$ is not equal to its preferred
%value $\zeta_T$, $\tau>0$ corresponds to a stretched membrane. 

%Incorporating the effect of the orientational field and external stresses,
%from considerations of symmetry,
%\textcolor{red}{
Based on symmetry considerations, complemented by a model-building 
derivation (presented at the end of the paper), external orientational and boundary stresses introduce new relevant elastic nonlinearities, with five new independent couplings, that by symmetry lead to a modified
effective Hamiltonian 
${\cal F}= {\cal F}_1 + {\cal F}_2$, where ${\cal F}_2$ breaks rotational invariance in the embedding space,
\bea \label{F2} 
&& {\cal F}_2[\vec h,u_\alpha] = \int d^D x \, \bigg( 
\frac{\gamma}{2} (\partial_\alpha \vec h)^2 \\
&& + \frac{\lambda_1}{2} \partial_\alpha u_\alpha (\partial_\beta \vec h)^2 + \frac{\lambda_2}{8} [(\partial_\alpha \vec h)^2]^2 \nonumber \\
&& + \mu_1 \partial_\alpha u_\beta (\partial_\alpha \vec h \cdot \partial_\beta \vec h) + \frac{\mu_2}{4} [\partial_\alpha \vec h \cdot \partial_\beta \vec h]^2  \bigg), \nonumber 
\eea
retaining in-plane isotropy and the $h \to - h$ invariance as a feature of
our geometry, preserving the equivalence between the two sides of the membrane. 

We now study the membrane with parameters tuned to the thermal buckling critical point
defined by the renormalized $\gamma_R=0$. Integrating over the in-plane phonon modes $u_\alpha$
and, rescaling for convenience all elastic constants by $1/d_c$, we obtain an
effective Hamiltonian for the height field,
\bea \label{Fh} 
&& {\cal F}[\vec h] = \int d^D x \, [ \frac{\kappa}{2}  (\partial^2 \vec h)^2 + \frac{\gamma}{2}  (\partial_\alpha \vec h)^2 ]
+ \frac{1}{4 d_c} \int d^D x \, d^D y  \nonumber \\
&& \,\,\,\, \times 
\partial_\alpha \vec h(\xb) \cdot \partial_\beta \vec h(\xb) \,
R_{\alpha \beta,\gamma \delta}(\xb-{\bf y}) \, \partial_\gamma \vec h({\bf y}) \cdot \partial_\delta \vec h({\bf y}), 
\eea 
with a non-local quartic tensorial interaction, which in Fourier space is given by\cite{footnoteFisherShankaNelsonRG} 
\be
R_{\alpha \beta,\gamma \delta}(\qb) = \sum_{i=1}^5 w_i \, (W_i)_{\alpha \beta,\gamma \delta}(\qb).
\ee 
The $W_i$ are five projectors in the space of four index tensors, equal to bilinear combinations of
longitudinal $P^L_{\alpha \beta}(\qb)= q_\alpha q_\beta/q^2$ and transverse $P^T(\qb)=\delta_{\alpha \beta}- P^L_{\alpha \beta}(\qb)$ projectors on the wave vector $\qb$. The five "bare couplings" $w_i$ 
are given in the Supplementary Material (SM) \cite{SM} 
in terms of the bare elastic moduli in \eqref{F1} and \eqref{F2}, 
together with the basis tensors $W_i$. \cite{footnote3}
The important features are the following. When rotational symmetry breaking is 
absent, $\gamma=0$, $\mu_1=\mu_2=\lambda_1=\lambda_2=0$, the couplings $w_2,w_4,w_5$ vanish and
\be \label{w13mu} 
w_1= \mu \quad , \quad w_3 = 
\mu + (D-1) \frac{\mu \lambda}{\lambda+ 2 \mu}, 
\ee
leading to (the $\qb$ dependence suppressed)
\be
R_{\alpha \beta,\gamma \delta} = (w_3 - w_1) P_{\alpha \beta}^T
P_{\gamma \delta} ^T + w_1 \frac{1}{2} (P_{\alpha \gamma}^T P_{\beta \delta}^T
+ P_{\alpha \delta}^T P_{\beta \gamma}^T),
\ee
which is the usual quartic coupling associated to ${\cal F}_1$.
When $\lambda_1$ and $\lambda_2$ are turned on, while
$\mu_1=\mu_2=0$, all $w_i$ are nonzero except $w_2=0$.
Finally, when all couplings in ${\cal F}_2$ are nonzero, 
all $w_i$ are nonzero.

{\em SCSA analysis.} The form \eqref{Fh} is suitable to apply the SCSA method,
which is exact in the limit of large $d_c$. The calculation is
performed in the SM \cite{SM} and parallels the one in Section IV. A of \cite{LRReview}.
Consider the two point correlation of the height field in Fourier space, $\langle h^i(\kb) h^j(\kb') \rangle
= {\cal G}(k) (2 \pi)^d \delta^d(\kb+\kb') \delta_{ij}$. If we neglect the quartic nonlinearities in \eqref{Fh} we find ${\cal G}(k)=G(k)=1/(\gamma k^2 + \kappa
k^4)$. The nonlinearities lead to a nonzero self-energy $\Sigma(k)={\cal G}(k)^{-1} - \gamma k^2 - \kappa k^4$.
Together with the renormalized interaction tensor, $\tilde R(\qb)$,
it satisfies the SCSA equations 
\bea \label{sigma} 
&& \Sigma(k) = \frac{2}{d_c} \int_q k_\alpha (k_\beta - q_\beta) 
(k_\gamma - q_\gamma) k_\delta \tilde R_{\alpha \beta,\gamma \delta}(\qb)
{\cal G}(\kb-\qb)  \nonumber \\
&& \\
&& \tilde R(\qb) = R(\qb) - R(\qb) \Pi(\qb) \tilde R(\qb)  \label{R}
\eea
where $\Pi(\qb)$ encodes the screening of the in-plane elasticity
by out of plane fluctuations
\be
\Pi_{\alpha \beta, \gamma \delta}(\qb) = \frac{1}{4} \int_p v_{\alpha \beta}(\qb,\qb-\pb) 
v_{\gamma \delta}(\qb,\qb-\pb)
{\cal G}(\pb) {\cal G}(\qb-\pb) 
\ee 
and $v_{\alpha \beta}(\pb,\pb')=p_\alpha p'_\beta + p'_\alpha p_\beta$. 
One can decompose $\Pi(\qb) = \sum_{i=1}^5 \pi_i(q) W_i(\qb)$
and $\tilde R(\qb)=\sum_{i=1}^5 \tilde w_i(q) W_i(\qb)$, where $\tilde w_i(q)$ are the momentum dependent renormalized 
couplings. Looking for a solution which behaves at small $k$ as
${\cal G}(k) \simeq Z_\kappa^{-1}/k^{4-\eta}$,
and evaluating the integrals $\pi_i(q)$ \cite{SM} one finds that they diverge
at small $q$ as $\pi_i(q) \simeq Z_\kappa^{-2} a_i q^{-(4-D - 2 \eta)}$
where $a_i=a_i(\eta,D)$. From \eqref{R} we find that the renormalized couplings are
softened at small $q$ as $\tilde w_i(q) \propto Z_\kappa^2 c_i q^{\eta_u}$, with 
$\eta_u=4-D - 2 \eta$ and $c_i=1/a_i$ for $i=1,2$ and 
\bea
&& \begin{pmatrix} c_3 & c_4 \\
c_4 & c_5 \end{pmatrix} 
\simeq \begin{pmatrix} a_3 & a_4 \\ a_4 &
  a_5 \end{pmatrix}^{-1}
\label{inverse2} 
\eea 
Inserting this into the self-energy equation \eqref{sigma} and
performing the integrals we find that the factors of $Z_\kappa$ cancel and the
self-consistent equation, which implicitly determines $\eta$ as a function of $D,d_c$ is given by
\begin{equation} \label{SCSA}
\frac{d_c}{2} = \sum_{i=1,2} {b_i \over a_i } + { b_3 a_5 -b_4 a_4
+b_5 a_3 \over {a_3 a_5 - a_4^2 }},
\end{equation}
where $b_i=b_i(\eta,D)$ are self-energy integrals, given 
with the $a_i(\eta,D)$ in the SM. Note that here we have considered
the case where all bare couplings $w_i$ are nonzero. For
a physical membrane, $D=2$, \eqref{SCSA} 
reduces to finding the root of a cubic equation
\bea \label{cr2} 
d_c= \frac{24 (\eta -1)^2 (2 \eta +1)}{(\eta -4) \eta  (2 \eta -3)}.
\eea
For $d_c=1$ we obtain our main result \eqref{eta}.
For large $d_c$ we find $\eta = 2/d_c + O(1/d_c^2)$. 
The roughness of a size $L$ membrane is characterized
by $h_{\rm rms} = \langle h^2 \rangle^{1/2} \simeq L^\zeta$
where $\zeta=(4-D-\eta)/2$. Hence for $d_c=1$ we find 
$\zeta=0.623$.

One can define renormalized amplitude ratio as
\bea
\lim_{q \to 0} \frac{\tilde w_i(q)}{\tilde w_j(q)} = \frac{c_i}{c_j} 
\eea 
for any pair $(i,j)$ such that the bare couplings $w_i,w_j$ are nonzero.
Near $D=4$ we find that these renormalized couplings take values such that
the interaction energy becomes $ \frac{v_1}{2} [(\partial_\alpha \vec h)^2]^2 + v_2 (\partial_\alpha \vec h
\cdot \partial_\beta \vec h)^2$, i.e., local in the fields $\partial_\alpha \vec h$.
This property however does not hold for $D<4$, e.g., one finds
$c_2/c_1= (D+\eta-2)/(2-\eta)$ instead of unity for $D=4$, $\eta=0$. 
Thus the critical point requires a fully non-local five coupling description.
The $c_i$ are given in \cite{SM}. In the physical case of $D=2$ and $d_c$=1 we find
\be
c_i = \{\frac{1}{2},0.302,0.338,-0.029,0.173\},
\ee
and the {\em universal} $\lambda/\mu=-0.978$ and the Poisson ratio
(not to be confused with external stress),
\begin{equation}
\sigma^{\rm anis}=-0.968,
\end{equation} 
to be contrasted with $\sigma = -1/3$ for an isotropic tensionless membrane\cite{LRprl,LRReview}

There are other fixed points that lie in the invariant subspaces
of the SCSA equations. The rotationally invariant membrane corresponds to bare couplings $w_2=w_4=w_5=0$. The corresponding renormalized couplings also vanish, which amounts to $b_2=b_4=b_5=0$ 
in \eqref{SCSA}, leading to 
\begin{equation} \label{SCSA2}
\frac{d_c}{2} = \frac{b_1}{a_1} + { b_3  \over {a_3   }},
\end{equation}
which is precisely the SCSA equation for the anomalous flat phase of
the isotropic membrane, leading for $D=2$ to $\eta=\frac{4}{d_c+ \sqrt{16-2 d_c + d_c^2}}$,
and $\eta \simeq 0.821$, $\zeta=0.590$ for $d_c=1$ \cite{LRprl,LRReview}. Near $D=4$
one recovers $\eta=\frac{12}{d_c+24} \epsilon+ O(\epsilon^2)$ from the Aronovitz-Lubensky's $\epsilon$-expansion\cite{AL}. Another fixed manifold is 
$w_2=0$, corresponding to a choice of bare couplings so that
$(\mu+\mu_1)^2=\mu(\mu+\mu_2)$, which includes the choice
$\mu_1=\mu_2=0$, leading to $\tilde w_2(q)=0$ and 
\begin{equation} \label{SCSA3}
\frac{d_c}{2} = {b_1 \over a_1 } + { b_3 a_5 -b_4 a_4
+b_5 a_3 \over {a_3 a_5 - a_4^2 }}.
\end{equation}
This leads to yet another fixed point with slightly different exponents. For 
$D=2$ and $d_c=1$ we find $\eta=0.854$ 
and $\zeta=0.573$. Near $D=4$ we find $\eta=\frac{18}{d_c+36} \epsilon+ O(\epsilon^2)$.
Universal amplitude ratios have $c_2=0$. 

{\em RG analysis.} As a nontrivial check and for further insight, we have complemented this SCSA calculation and results using an RG analysis, controlled by an $\epsilon=4-D$ expansion near $D=4$.
We have calculated the one-loop corrections to the Hamiltonian \eqref{Fh}
and obtained the RG equations for the five dimensionless couplings
$\hat w_i = w_i/\kappa^2 C_4 \Lambda_\ell^{-\epsilon}$ of the form
$\partial_\ell \hat w_i = \epsilon \hat w_i + a_{ijk} \hat w_j \hat w_k$,
where the $a_{ijk}$ and details of the calculation are given in \cite{SM}. 
The anomalous dimension of
the out-of-plane height field $h$ defines the exponent $\eta$ given by
\bea
\eta = \frac{1}{12} \left(10 \hat w_1-18 \hat w_2+5 \hat w_3+3 \hat w_5-6 \hat w_{44}\right),
\eea 
with $\hat w_{44} = \sqrt{3} \hat w_4$, and evaluated at the fixed point
of interest $\hat w_i^*$ (see below). 
The anomalous dimension of
the phonon field is given by
\be
\eta_u = \frac{1}{12} (\hat w_1- \hat w_2). 
\ee
The isotropic membrane 
corresponds to the space $\hat w_2=\hat w_4=\hat w_5=0$, which is 
preserved by the RG flow and along which
\bea 
&& \partial_\ell \hat w_1 = -\frac{1}{12} \hat w_1 \left((d+20) \hat w_1+10
   \hat  w_3\right), \\
&&   \partial_\ell \hat w_3 =
   -\frac{5}{24} \hat  w_3 \left((d+4) \hat w_3+8
   \hat  w_1\right). 
\eea
The isotropic membrane fixed point is 
$\hat w^*_1=\frac{12 \epsilon}{d+24}$, $\hat w^*_3=\frac{24 \epsilon}{5 (d+24)}$,
corresponding to $\hat \mu^*=\frac{12 \epsilon}{24 + d}$, $\hat \lambda^*=\frac{-4 \epsilon}{24 + d}$
\cite{AL}.
Diagonalizing the RG flow for $\hat w_i = \hat w_i^* + \delta \hat w_i$ 
around this fixed point in the larger space of five couplings 
shows that, in addition to the two negative eigenvalues $-1$ and $-\frac{d_c}{d_c+24}$
within the plane $\delta \hat w_{1,3}$ of the isotropic membrane,
(i) there is a marginal direction mixing $\delta \hat w_{1,3,4}$ (eigenvalue $0$)
(ii) there are two unstable directions with eigenvalues $\frac{d_c}{d_c+24}$ 
with $\delta w_{2,5}$ nonzero (in the large $d_c$ limit this eigenspace is
purely along $\delta w_{2,5}$). Hence, consistent with the SCSA findings, the isotropic membrane
fixed point is unstable to anisotropy of the orientational field and external boundary stress. 

To determine where the general flow goes we searched for attractive fixed points
of the RG equations. We found one such fixed point in the subspace of
couplings $\hat w_i$ at which, the interaction energy is fully local in the
gradients $\partial_\alpha \vec h$ and parameterized by two couplings 
$v_1,v_2$ as defined above. This subspace is preserved by the RG
and also arises in the study of the crumpling transition. In fact the
RG flow within this subspace is identical to the one obtained in
\cite{PKN} with $d$ replaced by $d_c$. It admits a stable FP for $d_c>219$. Here we demonstrated that
this FP is fully attractive in the space of the five couplings. 
Hence the RG approach is consistent, around $D=4$, with the SCSA (which is exact
for large $d_c$ and any $D$), predicting a new fixed point
for membrane in anisotropic embedding space. For the physical
membrane $D=2$ and $d_c=1$, while the SCSA
predicts this new "anisotropic buckling transition" to be continuous,
the RG, if extrapolated from $D=4$, suggests a weakly first order transition,
as argued for the crumpling transition \cite{PKN,Mouhanna1,MouhannaCrumpling}.

To reach the new anisotropic buckling critical point requires tuning $\gamma=\gamma_c$, so
that $\gamma_R=0$. Slightly away from criticality the  correlation
length is long but finite, $\xi \sim |\delta \gamma|^{-\nu}$, diverging with a vanishing $\delta \gamma=\gamma-\gamma_c$.
Linearizing the RG flow around the fixed point yields $\delta \gamma(L) \sim \delta\gamma L^\theta$,
where $\theta = - \frac{\epsilon}{d_c} (1 - \frac{66}{d_c} + O(\frac{1}{d_c^2}) )$, see the
%numerical values for the exponent $\theta$ are given in 
SM \cite{SM}.
By balancing $\kappa(\xi) \xi^{-4} \sim \delta \gamma(\xi)  \xi^{-2}$
and using that $\kappa(\xi) \sim \xi^{\eta}$ we obtain the correlation
length exponent as $\nu = 1/(2 + \theta - \eta)$.

%Discussion on the other exponents (response to
%rotation).{\blue P: what did we have in mind there?}.

{\em Model development.}
Until now we argued for the model (\ref{F1},\ref{F2}) based on symmetry considerations. Here, as illustrated in Fig.\ref{membraneBucklingTransitionFig},
%\ref{bucklingMembraneFig}, 
we develop
an explicit model of a membrane undergoing buckling in the absence of rotational invariance in the embedding space. We consider an elastic membrane in an external field $\vec E$ (taken
along the z-axis) that aligns the membrane's normal $\hat n$ along the
field. We thus expect the energy density to be a monotonic function of
$\hat n\cdot\vec E$, namely of the %tilt $\nabla h$ and the corresponding
small tilt angle $\theta$, %of the membrane's normal away from the preferred z-axis,
\begin{eqnarray}
{\cal H}_{\text orient} &=& \frac{\alpha_1}{2}\theta^2 +
\frac{\tilde \alpha_2}{4}\theta^4+\ldots,
%\\
%&\approx& \frac{\alpha_1}{2}(\nabla h)^2 +
%\frac{\alpha_2}{4}(\nabla h)^4+\ldots,
\end{eqnarray}
with $\alpha_1 > 0$, $\tilde\alpha_2 > 0$.
Combining this orientational field energy with the Hamiltonian for an
elastic membrane\cite{Jerusalem,LRReview}, subjected to an in-plane
compressional boundary stress $\sigma > 0$, isotropic in the membrane's xy plane, and using
that, to lowest order $\theta \sim |\partial_\alpha h|$, we
obtain,
\begin{eqnarray}
{\cal H} &=& \frac{\kappa}{2}(\partial^2 h) + \mu u_{\alpha\beta}^2 +
\frac{\lambda}{2} u_{\alpha\alpha}^2 + \sigma\partial_\alpha u_\alpha\nonumber\\
&& + \frac{\alpha_1}{2}(\partial_\alpha h)^2 + \frac{\alpha_2}{4}(\partial_\alpha h)^4+\ldots.
\end{eqnarray}

We note that the external stress, $\sigma$ is an in-plane boundary term, that induces a stress-dependent inward displacement of the membrane's edges. Observing that $\sigma \partial_\alpha u_\alpha = \sigma u_{\alpha \alpha} 
-\frac{1}{2} \sigma(\partial_\alpha h)^2$, the rotationally invariant
strain component $\sigma u_{\alpha\alpha}$  can be accommodated
by simply changing the preferred extension of the membrane without 
breaking the embedding space
rotational symmetry (i.e., it amounts to a redefinition of the parameter $\tau$ in 
\eqref{F1}, which determines the preferred membrane's projected area
\cite{footnote5}). %based on the structure of the nonlinear strain
%$u_{\alpha\alpha} = \nabla\cdot u + \frac{1}{2}(\nabla h)^2$,
%that, 
%with otherwise free boundary conditions, 
The negative in-plane strain $\partial_\alpha u_\alpha$ induced by positive $\sigma$ can be relieved by a membrane tilt, $(\partial_\alpha h)^2>0$, stress-free in the actual plane of the
membrane. The lowering of the energy associated with the membrane tilt
is then given by the second term, i.e., ${\cal H}_\sigma = -\frac{1}{2} \sigma(\partial_\alpha h)^2$, which, neglecting bending energy and boundary
conditions, is unbounded, since tilt is unconstrained in the absence of the
orientational field. Putting these ingredients together and rescaling xy
coordinate system, we obtain the Hamiltonian governing a buckling
transition of a membrane in an orientational field,
\begin{eqnarray}
{\cal H} &=& \frac{\kappa}{2}(\partial^2 h) + \mu u_{\alpha\beta}^2
+ \frac{\lambda}{2} u_{\alpha\alpha}^2 \nonumber\\
&& + \frac{\gamma}{2}(\partial_\alpha h)^2 + \frac{\alpha_2}{4}(\partial_\alpha h)^4+\ldots,
\end{eqnarray}
where $\gamma = \alpha_1 - \sigma$ is the critical parameter which can
be tuned to $\gamma_c$ to reach the buckling transition (with
$\gamma_c=0$ at $T=0$), studied in here.  As detailed in SM, we can
estimate the buckling stress $\sigma_c$ based on a model of
homeotropic alignment of a membrane in a nematic solvent (using
typical values of Frank elastic constants)\cite{LehenyDiskNematic} and a model of a
ferroelectric membrane aligned by an electric field. These give
$\sigma_c\sim 1 - 10 eV/\mu m^2$, with the thermal fluctuation
corrections to $\gamma$ that we show in SM to be subdominant.

%\begin{figure}[hb]
%%\includegraphics[angle=-90,width=0.8\linewidth]{density_left2.pdf}
%%\includegraphics[width=0.8\linewidth]{density_right.pdf}
%\includegraphics[width = \linewidth]{new_combined_fig.pdf}
%{\caption{{\it Left:} Solid line}\label{Fig_density}}  
%\end{figure}

{\em Conclusion.} To summarize, in this Letter, in contrast to
previous works on tensionless crystalline membranes, we studied a
thermal elastic sheet tuned by an external boundary stress to a
critical point of a buckling transition, stabilized by an
orientational field. We find that this breaking of embedding
rotational symmetry has profound effects, and leads to a new class of
anomalous elasticity, that we have explored in detail here using the
SCSA and RG analyses. With much recent interest in elastic sheets,
most notably graphene and other van der Waals monolayers, we hope that
our predictions will stimulate further experiments to probe the rich
universal phenomenology predicted here for an elastic membrane tuned
to a buckling transition in an anisotropic environment.  We also expect that ideas explored here can be extended to a richer class of anomalously elastic media.\cite{LRelastomer, footnoteFutureLR}

{\em Note Added:} We have recently became aware of an ongoing work by S. Shankar and D. R. Nelson on a membrane subjected to a boundary stress or strain, which, in contrast to our work only breaks embedding rotational symmetry at the boundary. 

{\it Acknowledgments.} We thank John Toner, David Nelson and Suraj Shankar for enlightening discussions. LR also acknowledges support by the NSF grants MRSEC
DMR-1420736, Simons Investigator Fellowship, and thanks
\'Ecole Normale Sup\'erieure for hospitality. 
PLD acknowledge support from ANR under the grant 
ANR-17-CE30-0027-01 RaMaTraF. Both authors thank KITP for hospitality.
This research was supported in part by the National Science 
Foundation under Grant No. NSF PHY-1748958.

{}

%\end{document}
\newpage

%\appendix

\newpage

\begin{widetext} 

\bigskip

\bigskip

\begin{large}
\begin{center}

Supplementary Material for {\it Thermal buckling transition of crystalline membranes}

\end{center}
\end{large}

\bigskip

We give the principal details of the calculations described in the main text of the Letter. 

\bigskip

\medskip
\begin{center}
{\bf A. Projectors and tensor multiplication}
\end{center}
\medskip

Here we consider four index tensors, such as $R_{\alpha \beta,\gamma \delta}(\qb)$ 
introduced in the text, which are 
symmetric in $\alpha \leftrightarrow \beta$, in $\gamma \leftrightarrow \delta$ and in 
$(\alpha,\beta) \leftrightarrow (\gamma,\delta)$. The product of such tensors
is defined as $(T \cdot T')_{\alpha \beta,\gamma \delta}=T_{\alpha \beta,\gamma' \delta'}
T'_{\gamma' \delta',\gamma \delta}$, the identity being $
I_{\alpha \beta,\gamma \delta} = {1\over 2}(\delta_{\alpha\gamma}\delta_{\beta\delta}
+\delta_{\alpha\delta}\delta_{\beta\gamma})$. We recall the definition \cite{LRReview}
of the five "projectors"
$W_i$, $i=1,\ldots,5$, which span the space of such four index tensors
\bea
&& (W_3)_{\alpha\beta,\gamma\delta}(\qb)
={1 \over {D-1}} P^T_{\alpha \beta} P^T_{\gamma \delta}\;,\;\;\;
(W_5)_{\alpha\beta,\gamma\delta}(\q)=P^L_{\alpha \beta}P^L_{\gamma\delta}\;,\\
&& 
(W_4)_{\alpha\beta,\gamma\delta}(\q) = (W_{4a})_{\alpha\beta,\gamma\delta}(\q) 
+ (W_{4b})_{\alpha\beta,\gamma\delta}(\q)\;,\\
&& (W_{4a})_{\alpha\beta,\gamma\delta}(\q) = {1 \over {\sqrt{D-1}}} P^T_{\alpha \beta} P^L_{\gamma \delta}\;,
\quad 
(W_{4b})_{\alpha\beta,\gamma\delta}(\q) ={1 \over {\sqrt{D-1}}} P^L_{\alpha \beta} P^T_{\gamma \delta}\;,\\
&& (W_2)_{\alpha\beta,\gamma\delta}(\q)=
{1 \over 2}(P^T_{\alpha \gamma} P^L_{\beta \delta } 
+ P^T_{\alpha \delta} P^L_{\beta \gamma} + P^L_{\alpha \gamma}
P^T_{\beta \delta } + P^L_{\alpha \delta} P^T_{\beta \gamma})\;, \\
&& 
W_1(\q)={1\over 2}(\delta_{\alpha\gamma}\delta_{\beta\delta}
+\delta_{\alpha\delta}\delta_{\beta\gamma})
- W_3(\q) - W_5(\q) - W_2(\q)\;,
\label{(3.6)} \\
&& W_1(\q) + W_3(\q) = {1\over 2}(P^T_{\alpha\gamma}(\q) P^T_{\beta\delta}(\q)
+P^T_{\alpha\delta}(\q) P^T_{\beta\gamma}(\q))
\eea 
where $P^T_{\alpha \beta}=\delta_{\alpha \beta} -
q_{\alpha}q_{\beta}/q^2$ and $P^L_{\alpha
  \beta}=q_{\alpha}q_{\beta}/q^2$ are the standard transverse and
longitudinal projection operators associated to $\q$. The first two
projectors $W_1, W_2$ are mutually orthogonal and orthogonal to the
other three. Note that while $R$, being symmetric, can be expressed in
terms of the symmetric tensors $W_i$, $i=1,..5$, we will need at some
intermediate stages of the calculations some products (such as $\Pi*R$
see below), which are not symmetric. Hence we introduced $W_4^a$ and
$W_4^b$, which together with $W_i$, $i=1,2,3$ and $W_5$ make the
representation complete under tensor multiplication. The rules for the
tensor multiplication $T''=T'*T$ of the tensors $T =
\sum_{i=1}^3 w_i W_i + w_{4a} W_{4a} + w_{4b} W_{4b} + w_{5} W_{5}$
and $T' = \sum_{i=1}^3 w'_i W_i + w'_{4a} W_{4a} + w'_{4b} W_{4b} +
w'_{5} W_{5}$ are 
\bea
w_1''=w'_1 w_1\;,\;\;\; w_2'' = w'_2 w_2\;,\;\;\;
\begin{pmatrix}
w_3''&w_{4a}''\\ w_{4b}''&w_5'' 
\end{pmatrix}
=
\begin{pmatrix}
w'_3&w_{4a}' \\
w_{4b}' &w'_5 
\end{pmatrix}
\begin{pmatrix}
w_3&w_{4a}\\
w_{4b}&w_5
\end{pmatrix}\;,
\label{3.5}
\eea 
with $T'' = \sum_{i=1}^3 w''_i W_i + w''_{4a} W_{4a} + w''_{4b}
W_{4b} + w''_{5} W_{5}$.  \\

\medskip
\begin{center}
{\bf B. Integration over in-plane deformations}
\end{center}
\medskip

The integration over the phonon fields $u_\alpha(x)$ of the Gibbs measure $\sim e^{-{\cal F}[\vec h,u_\alpha]/T}$, with ${\cal F}= {\cal F}_1 + {\cal F}_2$ given by \eqref{F1} and \eqref{F2} 
leads to the Gibbs measure $\sim e^{-{\cal F}[\vec h]/T}$ for the height fields
with an effective Hamiltonian of the form \eqref{Fh} in the text (we set $\tau=0$). 
To perform it we use
a method slightly different from the one in e.g. \cite{LRReview} Section III B. 
Let us introduce the elastic matrix
\be \label{elast} 
C^{\mu,\lambda}_{\alpha \beta, \gamma \delta} = \lambda \delta_{\alpha \beta} \delta_{\gamma \delta} + \mu( 
\delta_{\alpha \gamma} \delta_{\beta \delta}  + \delta_{\alpha \delta} \delta_{\beta \gamma} )
\ee 
and denote $\tilde u_{\alpha \beta}= \frac{1}{2} (\partial_\alpha u_\beta + \partial_\beta u_\alpha)$ and
$A_{\alpha \beta} = \frac{1}{2} \partial_\alpha \vec h \cdot \partial_\beta \vec h$.
We then rewrite the model ${\cal F}= {\cal F}_1 + {\cal F}_2$ as
\bea \label{FC0} 
{\cal F}[u,\vec h]= \int d^D x \, \, \bigg[ \frac{\kappa}{2} (\nabla^2 h)^2 + \frac{1}{2} C^{\mu,\lambda}_{\alpha \beta, \gamma \delta} \tilde u_{\alpha \beta} \tilde u_{\gamma \delta}
+ \tilde u_{\alpha \beta} C^{\mu+\mu_1,\lambda+\lambda_1}_{\alpha \beta, \gamma \delta} A_{\gamma \delta} + 
\frac{1}{2} C^{\mu+\mu_2,\lambda+\lambda_2}_{\alpha \beta, \gamma \delta} A_{\alpha \beta} A_{\gamma \delta} \bigg]
\eea 
We must treat separately the contributions of the in plane strains which are uniform (i.e. with zero momentum), and those with nonzero wavevector, i.e. the phonons.

\subsection{B.1 Phonon integration: nonzero wavevector}. 

We recall the phonon field propagator 
\be \label{D}
\langle u_{\alpha}(\q) u_{\beta}(\q') \rangle = T (2 \pi)^D \delta^D(\q+\q')  \left(\frac{P_{\alpha \beta}^T(\q)}{\mu q^2} 
+ \frac{P_{\alpha \beta}^L(\q)}{(2 \mu + \lambda) q^2}\right) 
\ee 
from which the in-plane strain correlator at nonzero wavevector is obtained as
\be
\langle \tilde u_{\alpha \beta}(\q) \tilde u_{\gamma \delta}(\q') \rangle = T (2 \pi)^D \delta^D(\q+\q') D_{\alpha \beta, \gamma \delta}(\q)
\ee 
with, for $\q \neq 0$,
\be
D_{\alpha \beta, \gamma \delta}(\q)= \frac{1}{4} \left[\hat q_\alpha \hat q_\gamma \left(\frac{P_{\beta \delta}^T(\q)}{\mu} 
+ \frac{P_{\beta \delta}^L(\q)}{2 \mu + \lambda}\right) + \text{3 permutations} \right] 
\ee
This tensor has a simple expression in terms of the projectors (suppressing the indices and the $q$ dependence)
\be \label{DW} 
D = \frac{1}{2 \mu} W_2  + \frac{1}{2 \mu+\lambda} W_5 
\ee
The integration over the phonon field in \eqref{FC0} using \eqref{D} is then a simple quadratic Gaussian integral leading to the form given in Eq. \eqref{Fh} in the main text
\be
{\cal F}_{\rm eff}(\vec h) = \frac{1}{d_c} \int_{q \neq 0}  R_{\alpha \beta, \gamma \delta}(\q) A_{\alpha \beta}(-\q) A_{\gamma \delta}(\q)
\ee
where the interaction tensor is 
\be
R = \frac{1}{2} C^{\mu+\mu_2,\lambda+\lambda_2}  - \frac{1}{2}  C^{\mu+\mu_1,\lambda+\lambda_1} \cdot D \, \cdot C^{\mu+\mu_1,\lambda+\lambda_1}
\ee 
Thanks to the projectors its explicit calculation is easy. One decomposes 
\be
C^{\mu,\lambda} = 2 \mu (W_1+W_2+W_3+W_5) + \lambda [ (D-1) W_3 + \sqrt{D-1} W_4 + W_5 ] 
\ee 
and uses the above multiplication rules for the $W_i$'s. One obtains 
\be
R_{\alpha \beta,\gamma \delta}(\q)  = \sum_{i=1}^5 w_i W_i(\q) 
\ee 
in terms of the five "elastic constants" $w_i$ 
\bea \label{wmu} 
&& w_1= \mu + \mu_2  \\
&& w_2= \mu + \mu_2 - \frac{(\mu+\mu_1)^2}{ \mu} \nn \\
&& w_3 = \mu + \mu_2 + \frac{(D-1)}{2} (\lambda + \lambda_2 - \frac{(\lambda + \lambda_1)^2}{\lambda+ 2 \mu} ) \nn \\
&& w_{44} = \frac{1}{2} (D-1) \left(\lambda+\lambda_2 - \frac{(\lambda + \lambda_1) (\lambda + \lambda_1+ 2 \mu + 2 \mu_1)}{\lambda+ 2 \mu} \right) \quad , \quad w_{44} = \sqrt{D-1} w_4 \nn \\
&& w_5 = \frac{1}{2}  \left(\lambda+\lambda_2 + 2 \mu + 2 \mu_2 - \frac{(\lambda + \lambda_1 + 2 \mu + 2 \mu_1)^2}{\lambda+ 2 \mu} \right) \nn
\eea 
Note that this is true under the condition that the phonon propagator is positive definite
i.e.
\be
\mu>0 \quad , \quad 2 \mu + \lambda >0 
\ee 
Note also that the interaction $R_{\alpha \beta,\gamma \delta}(q)$ 
given above is understood to explicitly exclude the zero-mode $q=0$,
which we address below. The stability of the zero-mode requires $\mu>0$ and $2 \mu + D \lambda >0$, 
which is a more stringent condition. 

Note that when $\mu_1=\mu_2=0$ one has $w_2=0$. When in addition
$\lambda_1=\lambda_2=0$ one has $w_2=w_4=w_5=0$ and 
\be
 w_1= \mu \quad , \quad w_3 = 
\mu+ (D-1) \frac{\mu \lambda}{\lambda+ 2 \mu} 
\ee 
as given in the text. Note that in general there are 5 couplings $w_i$ and 6 original couplings. Inversion
thus determines only the following five ratio as
\bea \label{5ratio}
&& \mu+\mu_2=w_1 \\
&& \frac{(\mu+\mu_1)^2}{\mu} = w_1-w_2 \nn \\
&& \frac{(\lambda+ \lambda_1)^2}{\mu} = \frac{4 \left(w_1-w_2\right)
   \left(w_1-w_3+w_{44}\right){}^2}{\left(D
   \left(w_1-w_5\right)-w_3+w_5+2 w_{44}\right){}^2} \nn \\
&& \lambda + \lambda_2 = \frac{2 \left(w_{44}^2-(D-1) \left(w_1-w_3\right)
   \left(w_1-w_5\right)\right)}{(D-1) \left(D
   \left(w_1-w_5\right)-w_3+w_5+2 w_{44}\right)} \nn \\
&& \frac{\lambda}{\mu} = \frac{2 (D-1) \left(w_1-w_2\right)}{D
   \left(w_1-w_5\right)-w_3+w_5+2 w_{44}}-2 \nn
\eea
Here $\mu+\mu_2$ and $\lambda+\lambda_2$ are the two $h^4$ vertex couplings
in the original $u,h$ theory (before
integrating phonons) and
and $\frac{(\lambda+ \lambda_1)^2}{\mu}$ and $\frac{(\mu+\mu_1)^2}{\mu}$ 
are the natural $uhh$ vertex couplings combination appearing in
perturbation theory. Finally $\lambda/\mu$ is a ratio of elastic constants. 
Hence the overall elastic constant scale, $\mu$, remains undetermined and must be calculated separately from
the $u,h$ theory. Note that combining the above equations, one also obtains the following ratio 
\be \label{newratio} 
\frac{\lambda+ \lambda_1}{\mu+\mu_1} = \frac{2 (w_3 - w_1 - w_{44})}{D
   \left(w_1-w_5\right)-w_3+w_5+2 w_{44}}
\ee

Finally, in the case $\mu_2=\mu_1=0$ one has $w_2=0$ and one can invert the above relations 
for all remaining couplings
\bea \label{invertmu} 
&& \mu = w_1 \quad , \quad \lambda = -\frac{2 w_1 \left(-D
   w_5+w_1-w_3+w_5+2 w_{44}\right)}{D
   \left(w_1-w_5\right)-w_3+w_5+2 w_{44}} \\
   &&  \lambda_1= \frac{2 w_1 \left(-D
   w_5+w_5+w_{44}\right)}{D
   \left(w_1-w_5\right)-w_3+w_5+2
   w_{44}}
    \quad , \quad 
   \lambda_2= \frac{2 w_{44} \left(2 (D-1)
   w_1+w_{44}\right)-2 (D-1) w_5 \left((D-2)
   w_1+w_3\right)}{(D-1) \left(D
   \left(w_1-w_5\right)-w_3+w_5+2
   w_{44}\right)}
\eea
consistent with the above result.

One can also ask about necessary conditions for the quartic form in the effective stretching energy \eqref{Fh} to be positive
definite. Positivity of the quartic form 
\be
k_1^\alpha (\q-\kb_1)^\alpha R_{\alpha \beta,\gamma \delta}(\q) k_3^\beta (\q-\kb_3)^\delta
\ee 
for any choice of $\kb_1,\kb_3,\q$ implies for instance: (i) choosing all $\kb_i$ aligned with $q$
\be
w_5 >0
\ee 
(ii) choosing $\kb_3=\kb_1$ and considering various limits we also find
\be
w_2 >0 \quad , \quad (D-2) w_1 + w_3 >0
\ee 
Finally, note that one must have $w_1 \geq w_2$ for the above equations \eqref{5ratio}
to make sense. 

%
%\medskip
%
%{\tiny Probably to be removed later:
%We checked that the following transformation leaves invariant all the $w_i$
%\be
%\left\{\text{d$\lambda $}= \frac{\text{d$\mu $} \lambda }{\mu
%   },\text{d$\mu $}_1 = \frac{\text{d$\mu $} \left(\mu _1-\mu
%   \right)}{2 \mu },\text{d$\mu $}_2 = -\text{d$\mu
%   $},\text{d$\lambda $}_1 =  \frac{\text{d$\mu $} \left(\lambda
%   _1-\lambda \right)}{2 \mu },\text{d$\lambda $}_2 =
%   -\frac{\text{d$\mu $} \lambda }{\mu }\right\}
%   \ee
%This invariance comes from rescaling of $u$, which is allowed since we do integration over $u$. 
%Hence we have only five couplings in the $h$ theory. }

\subsection{B.2 zero-mode} 

We must treat separately the uniform part of the nonlinear strain tensor, $u_{\alpha \beta}(\q=0)$.
It is the sum of the uniform part of the in-plane strain tensor, which we denote $\tilde u^0_{\alpha \beta}$ and of 
$A^0_{\alpha \beta}= \frac{1}{2} [(\partial_\alpha h)(\partial_\beta h)](\q=0)$. The energy per unit volume associated
to this zero-mode is
\be
f(\tilde u^0,A^0) =\mu (\tilde u^0_{\alpha \beta} + A^0_{\alpha \beta})^2 + \frac{\lambda}{2} 
(\tilde u^0_{\alpha \alpha} + A^0_{\alpha \alpha})^2
+ \lambda_1 \tilde u^0_{\alpha \alpha} A^0_{\alpha \alpha} + 2 \mu_1 
\tilde u^0_{\alpha \beta} A^0_{\alpha \beta} + \mu_2 (A^0_{\alpha \beta})^2 + \frac{\lambda_2}{2} (A^0_{\alpha \alpha})^2,
\ee 
which can be rewritten as
\be \label{zero1} 
 f(\tilde u^0,A^0) = \frac{1}{2} C^{\mu,\lambda}_{\alpha \beta, \gamma \delta} \tilde u^0_{\alpha \beta} \tilde u^0_{\gamma \delta}
+ \tilde u^0_{\alpha \beta} C^{\mu+\mu_1,\lambda+\lambda_1}_{\alpha \beta, \gamma \delta} A^0_{\gamma \delta} + (\mu+\mu_2) (A^0_{\alpha \beta})^2 + \frac{\lambda+ \lambda_2}{2} (A^0_{\alpha \alpha})^2.
\ee 
Minimizing the energy over the $D(D+1)/2$ independent components of the in-plane strain tensor
$\tilde u^0_{\alpha \beta}$ (or integrating the Gibbs measure, which is equivalent since the energy is quadratic
in the $\tilde u^0_{\alpha \beta}$) we obtain the minimum
\be \label{min} 
[\tilde u^0_{\rm min}]_{\alpha \beta} = - [C^{\mu,\lambda}]^{-1} _{\alpha \beta, \gamma' \delta'}  
C^{\mu+\mu_1,\lambda+\lambda_1}_{\gamma' \delta',\gamma,\delta} 
A^0_{\gamma,\delta} = - \frac{\mu + \mu_1}{\mu} A^0_{\alpha \beta} + 
\frac{\lambda \mu_1- \lambda_1 \mu}{\mu(2 \mu + D \lambda)} 
\delta_{\alpha \beta} A^0_{\gamma \gamma}\ ,
\ee
where we have used that 
\bea
[C^{\mu,\lambda}]^{-1} _{\alpha \beta, \gamma \delta} =  \frac{-\lambda}{2 \mu (2 \mu+ D \lambda)} 
\delta_{\alpha \beta} \delta_{\gamma \delta} + \frac{1}{4 \mu} 
( \delta_{\alpha \gamma} \delta_{\beta \delta}  + \delta_{\alpha \delta} \delta_{\beta \gamma} ).
\eea
Plugging back this minimum into the energy we find 
\bea \label{zm1} 
f_{\rm eff}[h]= f_0(u_{\rm \min}^0,A^0) = 
\frac{1}{2} \bar C_{\alpha \beta, \gamma \delta} A^0_{\alpha \beta} A^0_{\gamma \delta},
\eea
where
\be \label{zm2}
\bar C  = C^{\mu+\mu_2,\lambda+\lambda_2}_{\alpha \beta, \gamma \delta} 
- C^{\mu+\mu_1,\lambda+\lambda_1} \cdot [C^{\mu,\lambda}]^{-1} \cdot  C^{\mu+\mu_1,\lambda+\lambda_1}. 
\ee
Upon explicit calculation the final result is 
\be \label{zm3}
f_{\rm eff}[h]= (\mu_2 -2 \mu_1 - \frac{\mu_1^2}{\mu}) (A^0_{\alpha \beta})^2 
+ \frac{1}{2} (\lambda_2 -
\frac{D \lambda _1 \left(2 \lambda +\lambda
   _1\right) \mu - 2 \lambda  \mu _1^2+ 4 \lambda _1
   \mu  \left(\mu +\mu _1\right)}{  \mu  (D \lambda
   +2 \mu )} ) (A^0_{\alpha \alpha})^2.
\ee
Note that it vanishes when the new terms breaking rotational symmetry are absent i.e. when $\mu_1=\mu_2=\lambda_1=\lambda_2=0$.
These zero-mode terms are thus generated only by the bulk anisotropy since we are working in the fixed stress setting and
freely integrate over the zero-mode of the in-plane strain. We leave their study for the future \cite{footnoteFutureLR}. 

%In this work we will not consider the effect of these terms
%and leave their study for the future. However, we note that they may have some nontrivial effect due to Fisher renormalization\cite{Fisher68}, since similar %terms arise in the
%problem of the fixed boundary strain constraints recently studied by Shankar and Nelson\cite{ShankarNelsonUnpublished}. 
%The interplay of these terms and the effects of the bulk anisotropy studied here is an interesting problem for the future.

%\bea
%&& w_1= \frac{\mu + \mu_2}{4} \\
%&& w_2= \frac{\mu + \mu_2}{4} - \frac{(\mu+\mu_1)^2}{4 \mu} \\
%&& w_3 = \frac{\mu + \mu_2}{4} + \frac{(D-1)}{8} (\lambda + \lambda_2 - \frac{(\lambda + \lambda_1)^2}{\lambda+ 2 \mu} ) \\
%&& w_4 = \frac{1}{8} \sqrt{D-1} \left(\lambda+\lambda_2 - \frac{(\lambda + \lambda_1) (\lambda + \lambda_1+ 2 \mu + 2 \mu_1)^2}{\lambda+ 2 \mu} \right) \\
%&& w_5 = \frac{1}{8}  \left(\lambda+\lambda_2 + 2 \mu + 2 \mu_2 - \frac{(\lambda + \lambda_1 + 2 \mu + 2 \mu_1)^2}{\lambda+ 2 \mu} \right) 
%\eea }

\subsection{B.3 Stability}

Here we note that we can rewrite
\bea
&& {\cal F}_1 + {\cal F}_2 = \int d^D x \, \left[ \frac{\kappa}{2}  (\partial^2 \vec h)^2 +
\tau u_{\alpha \alpha} + \frac{\gamma}{2} (\partial_\alpha \vec h)^2 + f_{\rm el}\right].
\eea
Using the traceless tensors and the traces as
\bea
&& f_{\rm el}  = 
\mu \left( \tilde u_{\alpha \beta} - \frac{1}{D} \delta_{\alpha \beta} \tilde u_{\gamma \gamma} + 
\frac{\mu+\mu_1}{\mu} (A_{\alpha \beta} - \frac{1}{D} \delta_{\alpha \beta} A_{\gamma \gamma}) \right)^2 
+ \frac{2 \mu + D \lambda}{2 D} 
\left( \tilde u_{\alpha \alpha} + 
\frac{2 (\mu+\mu_1) + D(\lambda + \lambda_1)}{2 \mu+ D \lambda} A_{\alpha \alpha} 
\right)^2 \\
&& + \hat \mu_2 \left(A_{\alpha \beta} - \frac{1}{D} \delta_{\alpha \beta} A_{\gamma \gamma} \right)^2 
+ B_2 A_{\alpha \alpha}^2, 
\eea
with
\bea
&& \hat \mu_2 = \mu + \mu_2 - \frac{(\mu+\mu_1)^2}{\mu} \\
&& B_2 = \frac{1}{2 D} \left(
2 (\mu+\mu_2) + D(\lambda+\lambda_2) - \frac{(2 (\mu + \mu_1) + D (\lambda+\lambda_1))^2}{2 \mu + D \lambda} \right),
\eea 
where we recall that $A_{\alpha \beta} = \frac{1}{2} \partial_\alpha \vec h \cdot \partial_\beta \vec h$.
Let us set $\tau=0$. Note that $\hat \mu_2=w_2$ as defined in \eqref{wmu}. 
Hence we see that, since the traceless part and the trace
are independent, for $w_2>0$ and $B_2>0$ the last two square terms imply that at the minimum energy (which is zero) one must have $A_{\alpha \beta}=0$, and, in turn from the two first squares, $\tilde u_{\alpha \beta}=0$. Hence in that case
$u_\alpha=0$, $\vec h=0$ is indeed the stable ground state. We note that, in contrast, in the rotationally invariant case
(i.e., setting $\mu_1=\mu_2=\lambda_1=\lambda_2$), the same reasoning leads to the zero energy minimum condition, $u_{\alpha \beta}= \tilde u_{\alpha \beta}+ A_{\alpha \beta}=0$, instead of the above anisotropic condition of $\tilde u_{\alpha\beta}$ and $A_{\alpha\beta}$ vanishing separately. This is expected since in isotropic embedding space, rotations of the membrane do not change its energy.

%
%
%Note that one can rewrite 
%\bea
%&& {\cal F}_1 + {\cal F}_2 = 
%\int d^D x \, [ \frac{\kappa}{2}  (\partial^2 \vec h)^2 +
%\tau u_{\alpha \alpha} + \frac{\gamma}{2} (\partial_\alpha \vec h)^2] \\
%&& + \mu \left(\frac{1}{2} [\partial_\alpha u_\beta + \partial_\beta u_\alpha] +
%\frac{\mu+\mu_1}{2 \mu}  (\partial_\alpha \vec h) \cdot (\partial_\beta \vec h) \right)^2 
%+ \frac{\lambda}{2} \left( \partial_\alpha u_\alpha +
%\frac{\lambda+\lambda_1}{2 \lambda}  (\partial_\alpha \vec h)^2 \right)^2 \\
%&& + \frac{1}{4} (\mu+ \mu_2  -  \frac{(\mu+\mu_1)^2}{ \mu})
%[\partial_\alpha \vec h \cdot \partial_\beta \vec h]^2 
%+ \frac{1}{8} (\lambda+ \lambda_2 - \frac{(\lambda+\lambda_1)^2}{ \lambda}
%[(\partial_\alpha \vec h)^2]^2 
%\eea 

\bigskip
\begin{center}
{\bf C. SCSA analysis}
\end{center}
Below we present the details of the SCSA analysis that was outlined in the main text, following closely the calculation in Ref.\onlinecite{LRReview}. 

\subsection{C.1. SCSA equations} 

\medskip

The SCSA is given by the
pair of coupled equations \eqref{sigma} and \eqref{R} given in the text
for the self-energy $\Sigma(k)={\cal G}(k)^{-1} - \kappa k^4$ and
for the renormalized interaction $\tilde R_{\alpha \beta, \gamma \delta}(\q)$.
The equation \eqref{R} involves tensor multiplication. We can thus
decompose $\Pi(\q) = \sum_{i=1}^5 \pi_i(q) W_i(\q)$
and $\tilde R(q)=\sum_{i=1}^5 \tilde w_i(q) W_i(\q)$, as indicated in the text, where
$\tilde w_i(q)$ are the momentum dependent renormalized 
couplings and $\pi_i(q)$ are polarization integrals calculated below. 
The rules for the tensor multiplication were given in the previous section.
Since the tensors  
$R_{\alpha \beta,\gamma \delta}(\q)$, $\tilde R_{\alpha \beta,\gamma \delta}(\q)$
and $\Pi_{\alpha \beta, \gamma \delta}(\q)$ are 
symmetric in $\alpha \leftrightarrow \beta$, in $\gamma \leftrightarrow \delta$ and in 
$(\alpha,\beta) \leftrightarrow (\gamma,\delta)$, they can be parameterized in terms of five couplings 
(i.e., with $w_{4a}=w_{4b}=w_4$).

%old result (wrong)
%$$
%w_1= w_2 = 2 \mu_0~~~w_3=(D-1)\lambda_0+ 2 \mu_0~~~w_4=\sqrt{D-1} \lambda_0~~~w_5=\lambda_0+2\mu_0
%\label{(3.7)}
%$$

%For the crumpling transition the SCSA equations now take the form:
%%
%\bea
%\Gc(\kb)^{-1} - \kappa k^4 =: \Sigma(\kb)&=& {2 \over d} \int_q k_{\alpha}(k_{\beta}-q_{\beta})(k_{\gamma}-q_{\gamma})k_{\delta}
%\tilde{R}_{\alpha \beta, \gamma \delta}(\q) \Gc(\kb-\q)\;,
%\label{(3.8a)} \\
%\tilde{R}(\q) &=& R(\q) - R(\q) \Pi(\q) \tilde{R}(\q)\;,
%\label{(3.8b)}
%\eea
%%
%where $\Gc(\kb)$ is the propagator of the $\vec r$ field, i.e., $\langle
%r_i(\kb) r_j(\kb') \rangle = \Gc(\kb) (2 \pi)^D \delta(\kb + \kb')
%\delta_{ij}$, and in the second line tensor-product notation is
%implied.  
%
%The vacuum polarization tensor is also a symmetric tensor (with the
%symmetry ${\cal S}$ defined above)
%\be
%\Pi_{\alpha \beta, \gamma \delta}(\q)
%= {1 \over 4} \int_p \left(p_{\alpha} (q_\beta-p_{\beta}) + p_{\beta} (q_\alpha-p_{\alpha})\right)
%\left(p_{\gamma} (q_\delta-p_{\delta}) + p_{\delta} (q_\gamma-p_{\gamma})\right)
%\Gc(\pb) \Gc(\q-\pb)\;.
%\label{(3.9)}
%\ee
%Hence it can be written as
%\be
%\Pi(\q) = \sum_{i=1}^5 \pi_i(\q) W_i(\q)\;,
%\label{(3.10)}
%\ee
%where the $\pi_i(\q)$ are ``polarization bubble'' integrals in the
%$W_i$ basis.  The renormalized interaction $\tilde{R}(\q)$ also
%exhibits ${\cal S}$ symmetry, and can thus be written as
%\bea \label{renR} 
%\tilde R(\q) = \sum_{i=1}^5 \tilde w_i(\q) W_i(\q)\;,
%\eea
We can now solve the equation \eqref{R} and find the renormalized couplings $\tilde w_i(q)$ as
\bea \label{(3.11a)}
&& \tilde{w}_1(q)={ w_1 \over {1 + w_1 \pi_1(q)}}\;,\;\;\;
\tilde{w}_2(q)={ w_2 \over {1 + w_2 \pi_2(q)}}\;, \\
&& 
\begin{pmatrix} \tilde{w}_3(q) & \tilde{w}_4(q) 
\\ \tilde{w}_4(q) & \tilde{w}_5(q) \end{pmatrix} 
=
\begin{pmatrix}  w_3 & w_4 \\ w_4 & w_5 \end{pmatrix} 
\left( 
\begin{pmatrix}  1 & 0 \\ 0 & 1  \end{pmatrix}   +
\begin{pmatrix} \pi_3(q) & \pi_4(q) \\ \pi_4(q) & \pi_5(q) \end{pmatrix}
\begin{pmatrix} w_3&w_4 \\ w_4&w_5 \end{pmatrix}\right)^{-1}\;\;.
\label{(3.11)}
\eea
These can be substituted into \eqref{sigma} to express the self-energy as
\bea \label{sigma2} 
\Sigma(k)= {2 \over d_c} \sum_{i=1,5} \int_\q \tilde{w}_i(q) \Gc(\kb-\q)
k_{\alpha}(k_\beta-q_{\beta}) (W_i)_{\alpha\beta,\gamma\delta}(\q) k_{\gamma} (k_\delta-q_{\delta})\;,
\label{(3.12)}
\eea
The above equations form a closed set of SCSA
equations for the five renormalized elastic coupling constants $\tilde w_i(q)$,
together with the self energy $\Sigma(k)$. The complete Dyson equation for the self-energy contains an
additional UV divergent "tadpole" diagram contribution, which scales
as $k^2$. The integral in \eqref{sigma2} also contains a component
that scales as $k^2$ at small $k$. Both contributions have been substracted
by tuning the bare coefficient $\gamma$ in order to sit at the
critical point.

To solve the above SCSA equations at the critical point, we look for a solution 
with the long wavelength form 
$\Gc(k) \simeq Z^{-1}_\kappa/k^{4-\eta}$. The $\pi_i(q)$ integrals 
have been calculated in the Appendix B of  \cite{LRReview}. They diverge for small $q$ as:
\bea
\pi_i(q) \simeq Z_\kappa^{-2} a_i(\eta,D) q^{-(4-D-2\eta)}\;.
\label{(3.13)}
\eea
For the amplitudes $a_i(\eta,D)$ we find 
\begin{eqnarray} \label{a} 
&&a_1= 2A\;,\;\; a_2=A {2(2-\eta) \over{D+ \eta - 2}}\;,\;\;
a_3=A (D+1)\;,\;\; a_4= A \sqrt{D-1} (D+ 2\eta - 3)\;,\\
&&a_5={A \over {D-2+\eta}}(-22+31D-10D^2+D^3+43\eta-32D\eta+5 D^2\eta-24\eta^2+8D\eta^2+4 \eta^3)\;,\nonumber
\end{eqnarray}
with 
\begin{equation}
A = A(\eta,D)=
{\Gamma(2-\eta-D/2)\Gamma(D/2+\eta/2)
\Gamma(D/2+\eta/2)\over4(4\pi)^{D/2}\Gamma(2-\eta/2)\Gamma(2-\eta/2)
\Gamma(D+\eta)}\;. 
\label{(3.14)}
%={ \Gamma(\eta/2+D/2)^2 \Gamma(2-\eta-D/2) \over { 4 (4\pi)^{{D \over{2}}}  \Gamma(2-\eta/2)^2 \Gamma(\eta+D) }} 
\end{equation}

To compute the self-energy we define the amplitudes $b_i(\eta,D)$
through:
\bea
 \int_\q q^{4-D- 2\eta} |\kb - \q|^{-(4-\eta)}
k_{\alpha}(k_\beta-q_{\beta}) (W_i)_{\alpha\beta,\gamma\delta}(\q) k_{\gamma} (k_\delta-q_{\delta})
= b_i(\eta,D) k^{4-\eta}\;.
\label{(3.15)}
\eea
%where $Q_i$ are normalisation factors $Q_1=\frac{1}{2} (D-2)(D+1)$, $Q_2=D-1$, $Q_3=Q_5=1$, $Q_4=1/2$. 
The explicit calculation in the Appendix B of \cite{LRReview} 
gives
\begin{eqnarray}
&&b_1=B(D-2)(D+1)\;,\;\;
b_2=-B{(D-1)(D^2-4+2 \eta) \over {D-2+\eta}}\;,\;\;
b_3=B(D+1)\;,\\
&&b_4=2B\sqrt{D-1}(2 \eta-3)\;,\;\;
b_5={B \over {D-2+\eta}}
(-22+15D-2D^2+43\eta-16D\eta-24\eta^2+4D\eta^2+4 \eta^3)\;,\nonumber
\end{eqnarray}
where
%=
%{ \Gamma(\eta/2) \Gamma(2-\eta) \Gamma(D/2+\eta/2)
%\over { 4 (4\pi)^{D \over{2}}  \Gamma(2-\eta/2) \Gamma(\eta+D/2) \Gamma(D/2 - \eta/2 +2 )}}
\begin{equation}
B = B(\eta,D) = {\Gamma(\eta/2)\Gamma(D/2+\eta/2)\Gamma(2-\eta)
\over4(4\pi)^{D/2}\Gamma(2-\eta/2)\Gamma(D/2+\eta)\Gamma(D/2+2-\eta/2)} 
\; .
\label{(3.16)}
\end{equation}

\subsection{C.2. Anisotropic fixed point} 

\medskip

Let us first search for a solution to the SCSA equations when all the bare couplings $w_i$ are nonzero.
This corresponds to the "anisotropic fixed point" discussed in the text. In the limit $\q \to 0$ we find 
from \eqref{(3.11)}
\bea
&& \tilde w_1(q) \simeq \frac{1}{\pi_1(q)}\;,\quad \tilde w_2(q) \simeq \frac{1}{\pi_2(q)}\;, \quad , \quad  \begin{pmatrix} \tilde{w}_3(q) & \tilde{w}_4(q)\;,\\
\tilde{w}_4(q) & \tilde{w}_5(q) \end{pmatrix} 
\simeq \begin{pmatrix} \pi_3(q) & \pi_4(q) \\ \pi_4(q) &
  \pi_5(q) \end{pmatrix}^{-1}\;.
\label{inverse1} 
\eea 
independent of the bare values, as long as they are nonzero.
Substituting Eqs.\eqref{(3.11)},\eqref{(3.14)},\eqref{(3.16)}
into \eqref{(3.12)} we see that factors of $Z_\kappa$ cancel and we find the
self-consistent equation:
\begin{equation}
\frac{d_c}{2} = \sum_{i=1,2} {b_i(\eta,D) \over a_i(\eta,D) } + { b_3(\eta,D)a_5(\eta,D)-b_4(\eta,D)a_4(\eta,D)
+b_5(\eta,D)a_3(\eta,D) \over {a_3(\eta,D) a_5(\eta,D) - a_4(\eta,D)^2 }}\;.
\label{(3.17)}
\end{equation}

Putting everything together, after considerable simplifications, the
equation determining the exponent $\eta=\eta^{\rm anis}(D,d_c)$ is found to be:
\bea \label{eqSC} 
d_c={ { D(D+1)(D-4+\eta)(D-4+2\eta)(2D-3+2 \eta)\Gamma[{1\over 2}\eta] \Gamma[2-\eta] \Gamma[\eta+D] \Gamma[2-{1\over 2}\eta] }
\over {2(2-\eta)(5-D-2\eta)(D+\eta-1)\Gamma[{1\over 2}D + {1\over 2}\eta] \Gamma[2-\eta-{1\over 2}D] 
\Gamma[\eta+{1\over 2}D] \Gamma[{1\over 2}D+2-{1\over 2}\eta]}}\;,
\label{(3.18)}
\eea
which in $D=2$ reduces to 
\bea \label{cr2SM} 
d_c= \frac{24 (\eta -1)^2 (2 \eta +1)}{(\eta -4) \eta  (2 \eta -3)}\;.
\eea
as given in the main text, leading to $\eta^{\rm anis}(D=2,d_c=1)=0.753645..$.
In the limit of large $d_c$, the solution of \eqref{eqSC} behaves as 
\bea
&& \eta^{\rm anis}(D,d_c) \simeq \frac{C(D)}{d_c} + O(1/d_c^2) \quad , \quad C(D) = \frac{(D-4)^2 (2 D-3) \Gamma (D+2)}{2 (5-D) (D-1)
   \Gamma \left(2-\frac{D}{2}\right) \Gamma
   \left(\frac{D}{2}+2\right) \Gamma
   \left(\frac{D}{2}\right)^2}\;,\nonumber
\eea
with $C(2)=2$. As discussed in \cite{LRReview} the leading coefficient $C(D)$ in the $1/d_c$ expansion is an
exact result, while the higher orders are specific to the SCSA. 

We note that the above equation \eqref{eqSC} is the same as the one obtained for the
crumpling transition, replacing $d$ by $d_c$. Hence, studying our new fixed point amounts, formally,
to studying the crumpling transition fixed point in embedding space dimension $d=1$ instead of $d=3$. 
Not surprisingly then, the leading term in the large $d_c$ expansion above then coincides with the
one in the $1/d$ expansion for the crumpling transition of Ref.\;\onlinecite{AL}. 
We can also expand our SCSA prediction in $\epsilon = 4-D$, finding
\bea 
\eta^{\rm anis}(D,d_c) \simeq \frac{25}{3 d_c} (4-D)^3 + O((4-D)^3),
\eea
%x
consistent with the vanishing of the leading order $O(\epsilon)$ of
$\eta^{\rm anis}(D,d_c)$ found below in the section on the RG
calculation.

This new "anisotropic" membrane fixed point is characterized by several universal
amplitude ratio. As discussed in the text, from \eqref{inverse1} we obtain
\bea \label{renw} 
&& \tilde w_i(q) \simeq Z_\kappa^2 c_i q^{4-D - 2 \eta}/A  \\
&& c_i=1/a_i \quad \text{for} \quad i=1,2 \quad , \quad 
 \begin{pmatrix} c_3 & c_4 \\
c_4 & c_5 \end{pmatrix} 
\simeq \begin{pmatrix} a_3 & a_4 \\ a_4 &
  a_5 \end{pmatrix}^{-1} \nn
\eea 
Inserting the $\tilde w_i(q)$ into \eqref{5ratio}, we obtain the renormalized couplings of the
$u,h$ theory. More precisely we obtain the $h^4$ couplings 
$\tilde \mu(q)+\tilde \mu_2(q)$, $\tilde \lambda(q)+\tilde \lambda_2(q)$,
and the $u h^2$ couplings $(\tilde \mu(q)+\tilde \mu_1(q))^2/\tilde \mu(q)$
and $(\tilde \lambda(q)+\tilde \lambda_1(q))^2/\tilde \lambda(q)$.
These four couplings thus vanish as $q^{4-D-2 \eta}$ at small $q$.
In addition we obtain the ratio $\tilde \lambda(q)/\tilde \mu(q)$ which
has a finite limit at small $q$. The determination of $\tilde \mu(q)$ however requires an 
additional calculation (see below), with the result that $\tilde \mu(q) \sim q^{\eta_u}$
where $\eta_u$ is now an
independent exponent (at variance with the rotationally invariant case where one has $\eta_u=4-D-2 \eta$).
For this anisotropic fixed point, $\eta_u=0$, i.e. $\tilde \mu(0)$ is finite.
Hence we find that $\tilde \mu_2(q) \to - \tilde \mu(0)$ at small $q$, so 
that the $h^4$ coupling can vanish at small $q$ as $\tilde \mu(q)+\tilde \mu_2(q) \sim q^{4-D-2 \eta}$,
and similarly for $\tilde \lambda_2(q)$. A similar property holds for the $u h^2$ couplings. 

Let us now determine the amplitude ratio, which are universal at the fixed point.
From \eqref{renw} we obtain the amplitude ratio in the long wavelength limit as
\bea
\lim_{q \to 0} \frac{\tilde w_i(q)}{\tilde w_j(q)} = \frac{c_i}{c_j} 
\eea 
for any pair $(i,j)$, with, using \eqref{a} and \eqref{renw} we obtain
\bea \label{ci} 
&& c_1= \frac{1}{2} \quad, \quad c_2= -\frac{D+\eta -2}{2 (\eta   -2)}
\quad , \quad c_3= \frac{1}{4} \left(\frac{1-D}{(D+3) (D+\eta
   -1)}+\frac{D}{\eta -2}-\frac{8}{(D+3) (D+2 \eta
   -5)}+2\right) \\
&& c_4=   -\frac{\sqrt{D-1} (D+\eta -2) (D+2
   \eta -3)}{4 (\eta -2) (D+\eta -1) (D+2 \eta
   -5)} 
   \quad , \quad 
c_5=   \frac{(D+1)
   (D+\eta -2)}{4 (\eta -2) (D+\eta -1) (D+2 \eta
   -5)}
\eea 
Note that these values of the $c_i$ assume that all bare $w_i$ are nonzero
hence they are valid only at the anisotropic fixed point. Inserting the value of $\eta$ for $D=2$ and $d_c$=1 we find, at the anisotropic fixed point
\be
c_i(D=2,d_c=1) = \left\{\frac{1}{2},0.30234,0.338287,-0.0292957,0.173248\right\}
\ee
From this, using \eqref{5ratio},
we find $\lim_{q \to 0} \tilde \lambda(q)/\tilde \mu(q) = -0.978449$ 
and the Poisson ratio $\sigma_R(q)=\frac{\tilde \lambda(q)}{2 \tilde \mu(q) + (D-1) \tilde \lambda(q)}
= -0.957808$ 

Note that for $D=2$ and large $d_c$ we find, up to $O(1/d_c^2)$ terms
\be
c_i = \left\{\frac{1}{2},\frac{1}{2
   d_c},\frac{1}
   {3}+\frac{1}{36
   d_c},\frac{1}
   {12
   d_c},\frac{1}
   {4
   d_c}\right\}
\ee
Hence for $D=2$, the anisotropic membrane fixed point converges as $d_c \to +\infty$ to
the one of the isotropic membrane since $\tilde w_2$, $\tilde w_4$, $\tilde w_5$ 
are parametrically smaller in that limit than $\tilde w_1$ and $\tilde w_3$ (which span the couplings
of the isotropic membrane). However, from \eqref{ci} we can state that these two fixed points
are {\it different} at infinite $d_c$ for $D>2$. In this limit one can simply set $\eta \to 0$ in \eqref{ci}
and one finds for the anisotropic fixed point
\be
\lim_{d_c \to + \infty} c_i 
= \left\{\frac{1}{2},\frac{D-2}{4},\frac{D^2-9 D+22}{40-8
   D},\frac{(D-3) (D-2)}{8 (D-5)
   \sqrt{D-1}},\frac{-D^2+D+2}{8 \left(D^2-6
   D+5\right)}\right\}
\ee 
while for the isotropic one $c_i = \{ \frac{1}{2}, 0, \frac{1}{D+1},0,0\}$, see below.
Hence, for $d_c=+\infty$, the anisotropic fixed point leaves the
isotropic subspace as $D$ increases from $D=2$ to $D=4$.\\

As mentionned in the text, there is an interesting subspace of couplings which corresponds to
a purely local interaction between the gradient fields $\partial_\alpha \vec h$
\be R_{\alpha \beta,
  \gamma \delta}
=\frac{\mu_0}{2}(\delta_{\alpha\gamma}\delta_{\beta\delta}+\delta_{\alpha\delta}\delta_{\beta\gamma})
+ \frac{\lambda_0}{2} \delta_{\alpha\beta}\delta_{\gamma\delta}.
\label{(3.3)}
\ee
for some constants denoted $\mu_0$ and $\lambda_0$ (these are denoted $4 v_2$ and $4 v_1$ respectively
in the main text). It is realized by the choice
\bea 
\label{manif1} 
w_1= w_2 = \mu_0,\;\;
w_3= \frac{1}{2}(D-1)\lambda_0+ \mu_0,\;\;
w_4=\frac{1}{2} \sqrt{D-1}\lambda_0,\;\;
w_5=\frac{1}{2} \lambda_0+ \mu_0\;.
\label{(3.7)}
\eea
Note that the two eigenvalues of the matrix formed by the $w_i$,
$i=3,4,5$, are then $\mu_0$, and $\mu_0 + \frac{1}{2} D \lambda$.
Replacing $d_c$ by $d$ this is also the subspace corresponding to the
bare action of the Landau theory for the crumpling transition \cite{PKN}. 

This subspace is preserved by the one-loop RG in an expansion in $D=4$, as we will see in the next section.
However, for any fixed $D<4$, it is {\it not preserved} by the RG flow in the large $d_c$ limit (hence it is also not preserved
by the SCSA). In $D=4$ at large $d_c$ it is indeed preserved (consistent with the RG), since in that case one has
\bea
\lim_{d_c \to + \infty}  c_i =\left\{\frac{1}{2},\frac{1}{2},\frac{1}{4},-\frac{1}{4
   \sqrt{3}},\frac{5}{12}\right\}
\eea 
which indeed belongs to the subspace \eqref{manif1}.
However, from the above discussion, we expect the two-loop corrections in the RG to fail to
preserve this subspace. This indicates that the study of the RG of the crumpling transition
to higher order in $\epsilon$ will be qualitatively different from the one given in \cite{PKN}, a subject we leave for future investigation. 

\subsection{C 3. RG flow associated to the SCSA equations}

It is instructive to recast the SCSA equations into an RG flow. We start with large $d_c$,
and discuss general $d_c$ below. The SCSA equations allow one to obtain the exact RG beta function 
to leading order in $1/d_c$
in any dimension $D$. Indeed, taking a derivative $\partial_\ell=- q \partial_q$ 
on both sides of equations \eqref{(3.11a)} we obtain,
\be
\partial_\ell \tilde w_i(q) =  - \frac{w_i^2}{(1+ w_i \pi_1(q))^2} (- q \partial_q) \pi_i(q) 
=  - \tilde w_i(q)^2 (- q \partial_q \pi_i(q))  \simeq 
- \tilde w_i(q)^2 \kappa^{-2} \epsilon q^{-\epsilon} a_i(0,D)
\ee
where we have used \eqref{(3.13)} setting $\eta \to 0$, i.e., using the bare propagator
with $Z_\kappa=\kappa$. The natural dimensionless coupling for the RG is
\be \label{hatw} 
\hat w_i := \tilde w_i(q) \kappa^{-2} q^{-\epsilon} 
\ee
In terms of these couplings we obtain the RG equation for $d_c=+\infty$, exact for any $\epsilon=4-D$,
\bea \label{rglarge} 
&& \partial_\ell \hat w_i = \epsilon \hat w_i - \epsilon {\red a_i(0,D)} \hat w_i^2  \quad , \quad i=1,2 \\
&& \partial_\ell \begin{pmatrix} \hat{w}_3 & \hat{w}_4 \\
\hat{w}_4 & \hat{w}_5\end{pmatrix} = \epsilon  \begin{pmatrix} \hat{w}_3 & \hat{w}_4 \\
\hat{w}_4 & \hat{w}_5\end{pmatrix} -  \epsilon \begin{pmatrix} \hat{w}_3 & \hat{w}_4 \\
\hat{w}_4 & \hat{w}_5\end{pmatrix}
 \begin{pmatrix} a_3(0,D) & a_4(0,D) \\
a_4(0,D) & a_5(0,D) \end{pmatrix}
 \begin{pmatrix} \hat{w}_3 & \hat{w}_4 \\
\hat{w}_4 & \hat{w}_5\end{pmatrix}
\eea 
The fixed point of these RG equations which describes the anisotropic membrane for $d_c=+\infty$ is then $\hat w_i=\hat w_i^*$ with
\be \label{fp2} 
\hat w_i^* = \frac{1}{a_i(0,D)} \quad , \quad i=1,2 \quad , \quad \begin{pmatrix} \hat{w}^*_3 & \hat{w}^*_4 \\
\hat{w}^*_4 & \hat{w}^*_5\end{pmatrix} =  \begin{pmatrix} a_3(0,D) & a_4(0,D) \\
a_4(0,D) & a_5(0,D) \end{pmatrix}^{-1}
\ee
is consistent with the above analysis \eqref{renw}. The calculation of the exponent $\eta$ to leading order $O(1/d_c)$ is then as follows. If one calculates $- k \partial_k (\Sigma(k)/k^4)$ from 
\eqref{sigma2} one obtains a convergent integral. Replacing $\tilde w_i(q) = \kappa^2 q^\epsilon 
\hat w_i$ in \eqref{sigma2} and using \eqref{(3.15)} we can write the RG function $\eta=\eta(\hat w)$ as
\be
\eta = - k \partial_k (\Sigma(k)/k^4) =  {2 \over d_c}  \sum_{i=1,5} \tilde b_i(D) \hat w_i 
%+ \begin{pmatrix} \hat{w}_3 & \hat{w}_4 \\
%\hat{w}_4 & \hat{w}_5\end{pmatrix} 
%\begin{pmatrix} b_3(D) & b_4(D) \\
%b_4(D) & b_5(D) \end{pmatrix} \right)
 \quad , \quad \tilde b_i(D)=\lim_{\eta \to 0} \eta b_i(\eta,D), 
\ee 
where in the r.h.s we used $\eta$ as a regulator to obtain the needed (finite) integral. 
One can then easily check that at the fixed point \eqref{fp2} the exponent $\eta = \eta(\hat w^*)$ recovers the result $\eta \simeq C(D)/d_c$ predicted by the 
self-consistent equation \eqref{(3.17)}.

In the above RG equations \eqref{rglarge} we have neglected the renormalization
of $\kappa$ which is subdominant in $1/d_c$. We can now take it into account and
define accordingly the running RG couplings as $\hat w_i := \tilde w_i(q) \kappa^{-2} q^{2 \eta -\epsilon} = 
 \tilde w_i(q) \tilde \kappa(q)^{-2} q^{-\epsilon}$. This allows to 
write the SCSA equations as RG flow equations for any $d_c$ as follows
\bea \label{flowSCSAdc} 
&& \partial_\ell \hat w_i = (\epsilon - 2 \eta) \hat w_i - (\epsilon - 2 \eta) a_i(D,\eta) \hat w_i^2  \quad , \quad i=1,2 \\
&& \partial_\ell \begin{pmatrix} \hat{w}_3 & \hat{w}_4 \\
\hat{w}_4 & \hat{w}_5\end{pmatrix} = (\epsilon-2 \eta)  \begin{pmatrix} \hat{w}_3 & \hat{w}_4 \\
\hat{w}_4 & \hat{w}_5\end{pmatrix} -  (\epsilon-2 \eta) \begin{pmatrix} \hat{w}_3 & \hat{w}_4 \\
\hat{w}_4 & \hat{w}_5\end{pmatrix}
 \begin{pmatrix} a_3(D,\eta) & a_4(D,\eta) \\
a_4(D,\eta) & a_5(D,\eta) \end{pmatrix}
 \begin{pmatrix} \hat{w}_3 & \hat{w}_4 \\
\hat{w}_4 & \hat{w}_5\end{pmatrix},
\eea 
where the $\eta$ RG function, $\eta=\eta(\hat w)$, is defined as
\bea \label{etaeq} 
\eta := - k \partial_k \tilde \kappa(k) =
- k \partial_k (\Sigma(k)/k^4) =  {2 \over d_c} \eta \sum_{i=1,5} b_i(\eta,D) \hat w_i .
%+ \begin{pmatrix} \hat{w}_3 & \hat{w}_4 \\
%\hat{w}_4 & \hat{w}_5\end{pmatrix} 
%\begin{pmatrix} b_3(\eta,D) & b_4(\eta,D) \\
%b_4(\eta,D) & b_5(\eta,D) \end{pmatrix} \right)
\eea 
The fixed point of these RG equations, corresponding to all bare $w_i$ being nonzero, i.e., the anisotropic membrane fixed point, is given by
\be \label{fp3} 
\hat w_i^* = \frac{1}{a_i(D,\eta^*)} \quad , \quad i=1,2 \quad , \quad \begin{pmatrix} \hat{w}^*_3 & \hat{w}^*_4 \\
\hat{w}^*_4 & \hat{w}^*_5\end{pmatrix} =  \begin{pmatrix} a_3(D,\eta^*) & a_4(D,\eta^*) \\
a_4(D,\eta^*) & a_5(D,\eta^*) \end{pmatrix}^{-1},
\ee
where $\eta^*$ is determined by \eqref{etaeq} at the fixed point. Equivalence with 
the full SCSA equation \eqref{(3.17)} is then immediately follows.

\medskip

{\bf Other fixed points} 

\medskip

As discussed in the main text, there are a number of other subspaces which are preserved by renormalization
within the SCSA method (hence also at large $d_c$). These can be labeled
as $S_{i_1,\dots i_n}$, with $1 \leq i_1 < i_2 < \dots < i_n \leq 5$, where the only nonzero bare couplings $w_i$ are $w_{i_1}, \dots w_{i_n}$. For those with $w_4=0$, i.e., $i_1,\dots, i_n \in \{1,2,3,5\}$, there are 
four with $n=1$, five with $n=2$ (that is $S_{12},S_{13},S_{15},S_{23},S_{25}$) together
with $S_{123}$ and $S_{125}$ (note that $w_4=0$ is not preserved unless one has also $w_3=0$ or $w_5=0$). Then, one has
$S_{345},S_{1345},S_{2345},S_{12345}$ with $w_4 \neq 0$. In each of these subspaces there is a fixed point
denoted by $P_{i_1,\dots i_n}$. It is obtained from \eqref{fp3} by 
setting to zero the $\hat w_i^*$ not in the set $\{i_1,\dots,i_n\}$
(disregarding their corresponding equation, except $a_4$ which
must be set to zero when $\hat w_4=0$). Their associated SCSA equation
is obtained as $\frac{d_c}{2} = \sum_{i=1}^5 b_i(\eta,D) \hat w_i^*$. Let us give some
examples.
%
%\begin{enumerate}
%
% \item $S_i$ for $i=1,2,3,5$ where a single $w_i$ is nonzero.
% 
% \item $S_{i,j}$ where only two $w_i$ in $i < j \in {1,2,3,5}$ are nonzero.
% 
% \item $S_{i,j,k}$ where only three $w_i$ in $i < j < k \in {1,2,3,5}$ are nonzero.
% 
% \item $S_{1235}$ where all $w_1,w_2,w_3,w_5$ are nonzero but $w_4=0$.
% 
% \item $S_{345}$ where only $w_3,w_4,w_5$ are nonzero.
% 
% \item $S_{1345}$, $S_{2345}$ and $S_{12345}$. 
%
%\end{enumerate} 

\begin{enumerate}

\item The fixed point $P_{13}$ describes the isotropic flat phase. 
Its exponent $\eta$ is determined by $\frac{d_c}{2} = \frac{b_1}{a_1} + { b_3  \over {a_3   }}$ i.e., Eq.
\eqref{SCSA2} in the main text, leading to the well known value $\eta=4/(1+ \sqrt{15})=0.820852..$,
$\zeta=\frac{1}{7} (8-\sqrt{15}) = 0.589574..$ for the exponents describing out-of-plane fluctuations of the physical membrane $D=2,d_c=1$. The amplitudes are $c_1=A/a_1$ and $c_3=A/a_3$ which gives
$c_i = \{ \frac{1}{2}, 0, \frac{1}{D+1},0,0\}$, leading to $\lim_{q \to 0} \tilde \lambda(q)/\tilde \mu(q)=-\frac{2}{D+2}$ 
and to the universal Poisson ratio, $\sigma_R = -1/3$ within the SCSA. 

\item The fixed point $S_{1345}$ has $\hat w_2^*=0$ and describes the
case where $\mu_2=\mu_1=0$. The exponent $\eta$ is determined by \eqref{SCSA3} in the text,
which for $D=2$ gives 
\be
\frac{d_c}{2} = \frac{93}{5 (\eta -4)}+\frac{6}{\eta -2}+\frac{1}{\eta
   }+\frac{16}{15-10 \eta }+8.
\ee
For $d_c=1$ one finds $\eta=0.853967$, and $\zeta=0.573016$.
The amplitudes $c_i$ are then given by \eqref{ci}, where one sets $c_2=0$.
For $D=2$, $d_c=1$ inserting the above value of $\eta$ one finds
\be
c_i = \left\{\frac{1}{2},0,0.346325,-0.0550542,0.233302\right\}.
\ee
Note that the manifold $\hat w_2=0$ is however not preserved within the $\epsilon$-expansion
(see analysis in section below). 

\end{enumerate} 

\medskip

%{\bf Stability and flow around the fixed points} 

{\bf Remark}. One bonus of these RG equations, as compared to the original self-consistent equations, is that one can determine
the direction of the RG flow, the Hessian around each fixed point, and the various
crossovers in the flow. For instance, to determine the Hessian around a fixed point $\hat w_i^*$, with 
associated exponent $\eta^*$, we need the variation of $\eta$. Variation of \eqref{etaeq} around the fixed
point gives $\delta \eta = - \frac{\sum_{j=1}^5 b_j \delta \hat w_j}{\sum_{k=1}^5 b'_k \hat w^*_k}$,
where we have denoted $a_i \equiv a_i(D,\eta^*)$, $b_i \equiv b_i(D,\eta^*)$, 
$a'_i \equiv \partial_\eta b_i(D,\eta)|_{\eta=\eta^*}$,
$b'_i \equiv \partial_\eta b_i(D,\eta)|_{\eta=\eta^*}$. Using this, one can obtain
the Hessian, and the flow around the fixed point. We defer this study to the future\cite{footnoteFutureLR}.

\medskip
\begin{center}
{\bf D. Renormalization group calculation for the $h^4$ theory}
\end{center}
\medskip

Here we present the details of the one-loop RG calculation for the quartic model $h^4$ defined in Eq. \eqref{Fh}
of the main text. The power-counting is the same as in the standard $\phi^4$ $O(N)$ model with quartic nonlinearities
which are relevant for $D < D_{uc} = 4$. This allows us to control the RG
analysis by an expansion in $\epsilon = 4-D$ around $D=4$ \cite{WilsonFisher,ZinnJustin,CL} 
Here we will simply display the calculation using the momentum shell RG, i.e introducing a running UV cutoff $\Lambda_\ell = \Lambda e^{-\ell}$ and integrating the internal momentum in the shell $\Lambda_\ell e^{-d\ell} < q < \Lambda_\ell$. However, we have checked all of our formula also using dimensional regularization for $D<4$ with the external momentum as an IR cutoff. 

In the critical theory there are two types of relevant one-loop corrections, the correction to the $h^4$ vertex
$\delta R$ and the correction to the bending rigidity $\delta \kappa$. Away from criticality
one also needs to calculate the correction to $\gamma$. 

\subsection{D 1. Correction to the quartic interaction} 

Having constructed the generic vertex
$R_{\alpha\beta,\gamma\delta}(\q)$, the analysis of the diagrams is
then quite similar to that of the $O(N)$ model\cite{WilsonFisher,ZinnJustin,CL}.
There are three distinct channels contributing to the renormalization of
$R_{\alpha\beta,\gamma\delta}(\q)$, with only one of them taken into
account in the large $d_c$ and SCSA analysis. The corrections to the quartic coupling can be written as the sum
\be
\delta R=\delta R^{(1)} + \delta R^{(2)} + \delta R^{(3)}
\ee
depicted by the three diagrams in Fig.\ref{RdiagramsFig}.

The contribution from the first (vacuum polarization) diagram, proportional to $d_c$, is given by the following integral
%(symmetry factor $8$)
\bea
\delta R_{\alpha\beta,\gamma\delta}^{(1)}(\q)
&=&-  \frac{Td_c}{\kappa^2}R_{\alpha\beta,\gamma'\delta'}(\q)
R_{\gamma''\delta'',\gamma\delta}(\q)
\intp\frac{p_{\gamma'}(q_{\delta'} -
  p_{\delta'})p_{\gamma''}(q_{\delta''}-p_{\delta''})}{p^4|\q-\pb|^4},\\
&\approx&-   \frac{Td_c}{\kappa^2}R_{\alpha\beta,\gamma'\delta'}(\q)
R_{\gamma''\delta'',\gamma\delta}(\q)
\intp\frac{p_{\gamma'}p_{\delta'}p_{\gamma''}p_{\delta''}}{p^8}
\eea
where in the second line we have kept only the leading terms in $D=4$. 
%{\red does this require an explanation, it is not so trivial since $\hat q$ 
%appears in general $D$ is that correct?}

Similarly, the contribution from the second (vertex correction) diagram 
% (total symmetry factor $32$) 
is given by
\bea
\delta R_{\alpha\beta,\gamma\delta}^{(2)}(\q,\kb_3)
&=&  - 4 \frac{T}{\kappa^2} {\rm sym} R_{\alpha\beta,\gamma'\delta'}(\q)
\intp\frac{p_{\gamma'}(q_{\delta'} -
  p_{\delta'})p_{\gamma''}(q_{\delta''}-p_{\delta''})
R_{\gamma''\gamma,\delta''\delta}(\pb-\kb_3)}{p^4|\q-\pb|^4},\\
&\approx&- 4 \frac{T}{\kappa^2}R_{\alpha\beta,\gamma'\delta'}(\q)
\intp\frac{p_{\gamma'}p_{\delta'}p_{\gamma''}p_{\delta''}
R_{\gamma''\gamma,\delta''\delta}(\pb)}{p^8}
\eea
where ${\rm sym}$ denotes the symmetrization $(\alpha,\beta) \leftrightarrow (\gamma,\delta)$.
Finally, the contribution from the third (box) diagram is
\bea
\delta R_{\alpha\beta,\gamma\delta}^{(3)}(\q,\kb_1)
&=&- 4  \frac{T}{\kappa^2}
\intp\frac{p_{\gamma'}(q_{\delta'} -
  p_{\delta'})p_{\gamma''}(q_{\delta''}-p_{\delta''})
R_{\alpha\gamma',\gamma\delta'}(\kb_1-\pb)
R_{\gamma''\beta,\delta''\delta}(\kb_2 + \pb)}{p^4|\q-\pb|^4},\\
&\approx&- 4 \frac{T}{\kappa^2}
\intp\frac{p_{\gamma'}p_{\delta'}p_{\gamma''}p_{\delta''}
R_{\alpha\gamma',\gamma\delta'}(\pb)
R_{\gamma''\beta,\delta''\delta}(\pb)}{p^8}.
\eea
where we recall that $\q=\kb_1+\kb_2$. Note that one should symmetrize with the crossed diagram 
but at the level of the last step exchanging $\gamma''$ and $\delta''$ does
not make a difference.

%
%
%{\tiny The box diagram is
%\bea
%\delta R_{\alpha\gamma,\beta\delta}^{(3)}(\q,\kb_1,\kb_3)
%&=&- {\red 4}  \frac{T}{\kappa^2}
%\intp\frac{p_{\gamma'}(q_{\delta'} -
%  p_{\delta'})p_{\gamma''}(q_{\delta''}-p_{\delta''})
%R_{\alpha\gamma',\beta\delta'}(\kb_1-\pb)
%R_{\gamma''\gamma,\delta''\delta}(\pb-\kb_3)}{p^4|\q-\pb|^4},\\
%&\approx&- {\red 4} \frac{T}{\kappa^2}
%\intp\frac{p_{\gamma'}p_{\delta'}p_{\gamma''}p_{\delta''}
%R_{\alpha\gamma',\beta\delta'}(\pb)
%R_{\gamma''\gamma,\delta''\delta}(\pb)}{p^8}.
%\eea
%}

\begin{figure}[ht]
\includegraphics[width=0.6\linewidth]{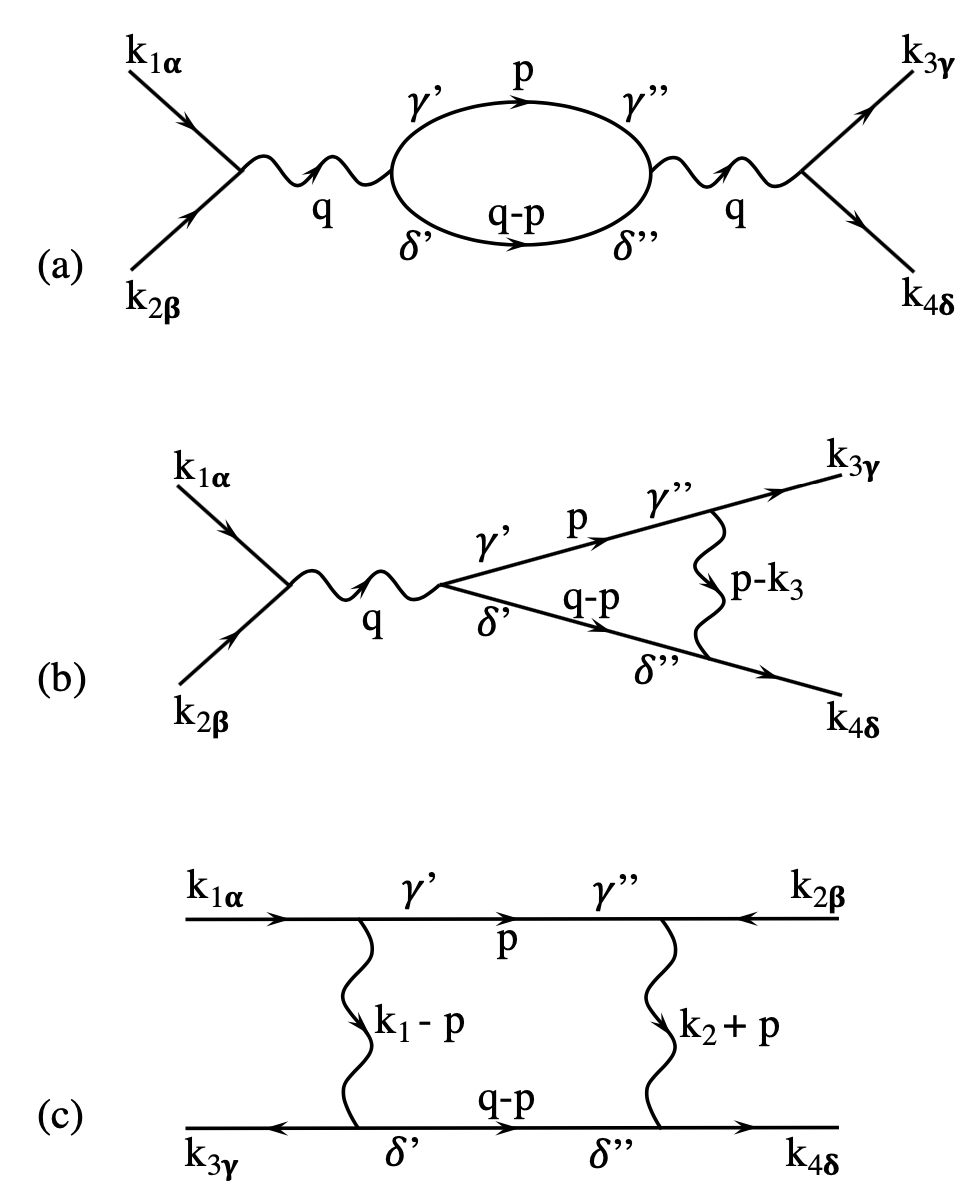}
{\caption{Feynman diagrams for one-loop corrections to the quartic vertex $R_{\alpha\beta,\gamma\delta}(\qb)$, with (a) "vacuum polarization" $\delta R^{(1)}_{\alpha\beta,\gamma\delta}(\qb)$, (b) "vertex correction" $\delta R^{(2)}_{\alpha\beta,\gamma\delta}(\qb,\kb_3)$, (c) "box diagram" $\delta R^{(3)}_{\alpha\beta,\gamma\delta}(\qb,\kb_1)$.}
\label{RdiagramsFig}}  
\end{figure}

To evaluate these integrals we now insert the decomposition $R_{\alpha \beta, \gamma \delta}(q)=
\sum_{i=1}^5 w_i (W_i(q))_{\alpha \beta, \gamma \delta}$ and use the definitions and the properties of the projectors
summarized in Section A. We further use the formula for the angular averages (denoted $\langle \dots \rangle$, where $\hat k=k/|k|$)
$\langle
\hat k_\alpha \hat k_\beta\rangle = \frac{1}{D}\delta_{\alpha\beta}$
and $\langle \hat k_\alpha \hat k_\beta \hat k_\gamma \hat
k_\delta\rangle =
\frac{1}{D(D+2)}\left(\delta_{\alpha\beta}\delta_{\gamma\delta}+\delta_{\alpha\gamma}\delta_{\beta\delta}+\delta_{\alpha\delta}\delta_{\gamma\beta}\right)$\cite{LRReview}.

Denoting $\delta w=(\delta w_1,\delta w_2,\delta w_3,\delta w_4, \delta w_5)$ the one loop corrections
to the couplings $w_i$ from the first diagram are
\bea \label{d1} 
&& \delta w  = \frac{- d_c}{D(D+2)} \bigg(2 w_1^2,2 w_2^2,(D+1) w_3^2+2 \sqrt{D-1} w_4 w_3+3 w_4^2,
w_4 (\sqrt{D-1} w_4+3 w_5)+w_3 (D w_4+\sqrt{D-1}
   w_5+w_4), \nn \\
   && ~~~~~~~~~~~~~~~~~~~~~~~~~~~~~~~~~~~~~~~ (D+1) w_4^2+2 \sqrt{D-1} w_5 w_4+3 w_5^2 \bigg)
   \times 
\frac{T}{\kappa^2}  \intp \frac{1}{p^4}
\eea
The contribution of the second diagram reads
\bea \label{d2}
&& \delta w  =  \frac{-4}{D(D+2)} \bigg(w_1 (2 w_5 - w_2) , w_2 (2 w_5-w_2), 
\frac{w_2}{2} ((D^2-3) w_3+\sqrt{D-1} (D+1)
   w_4)+ w_5 (D w_3+\sqrt{D-1} w_4+w_3), \nn \\
&&
~~~ \frac{w_2 ((D^2-1) w_3+\sqrt{D-1}
   (D^2+D-4) w_4+(D^2-1)
   w_5)+2 w_5 ((D-1) w_3+\sqrt{D-1} (D+4)
   w_4+(D-1) w_5)}{4 \sqrt{D-1} } ,
   \nn \\
   && ~~~~~~~~~~~~~
   w_5 (\sqrt{D-1} w_4+3 w_5)+\frac{w_2}{2}
   (\sqrt{D-1} (D+1) w_4+(D-1) w_5)
    \bigg)
   \times 
\frac{T}{\kappa^2}  \intp \frac{1}{p^4}
\eea
The contribution of the third diagram reads
\bea \label{d3} 
&& \delta w  =  \frac{-1}{D(D+2)} \bigg(\left(D^2-2\right) w_2^2+4 D w_5 w_2+8 w_5^2,
\left(D^2-2\right) w_2^2+4 D w_5 w_2+8 w_5^2,\nn \\
&& (D^2+D-3) w_2^2+4 (D+1) w_5^2+4 w_5 w_2,
\sqrt{D-1} (w_2-2 w_5){}^2,
(D^2-1) w_2^2+4 (D-1) w_5 w_2+12 w_5^2
\bigg)
\frac{T}{\kappa^2}  \intp \frac{1}{p^4} \nn
\eea
where we have kept the explicit factors $D$ in the geometric factors. 

\subsection{D 2. Correction to the bending rigidity $\kappa$} 

The correction to the self-energy to first order in perturbation
theory $O(R)$ can be read off from \eqref{sigma2} as \bea \label{calc}
\delta \Sigma(k)= \frac{2 T}{\kappa} k_\alpha k_\delta \int_q (k_\beta
+ q_\beta) (k_\gamma +q_\gamma) R_{\alpha \beta, \gamma \delta}(\qb)
\frac{1}{|\kb+\qb|^4} \eea from which we will identify the corrections
to $\kappa$ and $\gamma$ from the small external momentum $k$
expansion \bea \delta \Sigma(k)= \delta \gamma \, k^2 + \delta \kappa
\, k^4 + O(k^6) \eea

The calculation of \eqref{calc} proceeds by inserting again $R_{\alpha \beta, \gamma \delta}(\qb)=
\sum_{i=1}^5 w_i (W_i(\qb))_{\alpha \beta, \gamma \delta}$, performing the expansion at small $k$ of 
the numerator, and the resulting contractions of indices. In the course of the
calculation one needs the leading behavior near $D=4$ and expansion in $k$ of 
three integrals. One uses the expansion 
\be
\frac{1}{|\kb + \qb|^4 } = \frac{1}{q^4} ( 1 - 4 \frac{\kb \cdot \qb}{q^2} -2 \frac{k^2}{q^2} + 12 \frac{(\qb \cdot \kb)^2}{q^4} + O(k^3) )
\ee
The first integral is 
\be
\int \frac{d^D q}{(2 \pi)^D} \frac{1}{|\qb+\kb|^4} q_\alpha 
= \int \frac{d^D q}{(2 \pi)^D} \frac{q_\alpha}{q^4} \left( 1 - 4 \frac{\kb \cdot \qb}{q^2} + O(k^2) \right)
= - 4 k_\beta \frac{\delta_{\alpha \beta}}{D} \int \frac{d^D q}{(2 \pi)^D} \frac{1}{q^4} + O(k^2) 
\ee 
It can also be obtained by taking the ratio $\lim_{b \to 0, D \to 4} \frac{I_\alpha(a=2,b)}{I(a=2,b)} = - p_\alpha$
using Eqs. A34 and A43 in \cite{LRReview}. 

The second integral is
\bea
&& \int \frac{d^D q}{|\qb+\kb|^4} \frac{q_\alpha q_\beta q_\gamma}{q^2} =  \int \frac{d^D q}{q^4} \frac{q_\alpha q_\beta q_\gamma}{q^2} 
\left( 1 - 4 \frac{\kb \cdot \qb}{q^2} + O(k^2) \right) 
\\
&&
%&& = - \frac{4}{ D(D+2)} k_\delta (\delta_{\alpha \beta} \delta_{\gamma \delta} + 
%\delta_{\alpha \gamma} \delta_{\beta \delta} + \delta_{\alpha \delta} \delta_{\beta \gamma} ) 
%\int \frac{d^D q}{(2 \pi)^D} \frac{1}{q^4}   
= - \frac{4}{ D(D+2)} (\delta_{\alpha \beta} k_{\gamma} + 
\delta_{\alpha \gamma} k_{\beta} + k_{\alpha} \delta_{\beta \gamma} ) 
\int \frac{d^D q}{(2 \pi)^D} \frac{1}{q^4} + O(k^2)
\eea
One can check that this is also the result from A34 and A51 in \cite{LRReview},
i.e. $\lim_{b \to 0, D \to 4, D=2 a+2 b} \frac{I_{\alpha \beta \gamma}(a=2,b+1)}{I(a=2,b)}$,
being careful to obey the constraint $D=2 a + 2 b$ when taking the limits.

The third integral is
\bea
&& \int \frac{d^D q}{|\qb+\kb|^4} q_\alpha q_\beta = 
\int \frac{d^D q}{q^4} q_\alpha q_\beta 
\left( 1 - 4 \frac{\kb \cdot \qb}{q^2} -2 \frac{k^2}{q^2} + 12 \frac{(\qb \cdot \kb)^2}{q^4} \right)
\\
%&& = \frac{\delta_{\alpha \beta}}{D} \int \frac{d^D q}{(2 \pi)^D} \frac{1}{q^2} 
%-2 k^2 \int \frac{d^D q}{q^6} q_\alpha q_\beta 
%+ 12 \int \frac{d^D q}{q^8} q_\alpha q_\beta (q \cdot k)^2 \\
&& = \frac{\delta_{\alpha \beta}}{D} \int \frac{d^D q}{(2 \pi)^D} \frac{1}{q^2} 
-2 k^2 \frac{1}{D} \delta_{\alpha \beta} \int \frac{d^D q}{(2 \pi)^D} \frac{1}{q^4} 
+ \frac{12}{D (D+2)}  k_\gamma k_\delta 
(\delta_{\alpha \beta} \delta_{\gamma \delta} + 
\delta_{\alpha \gamma} \delta_{\beta \delta} + \delta_{\alpha \delta} \delta_{\beta \gamma} ) 
\int \frac{d^D q}{(2 \pi)^D} \frac{1}{q^4} \\
&& = \frac{\delta_{\alpha \beta}}{D} \int \frac{d^D q}{(2 \pi)^D} \frac{1}{q^2} 
+ \left( \frac{1}{D} \delta_{\alpha \beta} (-2 + \frac{12}{D+2}) k^2 + \frac{24}{D(D+2)} k_\alpha k_\beta\right)
 \int \frac{d^D q}{(2 \pi)^D} \frac{1}{q^4} 
\eea
It can also be obtained from $
\lim_{b \to 0, D \to 4} \frac{I_{\alpha \beta}(a=2,b)}{I(a=2,b)} = p_\alpha p_\beta$
from A34 and A48 in \cite{LRReview}.\\

We finally obtain the corrections $\delta \gamma$ and $\delta \kappa$ as
\bea
\delta \gamma = \frac{2 T}{\kappa} \frac{\left((D-1) w_2+2 w_5\right)}{2 D} \int \frac{d^D q}{(2 \pi)^D} \frac{1}{q^2} 
=|_{D\to 4} \frac{2 T}{\kappa}  \frac{1}{8} \left(3 w_2+2 w_5\right) \int \frac{d^D q}{(2 \pi)^D} \frac{1}{q^2} 
\eea 

\bea
&& \delta \kappa = \frac{2 T}{\kappa}  
\frac{-\left(D^2+D-2\right) w_2+(D-2) (D+1) w_1+D w_3-6
   \sqrt{D-1} w_4-2 D w_5+w_3+11 w_5}{D (D+2)}
   \int \frac{d^D q}{(2 \pi)^D} \frac{1}{q^4} \\
 &&  =|_{D\to 4}  \frac{2 T}{\kappa}  \frac{1}{24} \left(10 w_1-18 w_2+5 w_3-6 \sqrt{3} w_4+3
   w_5\right) \int \frac{d^D q}{(2 \pi)^D} \frac{1}{q^4} \label{dkappa} 
\eea 
where we will calculate the remaining integral using momentum shell $\int \frac{d^D q}{(2 \pi)^D} \frac{1}{q^4} 
\to \intqq \frac{1}{q^4}$. The correction $\delta \gamma$ (given by a UV divergent integral) obtained above
is analogous to the usual non universal 
shift in the critical temperature for $O(N)$ models, and of little interest to us since
we will tune the bare $\gamma$ so that the system is at its critical point $\gamma_R=0$ (i.e. $\gamma + \delta \gamma=0$). Said otherwise, the bare term in the model is $\frac{1}{2} (\gamma-\gamma_c) (\nabla h)^2$.

\subsection{D 3. Final RG equations}

We now use that the integral $\intp \frac{1}{p^4} = C_D \frac{1}{\epsilon} (e^{\eps d \ell}-1) \Lambda_\ell^{-\epsilon} = C_4 \Lambda_\ell^{-\epsilon} d\ell + O(\epsilon)$ with $\epsilon=4-D$ and $C_4=\frac{1}{8 \pi^2}$. We
define the scaled dimensionless coupling 
\be \label{wresc} 
\tilde w_i = \frac{T}{\kappa^2} w_i C_4 \Lambda_\ell^{-\epsilon}
\ee
To derive the flow equation we calculate $\partial_\ell \tilde w_i$ taking into account 
(i) the rescaling (ii) the sum of the three diagrams which correct $R$ (specifying $D=4$)
leading to $\delta \tilde w_i = \beta_i[\tilde w] d\ell = \sum_{j,k} c_{ijk} \tilde w_j \tilde w_k d\ell$ 
(iii) the extra term from the correction $\delta (\kappa^{-2}) = - \frac{2}{\kappa^3} \delta \kappa$
which leads to the $\eta$ function, $\eta[\tilde w]$.
This leads to the RG equation 
\bea \label{rgflow0} 
\partial_\ell \tilde w_i = \epsilon \tilde w_i + \beta_i[\tilde w] 
- 2 \eta[\tilde w] \tilde w_i   \quad , \quad \eta[\tilde w] = \frac{\partial_\ell \kappa}{\kappa}
\eea
where \eqref{dkappa} leads to (from now on for notational convenience we will suppress the tilde on $w$) 
\bea \label{etaw} 
\eta[w]= \frac{1}{12} \left(10 w_1-18 w_2+5 w_3+3 w_5-6 w_{44}\right)
\eea 
gives the $\eta$ exponent at the fixed point. Putting all together, the final RG equations are 
(with $w_{44} = \sqrt{3} w_4$)
\bea \label{rgflow} 
&& \partial_\ell w_1 = \epsilon w_1+ \frac{1}{12} \left(-(d_c+20) w_1^2+2 \left(19 w_2-5 w_3-5 w_5+6
   w_{44}\right) w_1-7 w_2^2-4 w_5^2-8 w_2 w_5\right) \\
 && \partial_\ell w_2 = \epsilon w_2+ \frac{1}{12}
   \left(-(d_c-31) w_2^2-20 w_1 w_2-2 \left(5 w_3+9 w_5-6 w_{44}\right)
   w_2-4 w_5^2\right)
  \nn  \\
   && \partial_\ell w_3 = \epsilon w_3+ \frac{1}{24} \bigg( -5 d_c w_3^2-d_c w_{44}^2-2 d_c
   w_3 w_{44}-17 w_2^2+46 w_3 w_2 \nn \\
   && -10 w_{44} w_2-20 w_3^2-20 w_5^2-40
   w_1 w_3+24 w_3 w_{44} -4 w_5 \left(w_2+8
   w_3+w_{44}\right)\bigg) \nn \\
   && \partial_\ell w_{44} = \epsilon w_{44}+ \frac{1}{24} \bigg(-w_{44} \left(5 (d_c+4)
   w_3+(d_c-24) w_{44}+40 w_1\right)-w_5 \left(3 (d_c+2) w_3+(3 d_c+28)
   w_{44}\right) \nn \\
   && -3 w_2^2+\left(-15 w_3-3 w_5+56 w_{44}\right) w_2-18
   w_5^2\bigg) \nn \\
   && \partial_\ell w_5 = \epsilon w_5+ \frac{1}{72} \left(-9 (d_c+12) w_5^2-6 w_5 \left((d_c-10)
   w_{44}+20 w_1-27 w_2+10 w_3\right)-5 \left(d_c w_{44}^2+9 w_2^2+6
   w_{44} w_2\right)\right) \nonumber \nn
\eea \\

{\bf Large $d_c$ limit}. In the above RG equations \eqref{rgflow} the couplings $w_i$ have not been rescaled
by $1/d_c$. If one rescales them, and then take the large $d_c$ limit one obtains
\bea \label{epsdc} 
&& \partial_\ell w_1 = \epsilon w_1  -\frac{w_1^2}{12}\ , \quad 
\partial_\ell w_2 = \epsilon w_2  -\frac{w_2^2}{12}\ , \quad
\partial_\ell w_3 = \epsilon w_3 - 
   \frac{1}{24} \left(5 w_3^2+2 w_{44} w_3+w_{44}^2\right)\ , \\
   && 
 \partial_\ell w_{44} = \epsilon w_{44} 
   -\frac{1}{24} \left(3 w_5 \left(w_3+w_{44}\right)+w_{44} \left(5
   w_3+w_{44}\right)\right)\ , \quad 
   \partial_\ell w_5 = \epsilon w_5 - \frac{1}{72} \left(9 w_5^2+6 w_{44} w_5+5
   w_{44}^2\right). \nn
\eea
Recall that $w_i$ here is in fact the rescaled coupling $\tilde w_i$ given in \eqref{wresc}.
Hence comparing with \eqref{hatw} (the factor $T$ being omitted there) we see that
we can identify $w_i \equiv \frac{1}{8 \pi^2} \hat w_i$. Inserting into \eqref{epsdc}
we obtain a set of RG equations for the $\hat w_i$ which, as one can check using
$\lim_{\epsilon=4-D \to 0} \epsilon a_i(D,0) = \frac{1}{192 \pi^2} \{2,2,5,\sqrt{3},3\}$
and $w_{44}=\sqrt{D-1} w_4$, agree exactly with the RG equations at large $d_c$ 
\eqref{rglarge} for $D=4$. Finally note that $\eta[w] = O(1/d_c)$ at large $d_c$ consistent
with the SCSA and large $d_c$ expansion.

\subsection{D 4. Analysis of the RG equations}

\medskip

{\bf Instability of the isotropic membrane fixed point}. The case of the rotationally invariant membrane is obtained setting $w_2=w_4=w_5=0$, which is a manifold preserved by the RG. The RG 
flow \eqref{rgflow} then reduces within this subspace $(w_1,w_3)$ to 
\bea 
&& \partial_\ell w_1 = \epsilon w_1 -\frac{1}{12} w_1 \left((d_c+20) w_1+10
   w_3\right)\ , \\
&&   \partial_\ell w_3 = \epsilon w_3
   -\frac{5}{24} w_3 \left((d_c+4) w_3+8
   w_1\right). 
\eea
We recall that in that subspace $(w_1,w_3)$ are related to $(\mu,\lambda)$ via 
$w_1= \mu$ and $w_3 = 
\mu + (D-1) \frac{\mu \lambda}{\lambda+ 2 \mu}$ as obtained from \eqref{wmu}, and given in \eqref{w13mu} in the text.
Using that relation one can derive RG equations for $\mu$ and $\lambda$ which can be checked to
be identical to the one in Ref. \cite{AL} (taking into account a difference by a factor of $4$ in the definition of $\mu,\lambda$ there).
There are four fixed points 
\be
\left\{w_1\to 0,w_3\to \frac{24 \epsilon}{5
   (d_c+4)}\right\},\left\{w_1\to \frac{12 \epsilon}{d_c+24},w_3\to
   \frac{24 \epsilon }{5 (d_c+24)}\right\},\left\{w_1\to 0,w_3\to
   0\right\},\left\{w_1\to \frac{12 \epsilon}{d_c+20},w_3\to
   0\right\}\ ,
\ee
which correspond to (in the same order \cite{footnote3})
\bea
(\mu,\lambda) = (0,0) ; ( \frac{12 \epsilon}{24 + d_c} , \frac{-4 \epsilon}{24 + d_c} ) ; ( 0, \frac{2 \epsilon}{d_c} ) ; ( \frac{12 \epsilon}{20 + d_c} ,  \frac{-6 \epsilon}{20 + d_c} )\ . 
\eea 
The second one is the standard fixed point which describes the isotropic flat membrane within the $\epsilon$-expansion \cite{AL}. The third one describes the fixed connectivity fluid (zero shear modulus), that is a model for nematic elastomer membranes\cite{LRelastomer}. The fourth one
is located on the line where the bulk modulus vanishes, i.e. 
$2 \mu + D \lambda = \frac{2 (D-1) w_1 w_3}{D w_1-w_3}=0$ which separates the thermodynamically stable and unstable regions of parameters, and controls the transition between these regions. The exponent $\eta$ is given by 
\be
\eta= \eta[w] = \frac{5}{12} (2 w_1 + w_3) = \frac{5 \mu  (\lambda +\mu )}{2 (\lambda +2 \mu )}
\ee 
and gives $\eta^{\rm iso}=\frac{12 \epsilon}{24 + d_c} $ for the isotropic membrane, as in \cite{AL}. 

Let us now discuss the stability of the isotropic membrane to the non-rotationally invariant terms (due to an external orienting field $\vec E$)
in the model. For this we calculate the eigenvalues and associated eigenvectors (represented as columns) of the Hessian around the isotropic fixed point, which are given by
\bea
&& \left\{0,-\frac{\epsilon d_c}{d_c+24},\frac{\epsilon d_c}{d_c+24},\frac{\epsilon d_c}{d_c+24},-\epsilon\right\} \\
%&& \left(
%\begin{array}{ccccc}
% \frac{16}{d+24} & 0 & -\frac{2 (d+8)}{5 (d+24)} & 1 & 0 \\
% -\frac{1}{2} & 0 & 1 & 0 & 0 \\
% -\frac{19 d+78}{5 (d+12)} & \frac{1}{5} (-d-2) & -\frac{23 d-24}{25
%   (d+12)} & 0 & 1 \\
% -\frac{19 d-6}{3 (d+12)} & -\frac{d}{3} & -\frac{26 (d-6)}{15
%   (d+12)} & 1 & 0 \\
% \frac{5}{2} & 0 & 1 & 0 & 0 \\
%\end{array}
%\right) \\
&& 
\left(
\begin{array}{ccccc}
 \frac{16}{d_c+24} & -\frac{1}{2} & -\frac{19 d_c+78}{5 (d_c+12)} & -\frac{19 d_c-6}{3 (d_c+12)} &
   \frac{5}{2} \\
 0 & 0 & \frac{1}{5} (-d_c-2) & -\frac{d_c}{3} & 0 \\
 -\frac{2 (d_c+8)}{5 (d_c+24)} & 1 & -\frac{23 d_c-24}{25 (d_c+12)} & -\frac{26 (d_c-6)}{15 (d_c+12)}
   & 1 \\
 1 & 0 & 0 & 1 & 0 \\
 0 & 0 & 1 & 0 & 0 \\
\end{array}
\right)\ .
\eea
The second and last columns are the two stable directions which are also obtained if one diagonalises
the flow inside the isotropic subspace. In the full space of
five couplings however, we see that the isotropic fixed point is {\it unstable} in two directions, with 
eigenvalue $\rho= \frac{\epsilon d_c}{d_c+24}$, and marginal
in a third direction. 

{\it Crossover for small anisotropy}. To discuss the effect of a small anisotropy let us first recall the analysis of the length scales in the isotropic
membrane. The dimensionless couplings $\tilde w_1, \tilde w_3$ (we temporarily restore the tilde) at scale $L$
are of order 
\bea 
&& \tilde w_{1,3} \sim  \frac{T K_0}{\kappa^2} L^{4-D} \quad , \quad  L < L_{\rm anh} \sim (\frac{\kappa^2}{T K_0})^{1/(4-D)}\ , \\
&& \tilde w_{1,3}  \simeq \tilde w^*_{1,3} \sim \frac{T K_0(L)}{\kappa(L)^2} L^{4-D} \quad , \quad  L > L_{\rm anh}\ ,
\eea 
where $L_{\rm anh}$ is the length scale below which the harmonic theory holds (and the elastic moduli and bending rigidity equal their bare values). For $L > L_{\rm anh}$ these are corrected and one has $\kappa(L) \sim \kappa (L/L_{\rm anh})^\eta$ and $K_0(L) \sim K_0 (L/L_{\rm anh})^{-(4-D-2 \eta)}$. The length $L_{\rm anh}$ is itself
determined when $\tilde w_{1,3}$ reach numbers of order unity, of order their value at the fixed point.

Consider now the model in presence of very small bare symmetry breaking couplings $\mu_1,\mu_2,\lambda_1,\lambda_2$ assumed to be of the same order. Then, from \eqref{wmu} the bare $w^0_2,w^0_5,w^0_{44}$ are linear combinations of those, hence small and of the same order. These couplings are relevant
and grow as 
\bea 
&& \tilde w_i \sim  \frac{T w_i^0}{\kappa^2} L^{4-D} \quad , \quad  L < L_{\rm anh}\ , \\
&& \tilde w_i  \sim \frac{T w_i^0}{\kappa^2} L_{\rm anh}^{4-D}  (\frac{L}{L_{\rm anh}})^{\rho} \quad , \quad  L > L_{\rm anh}\ ,
\eea 
where $w_i^0$ denote any linear combination of the bare symmetry breaking couplings ($i=2,4,5$) and $\rho$ was calculated 
above in the $\epsilon$ expansion. The length scale $L_{\rm anis}$ beyond which anisotropy will change the
property of the system is obtained when $\tilde w_i$ becomes of order unity, hence
\be
L_{\rm anis} \sim L_{\rm anh} (\frac{K_0}{w_i^0})^{1/\rho} \quad , \quad \rho = \frac{\epsilon d_c}{d_c+24} + O(\epsilon^2)\ , 
\ee
whenever $w_i^0 \sim \mu_{1,2}, \lambda_{1,2} \ll K_0$. \\

%
%An initially small anisotropy $w_i^0$ grows as $\tilde w_i \equiv w_i \sim e^{\rho \ell} C_4 \frac{T}{\kappa^2} \Lambda^{-\epsilon} w_i^0$, hence it will be felt only for length scales larger than 
%\be
%L_{\rm anis} \sim (C_4 \frac{T}{\kappa^2} \Lambda^{-\epsilon} w_i^0)^{1/\rho}
%\ee
%where the quantity in parenthesis is dimensionless, it is the bare dimensionless coupling.
%{\red P: Leo can one improve this estimate beyond the RG using SCSA and the $\eta_{AL}$ exponent?
%Or some clever dimensional argument? 
%what is $\Lambda$ there in practice? Can one define a crossover exponent as the ratio
%of the stable to unstable eigenvalues and how?
%For the isotropic membrane fixed point the dimensionless coupling is $T K_0(L)/\kappa(L)^2 L^\epsilon$
%It goes to a fixed point, i.e. a fixed universal amplitude and become $L$ independent. 
%For $L< L_{\rm anharm}$ the couplings have not been corrected yet and the
%dimensionless coupling is growing as $T K_0/\kappa^2 L^\epsilon$. It feels the nonlinearity
%of the RG when it reaches order unity. This provides the anharmonic scale $L_{\rm anharm}$.
%Maybe here one needs to write
%\be
%\frac{T \mu_2(L)}{\kappa(L)^2} L^\epsilon = \frac{T \mu_2}{\kappa^2} L_0^\epsilon (\frac{L}{L_0})^{\rho}
%\ee 
%and when it is of order unity things change.} \\
%
%where I replaced
%$\Lambda=1/L_c$. Saying that it is unity gives that $L_c= \kappa/\sqrt{T K_0}$ for $D=2$ which is
%the anharmonic length. 

{\bf Search for new fixed points}

We now study the RG flow \eqref{rgflow} in the five parameter space, for general codimension $d_c$.

For the physical case, $d_c=1$, we find 12 real fixed points. However all of them are repulsive, one with two
unstable directions, the others with even more. Hence around $D=4$ there is no perturbative fixed point
and we have a runaway RG flow. 

We find that an attractive fixed point exists only for high enough $d_c$. The situation is very similar
to the one for the crumpling transition, with $d$ replaced by $d_c$. For instance, for 
$d_c=220$ we find one, and only one, fully attractive fixed point
%\be
%w_i = \{0.0506278435265324, 0.0506278, 0.019124422987485, \
%-0.031503420539048, 0.0401267 \}
%\ee
\be
w_i = \{0.05063, 0.05063, 0.01912, \
-0.03150, 0.04012 \},
\ee
with eigenvalues ${-1., -0.86129, -0.86129, -0.46296, -0.08355}$
%${-1., -0.861292, -0.861292, -0.462961, -0.0835538}$. 
One can check that this fixed point lies in the manifold 
\bea \label{manif} 
w_1=w_2=\mu_0 \quad , \quad w_3 = \frac{1}{2}(D-1) \lambda_0 + \mu_0 \quad , \quad 
w_4 = \frac{1}{2} \sqrt{D-1} \lambda_0 \quad , \quad w_5 = \frac{1}{2} \lambda_0 + \mu_0\ ,
\eea 
with $\mu_0=0.050628$ and $\lambda_0=-0.021002$
%$\mu_0=0.0506278$ and $\lambda_0=-0.0210023$. 
This is the manifold mentioned in the text
which leads to a purely local interaction between the tangent fields, i.e. 
\be
R_{\alpha \beta,\gamma \delta}(q) = \frac{\mu_0}{2} (\delta_{\alpha \gamma} \delta_{\beta \delta} +
\delta_{\alpha \delta} \delta_{\beta \gamma} ) + \frac{\lambda_0}{2} \delta_{\alpha \beta} \delta_{\gamma \delta}\ .
\ee
One can check by inserting \eqref{manif} into \eqref{rgflow} that this manifold is preserved by the RG.
Furthermore, inside this manifold one can check inserting \eqref{manif} into \eqref{etaw} that $\eta[w]=0$ to the order
$O(\epsilon)$, and that the 
RG flow can be written as
\bea \label{flowmu0} 
&& \partial_\ell \mu_0 = \epsilon \mu_0 + \frac{1}{12} \left(-(d_c+21) \mu_0 ^2-\lambda_0 ^2-10 \lambda_0  \mu_0
   \right)\ , \\
      && \partial_\ell \lambda_0 = \epsilon \lambda_0 +
   \frac{1}{12} \left(-(6 d_c+7) \lambda_0 ^2-2 (3 d_c+17) \lambda_0  \mu_0
   -(d_c+15) \mu_0 ^2\right)\ .
\eea 
Defining $u=\mu$ and $v = \lambda/2 + \mu/4$ one can check that these equations
are identical to the Eqs. (5a,b) in Ref. \cite{PKN} (for their $u,v$) setting there $K_4=1/4$. 
Hence they are identical to those of the crumpling transition but with $d \to d_c$. 
From \cite{PKN} we know that this fixed point exists only for $d>219$. 
This fixed point, which we interpret here as describing the anisotropic membrane 
in its flat phase at the buckling transition found here within the RG in the $D=4-\epsilon$ expansion
is the one found within the SCSA (and large $d_c$) expansion described in the Section D4. 
While in the RG it disappears near $D=4$ for $d_c<219$, within the SCSA it survives for
the physical dimension $D=2$ and $d_c=1$. Hence while the RG suggests a fluctuation driven first order transition
in the physical dimension, the SCSA suggests a continuous transition. 
The question of which is the most accurate description is beyond the scope of the present
work and would presumably require numerical simulations, as was the case for the crumpling
transition (see e.g. \cite{MouhannaCrumpling} for discussion and references).

\bigskip
\begin{center}
{\bf E. Renormalization group for the $u,h$ theory}
\end{center}

Here we perform the one loop RG study on the $u,h$ theory given in \eqref{elast}, \eqref{FC0}, i.e. before integration over the phonons. It allows to obtain some extra information (the renormalization of $\mu$) and provides a useful check on the RG flow of the previous Section. 
We can rewrite the model as
\be \label{Fnew} 
{\cal F}[u,\vec h]= \int d^D x \, \, \frac{1}{2} (\nabla^2 h)^2 + \frac{1}{2} (G^{u})^{-1}_{\alpha \beta} u_{\alpha} u_{\beta}
+ u_\alpha C^{\mu+\mu_1,\lambda+\lambda_1}_{\alpha, \gamma \delta}  A_{\gamma \delta} + 
\frac{1}{2} C^{\mu+\mu_2,\lambda+\lambda_2}_{\alpha \beta, \gamma \delta} A_{\alpha \beta} A_{\gamma \delta}
\quad , \quad 
C^{\mu,\lambda}_{\alpha, \gamma \delta} = 
- \partial_\beta C^{\mu,\lambda}_{\alpha \beta, \gamma \delta}\ ,
\ee
where we have defined, in Fourier space, the $u h h$ vertex
\be \label{C1t} 
C^1_{\alpha, \gamma \delta}(\qb)= C^{\mu+\mu_1,\lambda+\lambda_1}_{\alpha, \gamma \delta}(\qb)
=- i \left( (\lambda+\lambda_1) q_\alpha \delta_{\gamma \delta} 
+ (\mu+ \mu_1) (\delta_{\alpha \delta} q_\gamma + \delta_{\alpha \gamma} q_\delta) \right)\ ,
\ee 
and the bare phonon propagator
\be
(G^{u})^{-1}_{\alpha \beta}(\qb) = \mu P_{\alpha \beta}^T(\qb) + (\lambda + 2 \mu) P_{\alpha \beta}^L(\qb)\ . 
\ee 

Here we calculate the corrections to the vertices, hence we evaluate to lowest order in the perturbation
theory in the nonlinearities, the vertices of the effective action $\Gamma_{uu}, \Gamma_{uhh}, \Gamma_{hhhh}$. 
These vertices will give us the corrections respectively to $(\mu,\lambda)$, $(\mu_1,\lambda_1)$ and
$(\mu_2,\lambda_2)$. The corresponding diagrams are shown in the Fig.\ref{huRGdiagramsFig}.

\begin{figure}[ht]
\includegraphics[width=0.7\linewidth]{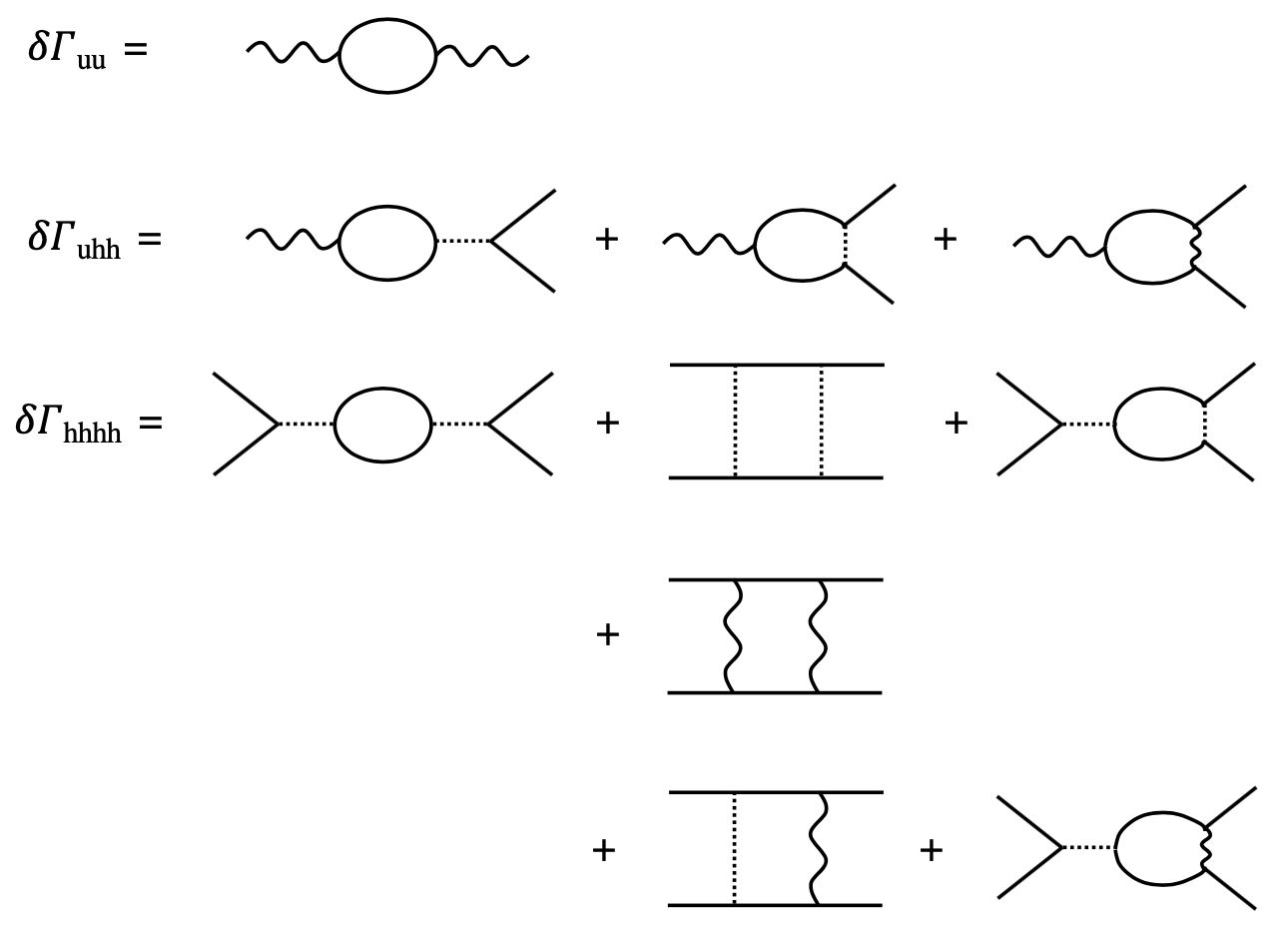}
\caption{Feynman diagrams for one-loop corrections to the renormalized vertices in 
the $u-h$ description of the critical buckling membrane.}
\label{huRGdiagramsFig} 
\end{figure}

\subsection{E. 1 Calculation of $\Gamma_{uu}$}

Let us calculate the one loop corrections to the phonon propagator, given by the single diagram in Fig.\ref{huRGdiagramsFig}. 
The effective action for the $u^2$ term is given, to one loop, as
\bea
&& \frac{1}{2} \int_\qb u(\qb) \cdot \Gamma_{uu}(\q) \cdot u(-\q) = \int_\q \frac{1}{2} u_{\alpha}(\q) u_{\beta}(-\q) [ G^{-1}_{\alpha \beta}(\q)
- C^{\mu+\mu_1,\lambda+\lambda_1}_{\alpha, \gamma \delta}(\q) 
C^{\mu+\mu_1,\lambda+\lambda_1}_{\alpha', \gamma' \delta'}(-\q)  \langle A_{\gamma \delta}(\q) A_{\gamma' \delta'}(-\q)
\rangle_0]\ ,\nn\\
\eea
where here and below $\cdot$ denotes index summations. 
We have the following average, performed with the quadratic action 
\bea
\langle A_{\gamma \delta}(\q) A_{\gamma' \delta'}(-\q) \rangle_0 = \frac{1}{2} d_c \Pi_{\gamma \delta,\gamma' \delta'}(\q) 
\quad , \quad 
\Pi_{\alpha \beta, \gamma \delta}(\q)
= {\rm sym} \int_p p_{\alpha} (q_\beta-p_{\beta}) p_{\delta} (q_\gamma-p_{\gamma})
G(\pb) G(\q-\pb)\ ,
\eea
which leads to 
\bea
(\Gamma_{uu})_{\alpha \beta}(\q) = G^{-1}_{\alpha \beta}(\q)
- \frac{d_c}{2} C^{\mu+\mu_1,\lambda+\lambda_1}_{\alpha, \gamma \delta}(\q)
C^{\mu+\mu_1,\lambda+\lambda_1}_{\alpha', \gamma' \delta'}(-\q)
\Pi_{\gamma \delta, \gamma' \delta'}(\q)\ . 
\eea 
%One defines
%\bea
%S_{\alpha \beta, \gamma \delta} = \frac{1}{D(D+2)} (\delta_{\alpha \beta} \delta_{\gamma \delta} 
%+ \delta_{\alpha \gamma} \delta_{\beta \delta}  + \delta_{\alpha \delta} \delta_{\beta \gamma} ) 
%\eea 
Within the Wilson RG and to leading order in $\epsilon$ one has
\bea
\Pi_{\alpha \beta, \gamma \delta}(\q) \simeq \frac{1}{\kappa^2} 
 \int_p \frac{p_{\alpha} p_{\beta} p_{\delta} p_{\gamma}}{p^8} 
= \frac{1}{\kappa^2} S_{\alpha \beta, \gamma \delta}
 \int_p \frac{1}{p^4}\ ,
\eea
i.e., the dependence in the external momentum $q \ll p$ is subdominant, where $p$ is the internal momentum in the loop.
We have defined 
\be
S^{(4)}_{\alpha \beta, \gamma \delta}  = S_{\alpha \beta, \gamma \delta} = \frac{1}{D(D+2)} (\delta_{\alpha \beta} \delta_{\gamma \delta} 
+ \delta_{\alpha \gamma} \delta_{\beta \delta}  + \delta_{\alpha \delta} \delta_{\beta \gamma} )\ .
\ee Hence we obtain
\bea
\Gamma_{uu}(q) = G^{-1}(q) - \frac{d_c}{2 \kappa^2} C^1(\q) \cdot S \cdot (C^1(-\q))^T    \int_p \frac{1}{p^4}\ .  
\eea 
Replacing $\int_p \frac{1}{p^4}  \to C_4 \Lambda_\ell^{-\epsilon}$ and
performing the contractions, one obtains the following corrections to $\mu$ and $\lambda$
\bea \label{corr4} 
&& \delta \mu= - \frac{d_c}{12 \kappa^2} (\mu+\mu_1)^2 C_4 \Lambda_\ell^{-\epsilon} d\ell\ , \\
%&& \partial_\ell (\lambda + 2 \mu) = - \frac{d_c}{24 \kappa^2} [ 12 (\lambda + \lambda_1)^2 + 6 (\mu+\mu_1)^2 
%+ 12 (\lambda+\lambda_1)(\mu+\mu_1) ] C_4 \Lambda_\ell^{-\epsilon} \\
&& \delta \lambda = - \frac{d_c}{24 \kappa^2} [ 12 (\lambda + \lambda_1)^2 + 2 (\mu+\mu_1)^2 
+ 12 (\lambda+\lambda_1)(\mu+\mu_1) ] C_4 \Lambda_\ell^{-\epsilon} d\ell\ .  \nn
\eea \\

{\it Exponent $\eta_u$}.
The first equation can be rewritten to obtain the anomalous dimension of the phonon field, i.e the
exponent $\eta_u$ defined by $\mu(L) \sim L^{-\eta_u}$,
\bea
&& \eta_u= - \frac{\delta \mu}{\mu d\ell } = \frac{d_c}{12}  g_\mu =  \frac{d_c}{12} (\tilde w_1 - \tilde w_2)\ , 
\eea 
where we have defined the proper dimensionless coupling 
$g_\mu = \frac{(\mu+\mu_1)^2}{\mu \kappa^2}  C_4 \Lambda_\ell^{-\epsilon}$
which we related to the $\tilde w_i$ using \eqref{wmu} and \eqref{wresc}.
At the fixed point this gives the exponent $\eta_u$:

- at the isotropic membrane fixed point $\tilde w_2=0$ and $\tilde w_1 = \frac{12 \epsilon}{d_c + 24}$,
leading to $\eta_u = \frac{d_c \epsilon}{d_c + 24}$. Since $\eta=\frac{12 \epsilon}{d_c + 24}$ we check,
to first order in $\epsilon$ the exact relation (to all orders), $\eta_u = \epsilon - 2 \eta$ guaranteed by
rotational invariance \cite{AL}. 

- at the anisotropic membrane fixed point $\tilde w_1=\tilde w_2$ hence $\eta_u=0$ to order $O(\epsilon)$. 
The relation $\eta_u = \epsilon - 2 \eta$ does not hold (since $\eta=0$ there, to $O(\epsilon)$). 

Note that one can also define the screening exponent $\eta_w$ for the coupling constants $w_i$
such that $w_i(L) \sim L^{-\eta_w}$. It is given by the graphical corrections 
$\beta_i = \frac{\delta w_i}{w_i}$. Since the RG equation for the scaled dimensionless coupling $\tilde w_i$ reads
$\partial_\ell \hat w_i = (\epsilon - 2 \eta(\hat w) - \beta_i[\tilde w] ) \tilde w_i$, 
at any fixed point one must have $\beta_i=\epsilon - 2 \eta$. In presence of anisotropy, $\beta_i$ becomes different
from $\eta_u$. The nonlinear interactions are still screened, since $\eta < \epsilon/2$ at the anisotropic fixed
point, but this screening is not directly related to the renormalization of $\mu$ and $\lambda$.

%
%One can also define an exponent $\beta_w$ as the anomalous dimension of $w_i$, 
%as 
%
%Let us define the exponent $\gamma$ such that 
%\bea
%\partial_\ell w_i = - \gamma w_i 
%\eea
%i.e. it means that $w_i$ has anomalous dimension $w_i \sim q^\gamma$. 
%To find $\gamma$ we write the RG flow of the dimensionless coupling $\hat w_i$
%\be
%\partial_\ell \hat w_i = (\epsilon - 2 \eta(\hat w) - \gamma(\hat w) ) \hat w_i 
%\ee 
%where $\gamma(\hat w)$ are the graphical corrections to $w_i$. 
%By definition of a fixed point we must thus have 
%\be
%\gamma = \epsilon - 2 \eta
%\ee
%which is now different from $\eta_u$.

\subsection{E 2. Calculation of $\Gamma_{uhh}$}

We now calculate the vertex corrections given by the three diagrams in
Fig.\ref{huRGdiagramsFig}. They are corrections to the term $u h h $ in \eqref{Fnew}, which we
write in the form
\be
\int_q u_\alpha(-\q) \, \delta V_{\alpha, \beta \gamma}(\q) \, A_{\beta \gamma}(\q)\ .
\ee 
%where $\delta V_{\alpha, \beta \gamma} \simeq q_\alpha$ and 
%$C^1_{\alpha \beta , \gamma \delta}(q)=- i q_\alpha 
%C^{\mu+\mu_1,\lambda+\lambda_1}_{\alpha \beta , \gamma \delta}$.
%
%Recall (up to a factor $-i$)
%\bea
%V^{(0)}_{\alpha, \beta \gamma}(q) = (\lambda+\lambda_1) q_\alpha \delta_{\gamma \delta} 
%+ (\mu+ \mu_1) (\delta_{\alpha \delta} q_\gamma + \delta_{\alpha \gamma} q_\delta) 
%\eea
One obtains for the first diagram
\bea
&& \delta V^{(1)}_{\alpha, \beta \gamma}(\q) = - \frac{d_c}{2}  (C^1(\q) \cdot S \cdot C^2)_{\alpha, \beta \gamma} \frac{1}{\kappa^2}\int_\p \frac{1}{p^4}\ , 
\eea
where $C^1(\q)$ is the three index tensor given in \eqref{C1t} (i.e., the bare $uhh$ vertex) and we denote here and below 
$C^2$ the four index tensor $C^{\mu+\mu_2,\lambda+\lambda_2}_{\alpha \beta, \gamma \delta}$
(entering the bare $h^4$ vertex) defined in \eqref{elast}. 
The second diagram gives the correction
\bea
&& \delta V^{(2)}_{\alpha, \beta \gamma}(\q) = -  C^1_{\alpha, \alpha' \alpha''}(\q)
S_{\alpha' \beta' \alpha'' \gamma'} C^2_{\beta' \beta, \gamma' \gamma} \frac{1}{\kappa^2} \int_\p \frac{1}{p^4} \ .
\eea
Finally, the third diagram gives, using that $C^1_{\beta,\gamma \delta}(\q) = - i q_\alpha C^1_{\alpha \beta, \gamma \delta}$ (where the momentum independent four index tensor $C^{\mu+\mu_1,\lambda+\lambda_1}_{\alpha \beta, \gamma \delta}$ is denoted $C^1$)
\bea
&&  \delta V^{(3)}_{\alpha, \beta \gamma}(\q) =  C^1_{\alpha,\beta' \gamma'}(\q)
\left[ S^{(8)}_{\beta' \gamma' \beta'' \gamma'' s' s'' \alpha' \alpha''} \left(\frac{1}{2 \mu + \lambda} - \frac{1}{\mu}\right) 
+ \frac{1}{\mu} S^{(6)}_{\beta' \gamma' \beta'' \gamma'' s' s''} \delta_{\alpha' \alpha''} \right]
C^1_{\alpha' s', \beta'' \beta} C^1_{\alpha'' s'', \gamma'' \gamma} \frac{1}{\kappa^2} \int_\p \frac{1}{p^4},\nn\\ 
\eea 
where we defined the 6 and 8 index symmetric tensors, schematically,
\bea
&& S^{(6)}_{\beta' \gamma' \beta'' \gamma'' s' s''} = \frac{1}{D (D+2) (D+4)} (\delta \delta \delta + 14\  \text{terms})\ , \\
&& S^{(8)}_{\beta' \gamma' \beta'' \gamma'' s' s'' \alpha' \alpha''} = \frac{1}{D (2 + D) (4 + D) (6 + D)}
(\delta \delta \delta \delta + 104\  \text{terms} )\ .
\eea 
Performing the contractions we obtain for $i=1,2,3$
\be
 \delta V^{(i)}_{\alpha, \beta \gamma}(\q) = [ A_i q_\alpha \delta_{\beta \gamma} + B_i 
(q_\beta \delta_{\alpha \gamma} + q_\gamma \delta_{\alpha \beta}) ] \frac{1}{\kappa^2} \int_\p \frac{1}{p^4}\ .
\ee
To display the results more compactly we define the new variables 
\be  \label{LaM} 
\Lambda_i = \lambda_i + \lambda \quad , \quad M_i = \mu_i + \mu \quad , \quad i=1,2\ .
\ee
In terms of these variables the coefficients $A_i,B_i$ read
\bea
&& A_1 = -\frac{d_c}{12}  \left(3 \Lambda _1 \left(2
   \Lambda _2+M_2\right)+M_1 \left(3 \Lambda
   _2+M_2\right)\right) \quad , \quad B_1= -\frac{d_c}{12} M_1 M_2 \ , \\
   && A_2 = 
\frac{1}{12} \left(-3 \Lambda _1 \left(\Lambda
   _2+5 M_2\right)-M_1 \left(\Lambda _2+7
   M_2\right)\right) \quad , \quad B_2=
    -\frac{1}{12} M_1 \left(\Lambda
   _2+M_2\right)\ , \nn \\
&& A_3 = \frac{3 \Lambda _1^3 \mu +\Lambda _1 M_1^2 (9
   \lambda +34 \mu )+M_1^3 (5 \lambda +14 \mu )+13
   \Lambda _1^2 \mu  M_1}{12 \mu  (\lambda +2 \mu
   )} \quad , \quad B_3=
       \frac{M_1 \left(\Lambda _1^2 \mu +M_1^2
   (-(\lambda -2 \mu ))+4 \Lambda _1 \mu 
   M_1\right)}{12 \mu  (\lambda +2 \mu )}\ . \nonumber
\eea 
From these coefficients we directly obtain the  corrections 
\be
 \delta \Lambda_1 = (A_1 + A_2 + A_3) \frac{1}{\kappa^2} \int_\p \frac{1}{p^4}  \quad , \quad
 \delta M_1 = (B_1 + B_2 + B_3) \frac{1}{\kappa^2} \int_\p \frac{1}{p^4} \ .
\ee
Putting all contributions together and replacing $\int_\p \frac{1}{p^4}  \to C_4 \Lambda_\ell^{-\epsilon}$, we obtain, from the vertex corrections 
\bea \label{corrvertex} 
&& \delta M_1 =    \frac{1}{12} M_1 \left(M_2
   \left(-\left(d_c+1\right)\right)-\Lambda
   _2+\frac{\Lambda _1^2 \mu +M_1^2 (-(\lambda -2 \mu
   ))+4 \Lambda _1 \mu  M_1}{\mu  (\lambda +2 \mu
   )}\right) \frac{1}{\kappa^2} C_4 \Lambda_\ell^{-\epsilon} d\ell\ , \\
&&  \delta \Lambda_1 = \frac{1}{12} \bigg(-d_c \left(3 \Lambda _1 \left(2
   \Lambda _2+M_2\right) +M_1 \left(3 \Lambda
   _2+M_2\right)\right) \nn \\
   && +\frac{3 \Lambda _1^3 \mu
    +\Lambda _1 M_1^2 (9 \lambda +34 \mu )+M_1^3 (5
   \lambda +14 \mu )+13 \Lambda _1^2 \mu  M_1}{\mu 
   (\lambda +2 \mu )}-3 \Lambda _1 \left(\Lambda _2+5
   M_2\right)-M_1 \left(\Lambda _2+7 M_2\right)\bigg) \frac{1}{\kappa^2} C_4 \Lambda_\ell^{-\epsilon} d\ell\ . \nn
%   && 
%\delta M_1 =    \frac{1}{12} M_1 \left(M_2
%   \left(-\left(d_c+1\right)\right)-\Lambda
%   _2+\frac{\Lambda _1^2 \mu +M_1^2 (-(\lambda -2 \mu
%   ))+4 \Lambda _1 \mu  M_1}{\mu  (\lambda +2 \mu
%   )}\right) \frac{1}{\kappa^2} C_4 \Lambda_\ell^{-\epsilon}
\eea
To recover the result for the isotropic membrane one sets $\Lambda_i=\lambda$ and $M_i=\mu$ and the above corrections reduce to 
\bea \label{corr5} 
&& \delta \mu = -\frac{d_c}{12 \kappa^2} \mu^2 C_4 \Lambda_\ell^{-\epsilon} d\ell\ , \\
&& \delta \lambda = -\frac{d_c}{12 \kappa^2}  \left(6 \lambda ^2+6 \lambda  \mu
   +\mu ^2\right) C_4 \Lambda_\ell^{-\epsilon} d\ell\ . \nn
\eea 
This simplification occurs because the second and third diagram exactly cancel due to rotational invariance. 
Indeed the corrections \eqref{corr5} coincide with \eqref{corr4} (setting $\mu_1=\lambda_1=0$ there).

\subsection{E. 3 Calculation of $\Gamma_{hhhh}$}

We now calculate the corrections to the $h^4$ vertex in \eqref{Fnew}.
They are given by the six diagrams in Fig.\ref{huRGdiagramsFig}. 
We recall that we denote $C^2$ the four index tensor $C^{\mu+\mu_2,\lambda+\lambda_2}_{\alpha \beta, \gamma \delta}$ which appears in the bare $h^4$ vertex. 

The first diagram gives the following correction to $C^2$
\bea
\delta C^2 = -  \frac{d_c}{2} C^2 \cdot S \cdot C^2 \frac{1}{\kappa^2} \int_\p \frac{1}{p^4}\ .
\eea
The second and third diagram give respectively  
\be
\delta C^2_{\alpha \beta, \gamma \delta} = - 2 {\rm sym} \, C^2_{\alpha \alpha', \gamma \gamma'} C^2_{\beta \beta', \delta \delta'} S_{\alpha' \beta' \gamma' \delta'} \frac{1}{\kappa^2} \int_\p \frac{1}{p^4}  \quad , \quad 
%\ee
%Third graph
%\bea
\delta C^2_{\alpha \beta, \gamma \delta} = -  2 \, C^2_{\alpha \beta, \alpha' \beta'} 
 S_{\alpha' \beta' \gamma' \delta'} C^2_{\gamma \gamma', \delta \delta'} \frac{1}{\kappa^2} \int_\p \frac{1}{p^4} \ .
\ee
The fourth diagram is more complicated
\bea
&& \delta C^2_{\alpha \beta, \gamma \delta} = -  2 C^1_{r_1 s_1,\alpha \alpha'} 
C^1_{r_2 s_2,\gamma \gamma'} 
C^1_{r_3 s_3,\beta \beta'} 
C^1_{r_4 s_4,\delta \delta'} 
\bigg[ \langle \hat p_{\alpha'} \hat p_{\beta'} \hat p_{\gamma'} \hat p_{\delta'} \hat p_{s_1} 
\hat p_{s_2} \hat p_{s_3} \hat p_{s_4} \hat p_{r_1} \hat p_{r_2} 
\hat p_{r_3} \hat p_{r_4} \rangle \left( \frac{1}{\lambda+ 2 \mu} - \frac{1}{\mu}\right)^2 
\\
&& + 
\left( \langle \hat p_{\alpha'} \hat p_{\beta'} \hat p_{\gamma'} \hat p_{\delta'} \hat p_{s_1} 
\hat p_{s_2} \hat p_{s_3} \hat p_{s_4} 
\hat p_{r_3} \hat p_{r_4} \rangle \delta_{r_1 r_2} +
\langle \hat p_{\alpha'} \hat p_{\beta'} \hat p_{\gamma'} \hat p_{\delta'} \hat p_{s_1} 
\hat p_{s_2} \hat p_{s_3} \hat p_{s_4} 
\hat p_{r_1} \hat p_{r_2} \rangle \delta_{r_3 r_4} \right) \frac{1}{\mu} \left( \frac{1}{\lambda+ 2 \mu} - \frac{1}{\mu}\right)
\nn \\
&& + \frac{1}{\mu^2} S^{(8)}_{\alpha' \beta' \gamma' \delta' s_1 s_2 s_3 s_4} \delta_{r_1 r_2} \delta_{r_3 r_4} \bigg]
\frac{1}{\kappa^2} \int_\q \frac{1}{q^4}\ , \nn
\eea
where $\langle \dots \rangle$ denote angular averages and $\hat \p=\p/p$ a unit vector. It was convenient to
use that notational trick, rather than the symmetric tensors of order 10 and 12, as it allows the contractions to
be taken more easily. This is equal to
\bea
&& \delta C^2_{\alpha \beta, \gamma \delta} = \bigg( -  2 C^1_{r_1 s_1,\alpha \alpha'} 
C^1_{r_2 s_2,\gamma \gamma'} 
C^1_{r_3 s_3,\beta \beta'} 
C^1_{r_4 s_4,\delta \delta'} 
\frac{1}{\mu^2} S^{(8)}_{\alpha' \beta' \gamma' \delta' s_1 s_2 s_3 s_4} \delta_{r_1 r_2} \delta_{r_3 r_4} \\
&& 
-2 \left( \frac{1}{\lambda+ 2 \mu} - \frac{1}{\mu}\right)^2  (\Lambda_1 + 2 M_1)^4 S_{\alpha \beta \gamma \delta} \nn \\
&& 
-2 \left( \frac{1}{\lambda+ 2 \mu} - \frac{1}{\mu}\right)\frac{1}{\mu} 
(\Lambda_1 + 2 M_1)^2 [2  ((\Lambda_1 + 2 M_1)^2 - M_1^2) S_{\alpha \beta \gamma \delta} 
+ \frac{M_1^2}{D} (\delta_{\alpha \gamma} \delta_{\beta \delta} + \delta_{\alpha \delta} \delta_{\beta \gamma})  ] 
\bigg) \frac{1}{\kappa^2} \int_\p \frac{1}{p^4}\ . \nn
\eea 

The fifth diagram leads to the correction
\bea
\delta C^2_{\alpha \beta, \gamma \delta} = {\rm sym} \, 4 C^1_{r_1 s_1,\alpha \alpha'} 
C^1_{r_2 s_2,\gamma \gamma'} C^2_{\beta' \beta,\delta' \delta} 
\bigg[ 
\left( \frac{1}{\lambda+ 2 \mu} - \frac{1}{\mu}\right) S^{(8)}_{\alpha' \beta' \gamma' \delta' s_1 s_2 r_1 r_2} 
+ \frac{1}{\mu} S_{\alpha' \beta' \gamma' \delta' s_1 s_2} \delta_{r_1 r_2} \bigg] \frac{1}{\kappa^2} \int_\q \frac{1}{q^4}\ ,
\eea 
and finally, the sixth diagram, to
\bea
\delta C^2_{\alpha \beta, \gamma \delta} = {\rm sym} \, 2 C^2_{\alpha \beta , \alpha' \beta'} 
C^1_{r_1 s_1,\delta \delta'}  C^1_{r_2 s_2,\gamma \gamma'}  
\bigg[ \left( \frac{1}{\lambda+ 2 \mu} - \frac{1}{\mu}\right) S^{(8)}_{\alpha' \beta' \gamma' \delta' s_1 s_2 r_1 r_2} 
+ \frac{1}{\mu} S_{\alpha' \beta' \gamma' \delta' s_1 s_2} \delta_{r_1 r_2} \bigg] \frac{1}{\kappa^2} \int_\q \frac{1}{q^4}\ .
\eea 

Performing the contractions, in total we find for the corrections to the $h^4$ vertex
\bea \label{corrh4} 
&& \delta M_2= \frac{1}{12} \bigg[\frac{-1}{\mu ^2 (\lambda +2 \mu )^2} 
\bigg(\mu ^2 M_2^2  \left(d_c+21\right) (\lambda +2 \mu )^2+\Lambda
   _1^4 \mu ^2+M_1^4 \left(7 \lambda ^2+44 \lambda 
   \mu +76 \mu ^2\right) \\
   && +8 \Lambda _1 \mu  M_1 \left(2
   M_1^2 (\lambda +4 \mu )-5 \mu  M_2 (\lambda +2 \mu
   )\right)+2 \Lambda _1^2 \mu  \left(2 M_1^2 (\lambda
   +8 \mu )-5 \mu  M_2 (\lambda +2 \mu )\right) \frac{1}{\kappa^2} C_4 \Lambda_\ell^{-\epsilon} \nn \\
   && -20 \mu
    M_2 M_1^2 (\lambda +2 \mu ) (\lambda +4 \mu )+8
   \Lambda _1^3 \mu ^2 M_1
 \bigg)
   -\Lambda _2^2+2 \Lambda _2 \left(\frac{\Lambda
   _1^2 \mu +2 M_1^2 (\lambda +4 \mu )+4 \Lambda _1
   \mu  M_1}{\mu  (\lambda +2 \mu )}-5
   M_2\right)\bigg] \frac{1}{\kappa^2} C_4 \Lambda_\ell^{-\epsilon} d\ell\ , \nn
\eea 
and
\bea
&& \delta \Lambda_2=
\frac{1}{12} \bigg[]-\Lambda _2^2 \left(6
   d_c+7\right)-2 \Lambda _2 M_2 \left(3
   d_c+17\right)+M_2^2
   \left(-\left(d_c+15\right)\right)  -\frac{\left(\Lambda _1^2 \mu +M_1^2 (-(\lambda -2 \mu ))+4 \Lambda
   _1 \mu  M_1\right){}^2}{\mu ^2 (\lambda +2 \mu
   )^2} \nn \\
   &&
   +\frac{4 M_2 \left(\Lambda _1^2 \mu +2 M_1^2
   (\lambda +4 \mu )+4 \Lambda _1 \mu  M_1\right)}{\mu
    (\lambda +2 \mu )}+\frac{8 \Lambda _2
   \left(\Lambda _1^2 \mu +2 M_1^2 (\lambda +4 \mu )+4
   \Lambda _1 \mu  M_1\right)}{\mu  (\lambda +2 \mu
   )} \bigg] \frac{1}{\kappa^2} C_4 \Lambda_\ell^{-\epsilon} d\ell\ .
\eea 
To recover the result for the isotropic membrane one sets $\Lambda_i=\lambda$ and $M_i=\mu$ and the above corrections reduce exactly, once again, to \eqref{corr5}. Here the simplification arises from the last five diagram cancelling due to rotational invariance. 

\subsection{E. 4 Final RG equations}

We can now put together $\delta \mu$, $\delta \lambda$ from \eqref{corr4}, 
$\delta M_1=\delta \mu + \delta \mu_1$, $\delta \Lambda_1=\delta \lambda + \delta \lambda_1$
from \eqref{corrvertex}, and 
$\delta M_2=\delta \mu + \delta \mu_2$, $\delta \Lambda_2=\delta \lambda + \delta \lambda_2$
from \eqref{corrh4}. This leads to the complete set of corrections to the six couplings, which
is bulky and which we will not display here in full (see below). Let us denote $m_i$, $i=1,\dots,6$ these couplings.
These corrections read schematically $\delta m_i = d_{ijk} m_j m_k$. To obtain the final RG flow one
defines scaled dimensionless couplings $\tilde m_i$, as in \eqref{wresc}, and take into account the 
corrections to $\kappa$ as we did in \eqref{rgflow0}, leading to
$\partial_\ell \tilde m_i = \epsilon \tilde m_i + d_{ijk} \tilde m_j \tilde m_k - 2 \eta[w[m]] \tilde m_i$.
Here we denote $w[m]$ the $w_i$ expressed as functions of the $m_i$ via the Eq. \eqref{wmu},
and we have used the same formula \eqref{etaw} for the $\eta[w]$ function. 

%{\red P: not quite correct because $\mu$ is not really a coupling, is it?} 

To check that these are consistent with the RG equations obtained via the quartic theory in
Section D, we simply need to compare the above corrections $\delta m$ and $\delta w$, i.e summing \eqref{d1},
\eqref{d2} and \eqref{d3} which can be written as $\delta w_i = \beta_i[w] d\ell = c_{ijk} w_j w_k d\ell$. 
We have performed the check as follows. We have evaluated in two ways
\bea
&& \delta w_i[m] = \frac{\partial w_i[m]}{\partial m_j} \delta m_j = \frac{\partial w_i[m]}{\partial m_j} d_{ijk} m_j m_k d\ell\ , \\
&& \delta w_i[m] = c_{ijk} w_j[m] w_k[m] d\ell\ ,
\eea
and using $w[m]$ from Eq. \eqref{wmu} we have shown using Mathematica that the two lines above are identical functions of the $m_i$. This provides a quite non trivial check of these two lengthy calculations.
Hence the RG equation for the 5 couplings $w_i$ can be deduced from the one for the 6 couplings $m_i$.
The reverse is not true however, there is, in the general case, additional information in the 6 coupling flow, as we discussed
above in Section E. 1 it allows to obtain $\delta \mu$ and from it we obtained there the exponent $\eta_u$, related to the anomalous dimension of the phonon field. Let us indicate for completeness the combination of couplings which enters the exponent $\eta$ 
\be
\eta = \frac{1}{12} \left(10 w_1-18 w_2+5 w_3+3 w_5-6 w_{44}\right) = \frac{M_1 \left(2 \Lambda _1 \mu +3 \lambda  M_1+5 \mu  M_1\right)}{2 \mu  (\lambda +2
   \mu )}\ .
\ee 

To express the RG flow it is natural to define the dimensionless ratio $r = \lambda/\mu$ and the 
four dimensionless coupling constants associated to the nonlinear terms in the action
\be \tilde M_1^2 = \frac{M_1^2}{\mu \kappa^2} C_4 \Lambda_\ell^{-\epsilon} \quad , \quad \tilde \Lambda_1^2 = \frac{\Lambda_1^2}{\mu \kappa^2} C_4 \Lambda_\ell^{-\epsilon} \quad , \quad  \tilde M_2 = \frac{M_2}{\kappa^2} C_4 \Lambda_\ell^{-\epsilon} \quad , \quad \tilde \Lambda_2 = \frac{\Lambda_2}{\kappa^2} C_4 \Lambda_\ell^{-\epsilon}\ , 
\ee
and then $\mu$ can still flow with eigenvalue $\eta_u$. Since the RG equations for these couplings are bulky let us only display them here to leading order in large $d_c$, and we have dropped the tilde for notational convenience
\bea
&& \partial_\ell r = \frac{1}{12} d_c \left(-6 \Lambda _1^2+r  M_1^2-6 \Lambda _1
   M_1-M_1^2\right) \quad , \quad     \partial_\ell M_1 = \frac{\epsilon}{2} M_1 + \frac{1}{24} M_1 \left(M_1^2-2 M_2\right) d_c\ , \\
&&    \partial_\ell \Lambda_1 = \frac{\epsilon}{2} \Lambda_1 + \frac{1}{24} d_c
   \left(-12 \Lambda _1 \Lambda _2+\Lambda _1 M_1^2-6 \Lambda _2 M_1-6 \Lambda _1 M_2-2
   M_2 M_1\right)\ , \\
   && 
   \partial_\ell M_2 = \epsilon M_2 -\frac{1}{12} M_2^2 d_c \quad , \quad    \partial_\ell \Lambda_2 = \epsilon \Lambda_2 + \frac{1}{12} d_c \left(-6 \Lambda _2^2-6
   \Lambda _2 M_2-M_2^2\right)\ . 
\eea
It is easy to see that the only attractive fixed point of these equations (and of the complete equations for any $d_c>219$) is such that 
\be \label{FP44} 
M_1 = 0 \quad , \quad \Lambda_1= 0 \quad , \quad M_2= \frac{12}{d_c} + O(\frac{1}{d_c^2}) \quad , \quad \Lambda_2= - \frac{4}{d_c} + O(\frac{1}{d_c^2})\ .
\ee
This is in agreement with the RG analysis using the $h^4$ theory presented above. Indeed this anisotropic fixed point lies in the manifold \eqref{manif} in the $w_i$ variables, which in the current variables imply the constraints $\mu+\mu_1=0$, $\mu_0=\mu+\mu_2$, $\lambda_0= \lambda + \lambda_2 - \frac{(\lambda + \lambda_1)^2}{\lambda + 2 \mu}$. The fixed point \eqref{FP44} obeys these constraints and one can check
that the values for $M_2$ and $\Lambda_2$ are consistent with those for the fixed point of 
\eqref{flowmu0} at large $d_c$ (and in fact, as one can check, for any $d_c > 219$).

Since the couplings $M_1$ and $\Lambda_1$ flow to zero exponentially
with $\ell$, at the anisotropic fixed point we see that the flow of
$r=\lambda/\mu$ and the flow of $\mu$, which is given (exactly) by
\bea \frac{1}{\mu} \partial_\ell \mu = -\frac{1}{12} M_1^2
d_c-\frac{M_1 \left(2 \Lambda _1+(3 r +5) M_1\right)}{r +2} \eea lead
to finite, but non-universal values for $\lambda$ and $\mu$. This is
consistent with the exponent $\eta_u=0$ as claimed above.

%Howewer this is very undetermined.
% So we need to study in more details the RG in the $u,h$ theory. 

%One way to set up the RG is to define {\red Leo do you think this is the correct/best way?}
%\be
%\lambda = \kappa \mu \quad , \quad \Lambda_1 = \kappa_1 M_1 \quad , \quad \Lambda_2 = \kappa_2 M_2
%\ee
%Then the dimensionless coupling constants are
%\be
%g_2= \frac{M_2}{\kappa^2} C_4 \Lambda_\ell^{-\epsilon} \quad , \quad \frac{M_1^2}{\mu \kappa^2} C_4 \Lambda_\ell^{-\epsilon}
%\ee 
%and $\mu$ can still flow with eigenvalue $\eta_u$. 

%
%
%In the AL manifold the last five diagram all cancel and one obtains
%\bea
%&& \delta \mu = -\frac{1}{12} \mu ^2 d_c \\
%&&\delta \lambda = 
% -\frac{1}{12} d_c \left(6 \lambda ^2+6 \lambda  \mu
%   +\mu ^2\right)
%\eea 
%In agreement with rotational invariance. 

\medskip
\begin{center}
{\bf F. Renormalization group flow of $\gamma$}
\end{center}
\medskip

Until now we have assumed $\gamma$ (and $\tau$) to be tuned so that the system is at the
critical point (the buckling transition), i.e. $\gamma_R=0$. Now we assume a small deviations
away and calculate the RG flow of $\gamma$, and the associated (independent) critical exponent $\nu$. To check consistency, we perform the calculation both in the $h^4$ theory and in the $u h^2 + h^4$ theory.

\subsection{F.1. Flow of $\gamma$ in quartic $h^4$ theory}

To obtain the flow of $\gamma$ to linear order in $\gamma$,
we expand the height field propagator at small $\gamma$ as
\be \label{propG} 
G(k)= \frac{1}{\kappa k^4 + \gamma k^2} = \frac{1}{\kappa k^4} - \frac{\gamma}{\kappa^2 k^6} + O(\gamma^2)\ . 
\ee 
Let us call here $\delta \Sigma(k) = \delta \gamma k^2 + O(k^4)$ 
the part of the self-energy proportional to $O(\gamma)$ at small
$\gamma$ (there is also a $O(1)$ part calculated in Section D.2 which determines
the shift in the critical point $\gamma_c$ (see discussion there) but which is of no interest to us here.
To lowest order in perturbation theory the 
self-energy is given by two diagrams, the sunset diagram in \eqref{calc}, leading to
$\delta \gamma^s$, and the tadpole diagram $\delta \gamma^t$, with $\delta \gamma=\delta \gamma^s+\delta \gamma^t$. From the sunset diagram one has from \eqref{calc}
\bea \label{sigma3} 
\delta \Sigma^s(\kb)  = - \frac{\gamma}{\kappa^2} k_{\alpha} k_{\gamma} 
{2 \over d_c} \sum_{i=1,5} w_i \int_q  \frac{1}{(k-q)^6} 
(k_\beta - q_\beta) (W_i)_{\alpha\beta,\gamma\delta}(\q) (k_\delta-q_{\delta}) 
=  \delta \gamma^s \, k^2 + O(k^4) \ .
\eea
Within Wilson RG, to lowest order in $\epsilon$ one can write
\bea
\delta \Sigma^s(\kb) 
= - \frac{\gamma}{\kappa^2} k_{\alpha} k_{\gamma} 
{2 \over d_c} \sum_{i=1,5} w_i \int_\q  \frac{q_\beta q_{\delta}}{q^6} 
 (W_i)_{\alpha\beta,\gamma\delta}(\q) =
 - \frac{\gamma}{\kappa^2} k^2
{2 \over d_c}  \left[ \left( \frac{1}{2} - \frac1{2 D} \right) w_2 + \frac{1}{D} w_5 \right] \int_\q \frac{1}{q^4}\ .  
\eea 

In addition there is the tadpole contribution, leading to the $O(\gamma)$ correction $\delta \gamma^t$
\be
\Sigma^{\rm t}(k)  = k_\alpha k_\gamma R^0_{\alpha \beta, \gamma \delta} 
\int_q q_\gamma q_\delta G(q) \quad \Rightarrow \quad \delta \gamma^t k^2 = - \gamma k_\alpha k_\gamma R^0_{\alpha \beta, \gamma \delta} \langle q_\gamma q_\delta \rangle \frac{1}{\kappa^2} \int_\q \frac{1}{q^4}\ ,  
\ee 
where $R^0$ is the $\kb=0$ component of the vertex. From \eqref{zm1} it is equal to 
$R^0=\frac{1}{2} \bar C$ where $\bar C$ given in \eqref{zm2}, and more explicitly,
from \eqref{zm3}
\be
R^0_{\alpha \beta, \gamma \delta} = \frac{1}{2} \left(M_2-\frac{M_1^2}{\mu }\right) (\delta_{\alpha \gamma} \delta_{\beta \delta} + \delta_{\alpha \delta} \delta_{\beta \gamma} )
   + 
   \frac{D \lambda  \Lambda _2 \mu -D \Lambda _1^2 \mu +2 \Lambda _2 \mu ^2+2 \lambda 
   M_1^2-4 \Lambda _1 \mu  M_1}{2 \mu  (D \lambda +2 \mu )} \delta_{\alpha \beta} \delta_{\gamma \delta}\ .
\ee 
Using $\langle q_\gamma q_\delta \rangle= \frac{1}{D} \delta_{\gamma \delta}$, performing the contractions, one finds,
for $D=4$ 
\be
\delta \gamma^{\rm t} = - \frac{1}{4} \gamma \left( 2 \Lambda _2+M_2 
- \frac{\left(2 \Lambda_1+M_1\right){}^2}{ (2 \lambda +\mu )} \right) \frac{1}{\kappa^2} \int_\q \frac{1}{q^4} \ .
\ee 
We can express the following combination using the $w_i$ 
\be
2 \Lambda _2+M_2 
- \frac{\left(2 \Lambda_1+M_1\right){}^2}{ (2 \lambda +\mu )} = \frac{3 w_2 \left(3 w_3+w_5+2 w_{44}\right)+4 \left(w_{44}^2-3 w_3 w_5\right)}{12 w_2-3
   w_3-9 w_5+6 w_{44}}\ .
\ee 
%
%The tadpole comes from the zero-mode free energy
%\be
%F = \frac{1}{L^d} \int d^D x A_{\alpha \beta}(x)
% \int d^D x' A_{\gamma \delta}(x') R^0_{\alpha \beta, \gamma \delta} 
%\ee 
%It leads to
%\bea
%2 \frac{1}{L^d} \int d^D x A_{\alpha \beta}(x)
% \int d^D x' \langle A_{\gamma \delta}(x') \rangle R^0_{\alpha \beta, \gamma \delta} 
% = \frac{1}{2} \int d^D x  \delta \gamma (\partial_x h)^2 
%\eea 
%Hence
%\bea
%\delta \gamma = 
%\eea 

Hence we obtain the flow for $\gamma$ in terms of the rescaled couplings defined
in \eqref{wresc}, dropping the tilde ($\tilde w_i \to w_i$) for simplicity
\bea \label{dgammafinal} 
&& \partial_\ell \gamma = 
%- \gamma \bigg[ 
%{1 \over d_c}  [  \frac{3}{4}  \tilde w_2 + \frac{1}{2} \tilde w_5 ] 
%+ \frac{1}{4}  \left( 2 \Lambda _2+M_2 
%- \frac{\left(2 \Lambda_1+M_1\right){}^2}{ (2 \lambda +\mu )} \right) \bigg] \\
%&& = 
- \gamma \bigg[ 
{1 \over d_c}  \left(\frac{3}{4}  w_2 + \frac{1}{2} w_5\right) 
+ \frac{1}{4}  \left( 
\frac{3 w_2 \left(3 w_3+w_5+2 w_{44}\right)+4 \left(w_{44}^2-3 w_3 w_5\right)}{12 w_2-3
   w_3-9 w_5+6 w_{44}} \right) \bigg]\ . 
\eea 
%\bea
%\partial_\ell \gamma = - \gamma \bigg[ 
%{2 \over d_c}  [ ( \frac{D-1}{2 D} ) \tilde w_2 + \frac{1}{D} \tilde w_5 ] 
%+ \frac{1}{4}  \left( 2 \Lambda _2+M_2 
%- \frac{\left(2 \Lambda_1+M_1\right){}^2}{ (2 \lambda +\mu )} \right) \bigg]
%\eea 
%
One can immediately check that for the isotropic membrane the right hand side vanishes exactly.
This arises from rotational invariance, there are no corrections to $\gamma$. Here the bare $\gamma$ is tuned
to the critical point $\gamma_c$ and the flow equation \eqref{dgammafinal} is, more properly,
the RG equation for the deviations to criticality $\gamma \to \gamma-\gamma_c$.

If one now inserts the values for the couplings at the anisotropic fixed point, or 
more generally of any couplings satisfying the constraints \eqref{manif}, one finds
that the ratio appearing in \eqref{dgammafinal} is of the form $0$ divided by $0$, i.e.
it is undetermined. We resolve this ambiguity in the next section by studying the $u-h$ theory. To this end we study the correlation length exponent related to the eigenvalue of $\gamma$.

\bigskip

{\bf Correlation length exponent $\nu$} 

From the propagator \eqref{propG} the bare correlation length is
$\xi_0 = \sqrt{\kappa/\gamma}$. Let us write \eqref{dgammafinal} as 
$\partial_\ell \gamma = \theta \gamma$. At the fixed point 
$\gamma(L) = \gamma_0 L^\theta$, where $\gamma_0$ is the bare value. The correlation length $\xi$
is defined by balancing $\kappa(\xi) \xi^{-4} \sim \gamma(\xi)  \xi^{-2}$. Taking into account
that $\kappa(\xi) \sim \xi^{\eta}$, we obtain
\be
\xi = \gamma_0^{-\nu} \quad , \quad \nu = \frac{1}{2 + \theta - \eta}\ .
\ee

\subsection{F.2. Flow of $\gamma$ in quartic $h,u$ theory}

We now calculate the corrections to $\gamma$ within the model described in \eqref{FC0},\eqref{elast},
and also in \eqref{Fnew}, whose RG was studied in Section E. The nonlinear terms are 
%Here we calculate the coarse-graining correction i.e., the RG flow of
%the anisotropic parameter (``tension'') $\gamma$, defined by the
%Hamiltonian, $H = H_1 + H_2$, that can be written as,
\begin{eqnarray}
%{\cal H} &=&\frac{1}{2}h_i G^{-1}_h h_i
%+ \frac{1}{2}u_\alpha G^{-1}_{\alpha\beta}u_\beta\\ 
%&+&
\frac{1}{2} C^1_{\alpha\beta\gamma\delta}\partial_\alpha
u_\beta\partial_\gamma\vec h\cdot\partial_\delta\vec h
+\frac{1}{8} C^2_{\alpha\beta\gamma\delta}
(\partial_\alpha\vec h\cdot\partial_\beta\vec h)
(\partial_\gamma\vec h\cdot\partial_\delta\vec h),\nonumber
\end{eqnarray} 
where we recall that 
\begin{eqnarray}
C^{1,2}_{\alpha\beta\gamma\delta} &=&
\Lambda_{1,2}\delta_{\alpha\beta}\delta_{\gamma\delta}
+ M_{1,2}(\delta_{\alpha\beta}\delta_{\gamma\delta}
+ \delta_{\alpha\beta}\delta_{\gamma\delta})
\end{eqnarray} 
in terms of the coupling defined in \eqref{LaM}. In Fourier space, we recall that the propagator of the
phonon field $u_\alpha$ is given by \eqref{D} and the propagator of the height field $h$ field by 
\eqref{propG}. 

%propagators are given by
%\begin{eqnarray}
%G_h(k) &=&\frac{1}{\kappa k^4 + \gamma k^2},\\ 
%G_{\alpha\beta}(k) &=&\frac{1}{\mu k^2}P^T_{\alpha\beta}
%+ \frac{1}{(2\mu + \lambda)k^2}P^L_{\alpha\beta}.
%\end{eqnarray}
%We have already calculated the coarse-graining RG corrections to other
%parameters, $\kappa$, $\mu$, \ldots. 

The contribution to $\delta\gamma = \delta\gamma_u + \delta\gamma_h$ is given by (i) two sunset diagrams, giving $\delta \gamma_u^{\rm s}$ and
$\delta \gamma_h^{\rm s}$: they correspond respectively to expansion to second order in the cubic phonon vertex and to first order expansion in the quartic vertex (ii) two tadpole diagrams 
$\delta \gamma_u^{\rm t}$ and
$\delta \gamma_h^{\rm t}$.

The "sunset" diagram involving phonons gives the following correction, evaluated to
lowest order in $\epsilon$ 
\begin{eqnarray}
\delta\gamma^{\rm s}_u &=& -\frac{1}{2!}\left(\frac{1}{2}\right)^2 2^3 
\hat k_\gamma\hat k_{\gamma'}
C^1_{\alpha\beta\gamma\delta}C^1_{\alpha'\beta'\gamma'\delta'}
\int^>_\q \frac{({\bf k}-{\bf q})_\delta({\bf k}-{\bf
    q})_{\delta'}q_\alpha q_{\alpha'}}
{\left[\kappa({\bf k}-{\bf q})^4 + \gamma({\bf k}-{\bf q})^2\right]q^2}
\left[\frac{P^T_{\beta\beta'}(\q)}{\mu} +
  \frac{P^L_{\beta\beta'}(\q)}{2\mu+\lambda}\right],\nonumber\\
&=& -\hat k_\gamma\hat k_{\gamma'}
C^1_{\alpha\beta\gamma\delta}C^1_{\alpha'\beta'\gamma'\delta'}
\int^>_\q \frac{q_\delta q_{\delta'} q_\alpha q_{\alpha'}}
{\left[\kappa q^2 + \gamma\right]q^4}
\left[\frac{P^T_{\beta\beta'}(\q)}{\mu} +
\frac{P^L_{\beta\beta'}(\q)}{2\mu+\lambda}\right],\nonumber\\
&=& -\hat k_\gamma\hat k_{\gamma'}
C^1_{\alpha\beta\gamma\delta}C^1_{\alpha'\beta'\gamma'\delta'}
\left[\frac{\Lambda^2 C_4 d\ell}{\kappa\mu}\left(\delta_{\beta\beta'}
\langle q_\delta q_{\delta'} q_\alpha q_{\alpha'}\rangle
-\frac{\mu+\lambda}{2\mu+\lambda}\langle q_\delta q_{\delta'} q_\alpha 
q_{\alpha'}q_\beta q_{\beta'}\rangle\right)\right.\nonumber\\
&&\left.-\frac{\gamma}{\kappa^2\mu}\left(\delta_{\beta\beta'}
\langle q_\delta q_{\delta'} q_\alpha q_{\alpha'}\rangle
-\frac{\mu+\lambda}{2\mu+\lambda}\langle q_\delta q_{\delta'} q_\alpha 
q_{\alpha'}q_\beta q_{\beta'}\rangle\right)\right] C_4 \Lambda_\ell^{-\epsilon} d\ell.
\end{eqnarray}
Using Mathematica and our spherical averages of product of
$q_\alpha$'s, we find,
\begin{eqnarray}
\delta \gamma^{\rm s}_u = - \frac{1}{4}\left(\frac{\Lambda^2}{\kappa}
-\frac{\gamma}{\kappa^2}\right)
\frac{\mu(\Lambda_1^2 + 4 \Lambda_1 M_1 + 10 M_1^2) + 3\lambda
  M_1^2}{\mu(2\mu+\lambda)} C_4 \Lambda_\ell^{-\epsilon} d\ell. \nonumber\\
\end{eqnarray}
The total correction $\delta\gamma^{\rm s}$ 
involving the $(\partial h)^4$
vertex is given by the sum of the sunset and tadpole diagram as
\begin{eqnarray}
\delta \gamma_h = \delta\gamma^{\rm s}_h + \delta\gamma^{\rm t}_h &=& \frac{2\times 2^2}{8}\hat k_\alpha\hat k_{\gamma}
C^2_{\alpha\beta\gamma\delta}
\int^>_\q \frac{q_\beta q_\delta}{\kappa q^4 + \gamma q^2}+ \frac{2\times 2d_c}{8}\hat k_\alpha\hat k_{\beta}
C^2_{\alpha\beta\gamma\delta}
\int^>_\q \frac{q_\gamma q_\delta}{\kappa q^4 + \gamma q^2},\nonumber\\
&=&\frac{1}{4}C^2_{\alpha\beta\gamma\delta}
\left(\frac{4}{D}\delta_{\beta\delta}\hat k_\alpha\hat k_{\gamma}
+\frac{2d_c}{D}\delta_{\gamma\delta}\hat k_\alpha\hat k_{\beta}\right)
\int^>_\q \frac{1}{\kappa q^2 + \gamma}.\nonumber
\end{eqnarray}
In $D=4$ we find,
\begin{eqnarray}
\delta \gamma_h = \frac{1}{4}\left(\frac{\Lambda^2}{\kappa}
-\frac{\gamma}{\kappa^2}\right)
\left[\Lambda_2 + 5 M_2 + d_c(2 \Lambda_2 + M_2)\right]
C_4 \Lambda_\ell^{-\epsilon} d\ell .\nonumber\\
\end{eqnarray}

We need to calculate the tadpole diagram involving the phonons.  It
arises from the term at zero momentum
$A^0_{\alpha \beta} C^1_{\alpha \beta , \gamma \delta } \langle \tilde
u^0_{\gamma \delta} \rangle$ in the energy \eqref{zero1}. The
expectation value $\langle \tilde u^0_{\gamma \delta} \rangle$ of the
in-plane strain field is given in \eqref{min} as
$\langle \tilde u^0 \rangle = - [C^{\mu,\lambda}]^{-1} C^1 \langle A^0
\rangle$. Hence we find

\be
 \delta \gamma^{\rm t}_u k^2 = \gamma \frac{d_c}{2} k_\alpha k_\beta [ C^1 \cdot [C^{\mu,\lambda}]^{-1} \cdot C^1]_{\alpha \beta,\gamma \delta} \langle q_\gamma q_\delta \rangle \frac{1}{\kappa^2} \int_\q \frac{1}{q^4} 
\ee
leading to 
\be
\delta \gamma^{\rm t}_u = \frac{\gamma}{4} \frac{d_c\left(2 \Lambda _1+M_1\right){}^2}{2 \lambda +\mu}
\frac{1}{\kappa^2} \int_\q \frac{1}{q^4} \ .
\ee

Putting all four contributions together we obtain the $O(\gamma)$ total correction as
\be \label{gg} 
\delta \gamma =  \frac{\gamma}{4 \kappa^2}  \bigg[
\frac{\mu(\Lambda_1^2 + 4 \Lambda_1 M_1 + 10 M_1^2) + 3\lambda
  M_1^2}{\mu(2\mu+\lambda)} - (\Lambda_2 + 5 M_2) + d_c \left(
\frac{\left(2 \Lambda _1+M_1\right){}^2}{2 \lambda +\mu} - (2 \Lambda_2 + M_2) \right)
  \bigg] C_4 \Lambda_\ell^{-\epsilon} d\ell\ ,
\ee 
which leads to the RG flow equation by defining the dimensionless scaled couplings. 
One can check using \eqref{5ratio}, \eqref{newratio} and \eqref{invertmu} that the
RG flow obtained here is formally identical to the one obtained in \eqref{dgammafinal}.

However, now one can check that the indeterminacy mentioned in the previous section is resolved.
Indeed in the expression \eqref{dgammafinal} there is a factor $M_1$ both in numerator and denominator, and since $M_1=0$ at the anisotropic fixed point this led to an ambiguous expression. However, above
these factors cancel and the exponent $\theta$ at the fixed point can be unambiguously determined from \eqref{gg}.
One finds, setting $M_1=\Lambda_1=0$
\be
\theta =  - \frac{1}{4} \left( d_c  (2 \Lambda_2 + M_2)  + \Lambda_2 + 5 M_2 \right)\ .
\ee
We can insert $M_2=\frac{\epsilon}{d_c}( 12 - \frac{640}{3 d_c} + O(\frac{1}{d_c^2}) )$ and 
$\Lambda_2= \frac{\epsilon}{d_c} ( - 4 - \frac{160}{3 d_c} + O(\frac{1}{d_c^2}) ) $, which can be obtained from the RG in the previous section, and obtain
\be
\theta = - \frac{\epsilon}{d_c} (1 - \frac{66}{d_c} + O(\frac{1}{d_c^2}) )\ .
\ee

%\subsection{Model building}

\section{G. Effect of the parameter $\tau$}

As indicated in the text, the parameter $\tau$ simply changes $\zeta$, such that
$\zeta^2$ is the ratio of the projected area of the membrane on its preferred plane
(here $xy$) to its internal size $L^2$. To see this, we rewrite the energy density in ${\cal F}_1$ in terms of
trace and traceless parts of the nonlinear stress tensor 
\be
\mu (u_{\alpha \beta} - \frac{1}{D} \delta_{\alpha \beta} u_{\gamma \gamma})^2 + B u_{\alpha \alpha}^2 + \tau u_{\alpha \alpha} \ ,
\ee 
where $B=\frac{2 \mu+D \lambda}{2 D}$. Completing the square and defining
%One can complete the square
%\be
%\mu (u_{\alpha \beta} - \frac{1}{D} \delta_{\alpha \beta} u_{\gamma \gamma})^2 + B (u_{\alpha \alpha} + \frac{1}{2 B} \tau)^2 - \frac{\tau^2}{2 B} 
%\ee 
%One can then write
$u_{\alpha}= \tilde u_{\alpha} -  \frac{1}{2 B D} \tau x_\alpha$, the energy density becomes
\be
\mu (\tilde u_{\alpha \beta} - \frac{1}{D} \delta_{\alpha \beta} \tilde u_{\gamma \gamma})^2 + B (\tilde u_{\alpha \alpha})^2 - \frac{\tau^2}{2 B}\ . 
\ee
Here $\tilde u_\alpha$ is the "centered" phonon field and $\tilde u_{\alpha\beta}= \frac{1}{2} (\partial_\alpha u_\beta + \partial_\beta \tilde u_\alpha +  \partial_\alpha \vec h \cdot \partial_\beta \vec h)$ its associated nonlinear strain. The new parameterization for the positions in the embedding space is thus 
\be
\vec r_{\alpha} = [\zeta x_\alpha + \tilde u_\alpha] \vec e_\alpha
+ \vec h   \quad , \quad \zeta = 1 -   \frac{1}{2 B D} \tau\ .
\ee  
In fact $\zeta$ is also the order parameter of the crumpling transition, and
the term $\tau u_{\alpha \alpha}$ is identical to the term $\frac{1}{2} t (\partial_\alpha \vec r)^2$
at the crumpling transition\cite{PKN}. 

%$\tau$ changes $T_c$ 

%Take a membrane at $T$ one with $T_{c_1}$ the other with $T_{c_2}$. 

\section{H. Estimate of the bare critical buckling stress, $\sigma_c$}

As discussed in the main text, the critical value of the bare buckling
stress $\sigma_c$ is determined by the parameter $\alpha_1$, and in
the presence of broken rotational symmetry of the embedding space the
coupling $\alpha_1$ and thus critical stress $\sigma_c$ are nonzero in
thermodynamic limit. This constrasts qualitatively with the the
critical buckling stress of Euler buckling, that is set by the finite
system size and thus vanishes in the thermodynamic limit.
To estimate $\alpha_1$, we can consider two models of breaking
embedding space rotational symmetry.

For model A, we consider a membrane in a nematic solvent with
homeotropic nematic alignment of the director $\hat n$, with the
membrane's normal $\hat N$, given by energy density (per unit of
membrane's area) $\varepsilon = c (\hat n\cdot \hat N)^2$. Now,
tilting of the membrane normal relative to the far field director
field $\hat n_\infty = \hat z$, will create a long range power-law
distortion\cite{LehenyDiskNematic}. Generically the distortion at
angle $\theta$ will be on the scale of membrane's linear dimension
$L$, controlled by the Frank free energy with elastic Frank constant
$K$ (with units of energy/length) and proportional to
$\cos^2\theta$. The associated coefficient $c$ is thus obtained by
integrating the nematic distortion strains $(\theta/L)^2$ over
associated volume $L^3$. The corresponding energy density (per unit of
membrane area $L^2$) is given by
$\varepsilon = \frac{1}{2} (K/L) \theta^2$. Thus
$c = \alpha_1 = \sigma_c = K/L$. A typical scale for
$K \sim 1 \text{pico-Newtons} = 10 \text{eV/micron}$, which for a $10$
micron membrane (e.g., graphene flake) gives,
\begin{equation}
  \sigma_c \sim 1 \text{eV/micron}^2.
\end{equation}

In model B, we consider an alignment of ferroelectric membrane with an
external electric field $\vec E$. This corresponds to energy density
$\vec p\cdot \vec E$, where $\vec p$ is electric dipole 2D density. In
a ferroelectric crystal 3D dipole density magnitude $P$ is roughly
given by $P = 10$ micro-Coulombs/cm$^2$ = $10^{-1}$
Coulomb/m$^2$\cite{grapheneNematicClark}. For an Angstrom thick
membrane (like graphene) this gives $p = P \times 10^{-10}$m = $10^{-11}$
Coulomb/m = $10$ e/micron. For a typical switching field of
$E \sim 10^6$ V/m, this gives $\vec p \cdot\vec E = 10$ eV/micron$^2$,
about $10$ times larger $\sigma_c$ than for model A estimate above.

One may worry that this critical stress value is shifted by the
thermal fluctuation correction $\delta\gamma$, that we computed in
Sec. F.2, and estimate to be given by
$\delta\gamma \sim T \frac{\Lambda^{d-2}}{\kappa} \lambda_2$. Noting
that like $\alpha_1$, estimated above, $\lambda_2$ is associated with
the rotational symmetry breaking of the embedding space, we thus
expect $\lambda_2\approx\alpha_1$. We then estimate fluctuation shift
in $\gamma$ in a 2D graphene membrane (characterized by
$\kappa\approx 1$ eV) to be,
\begin{equation}
  \delta\gamma = \alpha_1 T/\kappa\approx \alpha_1/40\ll \alpha_1.
  \end{equation}
We thus conclude that we can neglect the fluctuations shift in
$\gamma$ in estimating the critical value of the buckling stress
$\sigma_c$ given above and in the main text.

%Let us focus here on $D=2$. The total area of the membrane is
%\bea
%A= \int d^D x \sqrt{g} \quad , \quad g= \det( \partial_\alpha \vec r \cdot \partial_\beta \vec r) 
%\eea 
%One has, using the definition of the nonlinear strain tensor,
%\bea
%g =  \det( \delta_{\alpha \beta} + 2 u_{\alpha \beta}) 
%\eea 
%Hence
%\bea
%A = \int d^D x  
%\eea 

%Consider the projected area of the membrane in some direction $\vec a$. 

%Let us specify $D=2$. The normal vector is
%\be
%\vec n = \frac{ \partial_1 \vec r \times \partial_2 \vec r }{|| \partial_1 \vec r \times \partial_2 \vec r ||}
%\ee 

\end{widetext}


\begin{thebibliography}{}
\bibitem{gentle} 
Note that this orientational field is a softer 
perturbation than the pinning effect, that may be induced by a substrate.
\bibitem{footnote2}
This finite value can be tuned to zero at a multicritical
point, where the anomalous elasticity is restored.
\bibitem{footnote5}
It is important to contrast this rotational symmetry-breaking stress
applied in the $xy$ plane with the ``stress'' $\tau$ imposed in the
actual plane of the membrane, without breaking the rotational symmetry
of the embedding space\cite{CrumplingBucklingGuitter}, by e.g., confining the membrane in a spherically symmetric potential.  Namely, we note that the
added $\tau u_{\alpha\alpha}$ (instead of $\sigma\partial_\alpha u_\alpha$) can
be eliminated by absorbing it into $u_{\alpha\alpha}^2\rightarrow
(u_{\alpha\alpha} + \sigma/\lambda)^2$ and redefining the crumpling
order parameter $\zeta$.
\bibitem{footnote3} 
Since the map from $(\mu,\lambda)$ to
$(w_1,w_3)$ is not bijective for $w_1=\mu=0$, to distinguish the 
first and third fixed points one needs to first write the RG flow using $\mu,\lambda$
and, second, look for fixed points or, alternatively, to carefully take limits.
\bibitem{GaussCurv} The other geometrical invariant is the intrinsic
  Gaussian curvature. However the integral of the Gaussian curvature
  can be shown to be a topological invariant proportional to the Euler
  character or the genus of the manifold and therefore will not be
  considered in this paper.
\bibitem{ModeCount} In the isotropic description the theory is
  $O(d)\times O(D)$ invariant and there are a total of $Dd$
  independent deformational modes.  In the spontaneously broken flat
  phase the symmetry is reduced to $O(d_c)\times O(D)\times O(D)$. The
  general principles require that when a group G is broken down to a
  group H, the number of massless Goldstone modes is equal to the
  dimension of the coset space G/H, here $O(d)/O(D)/O(d_c)$. We
  therefore expect $Dd_c$ massless modes and $D^2$ massive ones
  corresponding to $\pt\as\h$ Goldstone modes and $\pt\as u\bs$
  elastic deformations, respectively.
\bibitem{CommentDist} As usual at long wavelengths the higher moments
  are expected to be irrelevant in renormalization group sense.
\bibitem{comment4dc} It is not clear whether there is an underlying
  physical reason for this $d_c\rightarrow 4d_c$ property or that it
  is just an accident special to the SCSA.
\end{thebibliography}
\end{document}